\documentclass[aip,reprint]{revtex4-2}

\usepackage[separate-uncertainty=true,multi-part-units=single]{siunitx}
\usepackage{todonotes}
\usepackage{chemformula}
\usepackage[caption=false]{subfig}

\newcommand{\layer}[2]{\allowbreak\ch{#1}(#2)}
\newcommand{\HEB}[1]{$H_{\mathrm{EB}}#1$}
\newcommand{\mytilde}{\raise.17ex\hbox{$\scriptstyle\mathtt{\sim}$}}

\begin{document}

\title{Understanding field-free single-shot laser-induced reversal of exchange bias}

\author{Floris J.F. van Riel}
\email{f.j.f.v.riel@tue.nl}
\author{Stoyan M. Vercruysse}
\author{Bert Koopmans}
\author{Diana C. Leitao}
\affiliation{Department of Applied Physics, Eindhoven University of Technology, The Netherlands}

\date{\today}

\begin{abstract}
Exchange bias is applied ubiquitously throughout the spintronics industry for pinning magnetic moment in an orientation robust against external magnetic field disturbances. Enabling efficient manipulation of the exchange-biased pinning direction is key to add or advance device functionality. Conventional methods for achieving this reorientation commonly require long heating times and assisting magnetic fields, thus limiting speed and versatility. Here, we demonstrate a field-free \SI{180}{\degree} reorientation of perpendicular exchange bias in a thin film system integrating \ch{Co}/\ch{Gd} layers with \ch{IrMn}. Single femtosecond laser pulses trigger a reversal of the pinning direction, while preserving the unidirectionality required for referencing. We formulate a theoretical microscopic framework involving a wide range of timescales that accurately reproduces the experimental outcomes. The understanding of the underlying mechanisms delivers guiding principles for designing reprogrammable spintronic devices.
\end{abstract}


\maketitle

\section*{Introduction}
State-of-the-art magnetic memory \cite{Du2023} and spintronic sensors \cite{Dieny2020,Becker2022,Becker2019} fundamentally utilize exchange bias for referencing. Arising from exchange coupling at an antiferromagnetic/ferromagnetic thin film interface, exchange bias induces a unidirectional anisotropy that pins the magnetization in the ferromagnet and makes it robust against external magnetic field perturbations \cite{Freitas2016,Meiklejohn1956,Nogues1999}. The exchange bias-induced reference axis is programmed during fabrication by field-cooling \cite{Meiklejohn1956}, i.e., heating above the blocking temperature $T_b$ (where exchange bias vanishes) and cooling down in the presence of a magnetic field, saturating the ferromagnetic layers and uniformly setting the exchange bias. In many applications though, local control of the exchange bias orientation is required, such as for memory \cite{Du2023}, neuromorphic computing \cite{Hasan2023}, 2D/3D vector and low-noise sensors \cite{Becker2022,Ueberschar2015b,Fujiwara2018} or terahertz emitters \cite{Wang2023a}.

In this work, we locally reorient the exchange bias with pulsed laser light. Focusing light allows for targeted exchange bias setting in specific elements \cite{Berthold2014,Almeida2015,Ueberschar2015b}. Furthermore, our approach reorients the exchange bias without the assistance of an external magnetic field, by employing all-optical helicity-independent magnetization switching (AOS) with individual femtosecond laser pulses \cite{Radu2011,Ostler2012,Lalieu2017}. Unlike electrical approaches (e.g., based on Joule heating \cite{Cao2010,Papusoi2008,Wang2024}, spin-orbit torques \cite{Du2023,Wang2022b,Lin2019,Yun2024,Guo2024a,Liu2023,Chen2023} or strain \cite{Qi2024a,Su2020,Wang2023}) or self-assembling structures \cite{Becker2019,Becker2022}, laser illumination does not impose any extra connections and maintains compatibility with monolithic integrated circuits.

AOS was first observed in perpendicularly magnetized amorphous ferrimagnets of transition metals (\ch{Fe} and/or \ch{Co}) alloyed with the rare-earth element \ch{Gd}, where an ultrashort femtosecond laser pulse toggle-switches the magnetization for sufficiently energetic pulses \cite{Radu2011,Ostler2012}. It was discovered by Guo et al. \cite{Guo2024} that when such alloys are exchange biased by antiferromagnetic \ch{Ir_{0.2}Mn_{0.8}}, a single laser pulse reverses both the magnetization and the exchange bias, although crucially losing the unidirectionality of the pinned layer. Our stacks utilize synthetic ferrimagnets (i.e. multilayers containing elemental \ch{Co} and \ch{Gd}) displaying AOS \cite{Lalieu2017,Beens2019}, designed to be compatible with multilayer spintronic devices \cite{Wang2022a,Li2023}. Multilayers also allow for enhanced composition tunability compared to their alloy counterparts \cite{Beens2019}.

Experimentally, we characterize the laser-induced exchange bias reversal process with a data set of millions of magnetic hysteresis loops that map out a state space of three key parameters: laser pulse energy, \ch{IrMn} layer thickness and externally applied magnetic field. As our results demonstrate, \ch{Co}/\ch{Gd} multilayers show performance benefits for laser-induced exchange bias reversal compared with literature on alloys \cite{Guo2024}. The differences are explained by a modeling approach involving a wide range of time scales, describing magnetization dynamics on the picosecond timescale and thermally assisted reversal of antiferromagnetic grains up to the millisecond timescale, revealing the importance of critical temperatures and thermal relaxation times. Combining the experimental and simulated results, we give a brief perspective on the capabilities and limitations for future applied research on reprogrammable devices.

\begin{figure*}%
	\centering%
	\includegraphics[width=\textwidth]{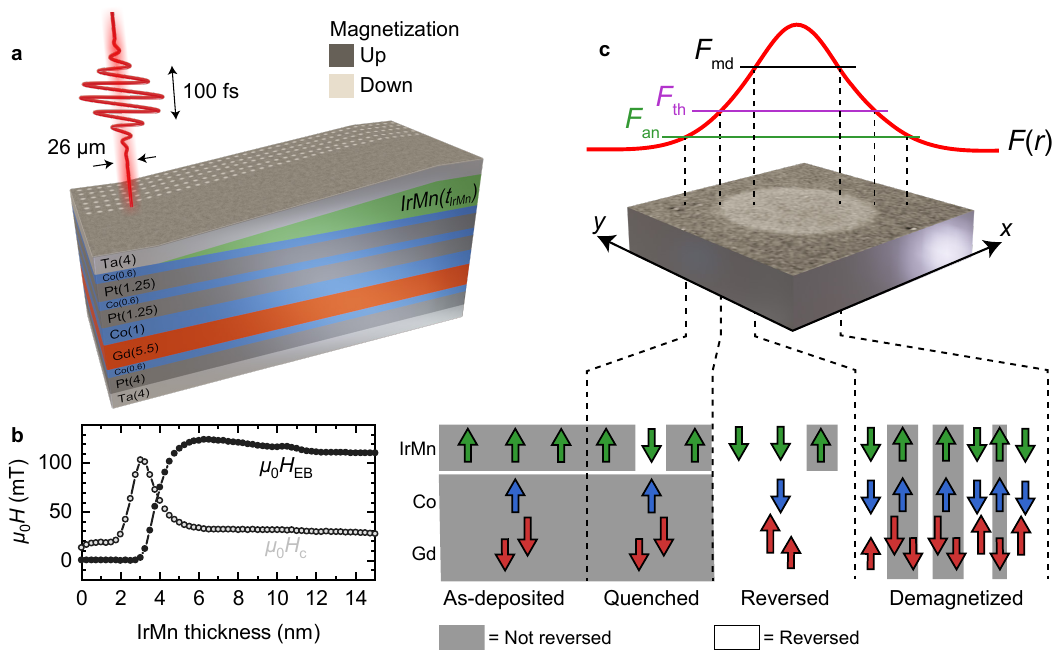}%
	\caption{\label{fig:1}\textbf{Design of samples and experiments. a} Overview of the layer stacking and thicknesses (in nanometers) as used in the experiments. The antiferromagnetic \ch{IrMn} layer is grown with a linearly varying thickness $t_{\ch{IrMn}}$. A pulsed laser is swept over the sample writing regions with reversed exchange bias under various conditions (pulse energy, external magnetic field). \textbf{b} Mapping of exchange bias (\HEB{}) and coercivity ($H_c$) as a function of \ch{IrMn} thickness, measured via the magneto-optic Kerr effect by sweeping a laser bundle along the wedge and measuring $M$-$H$ loops at each position. The sample used for the measurement is identical in structure to the one schematically shown in \textbf{a}, but with $t_{\ch{IrMn}}$ ranging to \SI{25}{\nano\meter} to eliminate wedge edge effects. \textbf{c} Close-up of a region of the sample that was illuminated by a laser pulse cf. \textbf{a}. The red bell curve illustrates the fluence profile $F(r)$ as a function of distance $r$ from the peak center. The horizontal lines correspond to various thresholds (annealing threshold $F_{\mathrm{an}}$, AOS threshold $F_{\mathrm{th}}$ and demagnetizing threshold $F_{\mathrm{md}}$) that delimit regions of different amounts of \HEB{} reversal, ranging from no \HEB{} reversal (as-deposited) to only quenched \HEB{} but not yet reversed, to reversed and finally to thermally demagnetized.}%
\end{figure*}%

\section*{Results}
\subsection*{Demonstration of laser-induced exchange bias reversal}
The sputter-deposited multilayer stack as used in the experiments is schematically shown in Fig.~\ref{fig:1}a (see Methods for details). The antiferromagnetic \ch{IrMn} layer is deposited with a continuous thickness gradient $t_{\ch{IrMn}}$, such that thicknesses in the range between \SIrange[range-units=single,range-phrase=\textrm{ and }]{0}{15}{\nano\meter} may be investigated. Exchange bias is characterized by the effective perpendicular exchange bias field \HEB{}, extracted from the $M$-$H$ loop offset. Figure~\ref{fig:1}b depicts how \HEB{} varies with $t_{\ch{IrMn}}$, with a clear onset of exchange bias at around \SI{4}{\nano\meter}, preceded by a characteristic peak in the coercivity \cite{Yang2019}. As Fig.~\ref{fig:1}a also illustrates, single femtosecond laser pulses are fired at various positions (i.e., various $t_{\ch{IrMn}}$) and under different conditions (pulse energy and externally applied perpendicular magnetic field), allowing for the exploration of a three-parameter state space (see Methods).

\begin{figure}%
	\centering%
	\includegraphics[width=\columnwidth]{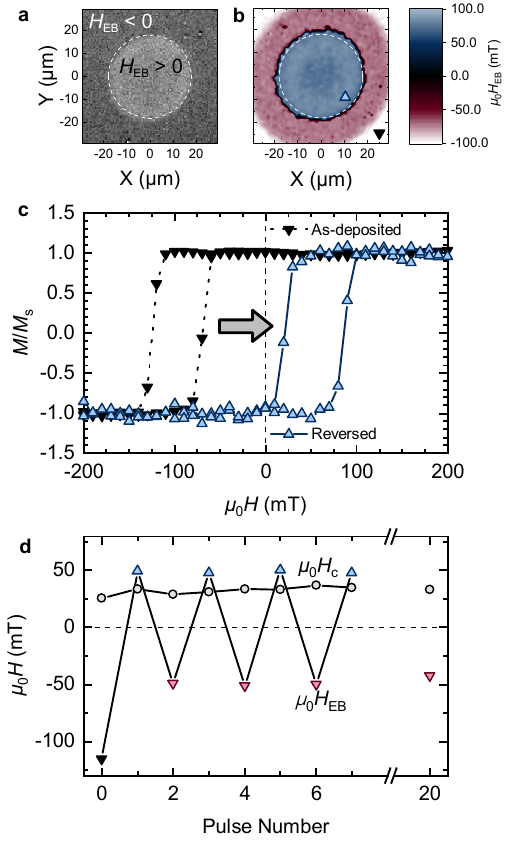}%
	\caption{\label{fig:2}\textbf{Demonstration of laser-induced field-free exchange bias reversal. a} A Kerr microscopy image of a particular area of the sample at $t_{\ch{IrMn}}=\SI{4.0}{\nano\meter}$ that was illuminated with a single \SI{600}{\nano\joule} laser pulse. \textbf{b} Color map of the spot in \textbf{a}, with colors indicating the value of \HEB{} throughout the spot. The triangles indicate the locations where the respective loops in \textbf{c} are extracted from. \textbf{c} Demonstration of laser-induced \HEB{} reversal at $\SI{28}{\milli\joule\per\square\centi\meter}$ fluence, visible in the hysteresis curve as a sign reversal of the loop shift from the as-deposited state (in black) to the reversed state (in blue). \textbf{d} \HEB{} after \numrange[range-phrase=\textrm{ to }]{0}{7} pulses and after $20$ pulses, showing a reproducible toggling behavior after the first pulse between two levels of \HEB{} with opposite sign.}%
\end{figure}

The Gaussian spatial fluence profile (energy per unit area, see Fig.~\ref{fig:1}c), gives rise to regions with different amounts of \HEB{} reversal in the illuminated area. An experimental example of this is shown in Fig.~\ref{fig:2} %
for our optimized stacks. Figure~\ref{fig:2}a depicts a Kerr microscopy image of the magnetic state after \SI{600}{\nano\joule} laser excitation in field-free conditions. The same spot is shown in Fig.~\ref{fig:2}b, where the exchange bias field is extracted at each $(x,y)$ location from the local hysteresis loop. A clear non-uniform $(x,y)$-dependence of \HEB{} is observed, the interpretation of which is depicted in Fig.~\ref{fig:1}c. For fluence levels above the annealing threshold, $F_{\mathrm{an}}$, the temperature after excitation exceeds the antiferromagnetic N\'{e}el temperature $T_N$ before the exchange bias is restored. Due to heat-driven stochastic effects on the granular polycrystalline texture of the \ch{IrMn} layer \cite{Jenkins2019,Jenkins2020,Jenkins2021,Szunyogh2011}, smaller more mobile grains may relax in a state where the uncompensated interfacial spins favor \HEB{} reversal, while larger more stable grains are more likely to still favor the initial \HEB{}. This results in a reduction of the \HEB{} magnitude compared to the as-deposited value. For fluence levels above the AOS threshold $F_{\mathrm{th}}$ where the ferromagnetic Curie temperature $T_C$ is exceeded and AOS takes place, the reversed magnetization is imprinted into the antiferromagnet. In this fluence regime the sign of \HEB{} is reversed, but it maintains the same reduction in magnitude as between $F_{\mathrm{an}}$ and $F_{\mathrm{th}}$. The subsections below further explore these claims.

$M$-$H$ loops from Fig.~\ref{fig:2}b at two distinct $(x,y)$ positions are shown in Fig.~\ref{fig:2}c: one with laser excitation measured within the spot at a radius corresponding to a fluence of $\SI{28}{\milli\joule\per\square\centi\meter}$, and one without laser excitation measured far away from the spot. From Fig.~\ref{fig:2}c it is readily observed that the exchange bias sign was reversed from negative to positive. More notably and crucially for application in referencing, the remanent state at $H=0$ remains unidirectional, i.e. $|H_{\mathrm{EB}}|>H_{c}$ such that only one orientation of $M$ (up or down) is stable at zero field, and the stable state toggles between up and down with each successive laser pulse (Fig.~\ref{fig:2}d). Moreover, we confidently state that indeed AOS in the \ch{Co}/\ch{Gd} layer successfully propagates to the \ch{Co}/\ch{Pt} layers and ultimately drives the exchange bias reversal.

Our interpretation from Fig.~\ref{fig:1}c predicts a repeatable reversal process without degradation upon successive laser pulses. As a function of the number of pulses, it was indeed observed that only the first pulse induces a loss of \HEB{}, after which successive pulses toggle between two stable states (see Fig.~\ref{fig:2}d). For this sample we tested the stability up to $20$ successive switches and it always retains the unidirectionality condition, with $|H_{\mathrm{EB}}/H_{c}|\approx1.4$. This is an important requirement for applications whose operation may rely on continuous repeated toggle switches.

\subsection*{Fluence dependence of \HEB{} reversal}
One key parameter we can control to shed more light on the reversal mechanism is the intensity of the laser pulse, characterized by its fluence. At constant pulse length, the fluence determines the temperature evolution in the layers. Whether or not the system temperature then exceeds certain critical temperatures (like Curie, blocking or N\'{e}el temperatures) and for how long those temperatures are exceeded influences how the magnetic state dynamically evolves. For exchange bias reversal, where various critical temperatures play a role, a very rich fluence dependence can then be expected.

\begin{figure*}%
	\centering%
	\includegraphics[width=\textwidth]{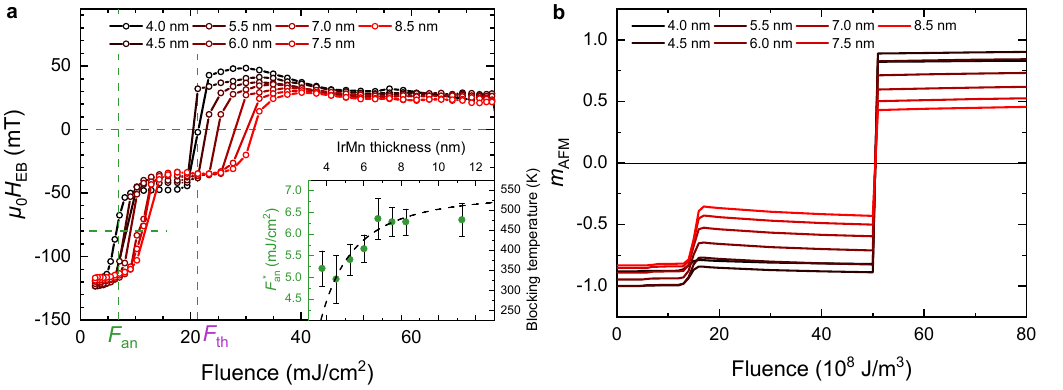}%
	\caption{\label{fig:3}\textbf{Controlling exchange bias reversal with fluence and \ch{IrMn} thickness. a} Dependence of the reversed exchange bias on the laser fluence, for various thicknesses of the \ch{IrMn} layer and for a fixed pulse energy of \SI{600}{\nano\joule}. The green and purple dashed lines respectively indicate the criteria used for determining $F_{\mathrm{an}}$ and $F_{\mathrm{th}}$, specifically for \SI{4.0}{\nano\meter}. The inset shows the annealing fluence $F_\mathrm{an}^{*}$ (green circles) analogous to \textbf{a}, but background-corrected for the attenuation in the \ch{IrMn} (\ref{sec:skindepth}). The black dashed line is a phenomenological fit of experimental blocking temperature data (\ref{sec:block}), whose behavior displays a similar trend as a function of \ch{IrMn} thickness. Error bar widths are explained in \ref{sec:skindepth}. \textbf{b} Simulations (M3TM + Arrhenius) reproducing the experimental data in \textbf{a} (note that the fluence is expressed per unit of volume instead of area).}%
\end{figure*}

From color maps such as in Fig.~\ref{fig:2}b, the dependence of exchange bias reversal on laser fluence can be extracted. This is shown in Fig.~\ref{fig:3}a %
for various thicknesses of the \ch{IrMn} layer. Over the range of fluences investigated we did not observe a monotonic increase of the exchange bias magnitude with increasing fluence. Rather, regions of both increasing and decreasing trends appear in the fluence dependence of \HEB{}. N.B., we verified that the regions indeed scale with the local fluence level (see Methods).

A number of important conclusions can be drawn from the fluence dependence in Fig.~\ref{fig:3}a. First of all, two steps can be distinguished: one located around \SI{6}{\milli\joule\per\square\centi\meter} and the other appearing for fluences above \SI{20}{\milli\joule\per\square\centi\meter}. This behavior can be understood as follows. The critical temperature of bulk \ch{IrMn} is approximately $T_N=\SI{700}{\kelvin}$, while for bulk \ch{Co} it is $T_C=\SI{1388}{\kelvin}$. This means that as the fluence is increased, the first critical temperature to be overcome by the system is that of the antiferromagnet, resulting in a quasi-annealing effect. In Fig.~\ref{fig:3}a this is represented by the threshold $F_{\mathrm{an}}$ and a decrease in magnitude of \HEB{} from around \SI{-120}{\milli\tesla} to \SI{-50}{\milli\tesla}. Only when the fluence is further increased the exchange bias will be reversed through zero, since at that point the threshold fluence $F_{\mathrm{th}}$ for switching the ferromagnetic layers is overcome.

We performed simulations bridging processes happening at different timescales: picosecond timescale magnetization dynamics is modeled by the layered microscopic three-temperature model (M3TM) \cite{Koopmans2010,Beens2019}. The model solves for the temperatures of the electron ($T_e$) and phonon ($T_p$) systems to calculate the magnetization dynamics of each atomic layer with equation (\ref{eq:m3tmmain}). This is followed by a slow millisecond temperature relaxation \cite{Khamtawi2023}. The simulations predict the magnitude of \HEB{} after relaxation for a particular grain size distribution with the Arrhenius law in equation (\ref{eq:transrates}), and it captures the outcome in a parameter $m_{\mathrm{AFM}}$ with $m_{\mathrm{AFM}}=\pm1$ corresponding to the largest positive/negative \HEB{} value. All features from the fluence dependence in Fig.~\ref{fig:3}a are reproduced, as shown in Fig.~\ref{fig:3}b. Furthermore, based on the simulations we understand the broadening of the transition in \HEB{} around $F_{\mathrm{an}}$ to be a consequence of the grain size distribution, allowing smaller grains to reverse already at lower fluences (when the system temperature exceeds $T_b$ but not yet $T_N$) than larger grains. No such broadening is observed around $F_{\mathrm{th}}$ since AOS only has a binary outcome. Note also that the magnitude of \HEB{} above and below $F_{\mathrm{th}}$ is always equal, for both the experiment as well as the simulation. Equation (\ref{eq:transrates}) indirectly implies that the ratio between attempt frequency and temperature relaxation time is of great importance, predicting that a slower temperature relaxation will result in a larger fraction of \HEB{} to be reset, analogous to conventional field-cooling. This notion hints at heat management strategies that may lead to further enhancement of the reversed \HEB{} fraction.

The results from Guo et al. \cite{Guo2024} fit the picture that our interpretation sketches, with a clear threshold fluence required to induce the \HEB{} switch. Their results lack the plateau between $F_{\mathrm{an}}$ and $F_{\mathrm{th}}$ though. One explanation for this discrepancy is that their systems contain only a ferrimagnet with a lower expected Curie temperature than our \ch{Pt}/\ch{Co} multilayers and their stacks are reversed with \ch{IrMn} on the bottom, leading to a situation where $F_{\mathrm{an}}>F_{\mathrm{th}}$. To further support our findings, we make use of static externally applied magnetic fields to separate contributions from the ferromagnetic and antiferromagnetic layers, as will be discussed in the next subsection.

A second observation deduced from Fig.~\ref{fig:3}a is that the absolute threshold fluences are shifted as the antiferromagnetic thickness is increased. Since we illuminate from the top, it is expected that the threshold fluence for switching the ferromagnets increases exponentially with antiferromagnetic thickness due to attenuation in the \ch{IrMn}. By fitting this exponential increase we corrected the data of the antiferromagnetic threshold to filter out the screening contribution (see \ref{sec:skindepth}). The attenuation-corrected data, shown in the inset of Fig.~\ref{fig:3}a, is clearly not monotonic, but rather it saturates at around $t_{\ch{IrMn}}=\SI{8}{\nano\meter}$. This behavior is in agreement with the dependence of blocking temperature on \ch{IrMn} thickness, that we separately measured (\ref{sec:block}) and also plotted in the inset of Fig.~\ref{fig:3}a. This further indicates that indeed $F_{\mathrm{an}}$ is related to a transition in the antiferromagnet.

\subsection*{Magnetic field dependence of \HEB{} reversal}\label{sec:field}%
\begin{figure}%
	\centering%
	\includegraphics[width=\columnwidth]{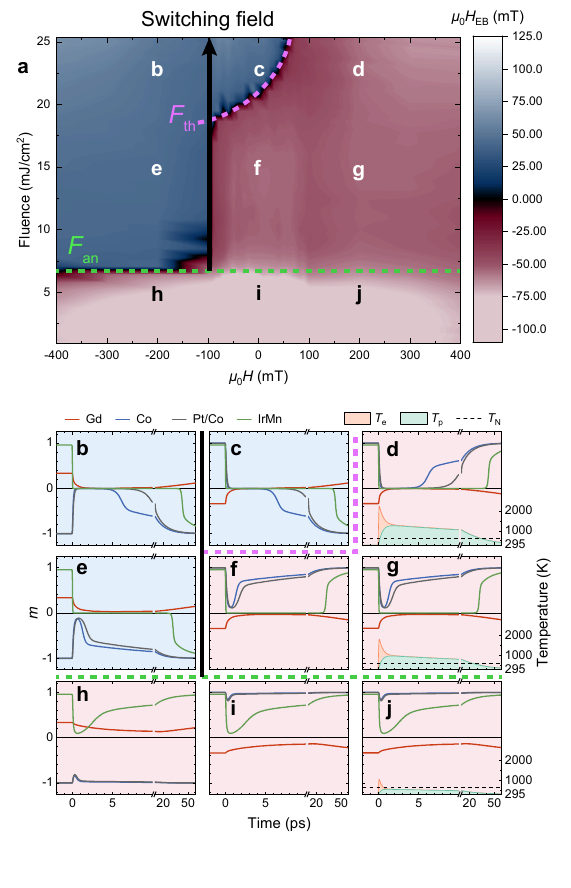}%
	\caption{\label{fig:4}\textbf{Controlling exchange bias reversal with fluence and external magnetic fields. a} Measured state diagram of \HEB{} as a function of the laser fluence and the externally applied field for $t_{\ch{IrMn}}=\SI{4}{\nano\meter}$. The threshold fluences $F_{\mathrm{an}}$ and $F_{\mathrm{th}}$ are indicated with the green and pink dashed lines, respectively. The switching field is indicated by the black arrow. \textbf{b}-\textbf{j} Simulated time evolutions of the \ch{Gd} (red), \ch{Co} (blue), \ch{Pt}/\ch{Co} (gray) and \ch{IrMn} (green) layer magnetization, for the conditions corresponding to the letters indicated in \textbf{a} and where the background colors use the same mapping. The green and pink dashed lines and black solid line delimit the same regions as in \textbf{a}. For the three different fluence values the temperature evolution of the electron temperature $T_e$ and the phonon temperature $T_p$ are shown by the shaded areas in panels \textbf{d}, \textbf{g} and \textbf{j}. The black dashed lines in those panels represent the N\'{e}el temperature of \ch{IrMn}.}%
\end{figure}%
A second key parameter in our investigation is the influence of an external magnetic field on the exchange bias reversal. Even though field-free reversal is the ultimate benchmark, magnetic fields are used here as a tool to manipulate the reversal process and deepen our understanding of the underlying mechanisms. We distinguish here between assisting fields that promote the reversal of \HEB{}, and hindering fields that stabilize the as-deposited \HEB{}. In a study by Peeters et al. \cite{Peeters2022} on \ch{Co}/\ch{Gd} bilayers it was shown that magnetization reversal always takes place within a few picoseconds after laser excitation independent of the applied field, but also that a hindering field shortens the time spent in the reversed state before remagnetizing to the initial state.

Figure~\ref{fig:4}a shows how experimentally the external field influences the exchange bias reversal for $t_{\ch{IrMn}}=\SI{4}{\nano\meter}$. As is expected for fluences above $F_{\mathrm{an}}$, the exchange bias is quasi-annealed by the external field. The point at which the transition between positive and negative \HEB{} takes place is determined by the switching field, i.e., the field required to overcome \HEB{} and the coercivity. For some interesting points in the state diagram we simulated the magnetization dynamics as marked by Fig.~\ref{fig:4}b-j using the layered M3TM approach. Note that here we only aim at reproducing the range of switching conditions and not the magnitude of \HEB{}. The latter includes the slower thermally assisted processes as was done to generate Fig.~\ref{fig:3}b.

Using magnetic fields allows us to distinguish between thresholds relating to processes in the ferromagnet or in the antiferromagnet. As is readily observed in Fig.~\ref{fig:4}a, the threshold indicated by $F_{\mathrm{an}}\approx\SI{6}{\milli\joule\per\square\centi\meter}$ is independent of the field, confirming that this identifies the peak temperature exceeding the N\'{e}el temperature of the antiferromagnet. This also follows from simulations (Fig.~\ref{fig:4}h, \ref{fig:4}i and \ref{fig:4}j), where below this threshold the phonon temperature will not exceed the N\'{e}el temperature of the antiferromagnet, which thus remagnetizes in the original direction (green lines in Fig.~\ref{fig:4}b-j). On the other hand, the threshold boundary at \SI{20}{\milli\joule\per\square\centi\meter} is highly dependent on the field. Its shape corresponds to what was predicted by Peeters et al. \cite{Peeters2022}, i.e., an increase of the AOS threshold fluence for large enough hindering fields. This provides additional evidence for our interpretation of $F_{\mathrm{an}}$ being an annealing-like threshold and $F_{\mathrm{th}}$ being an AOS threshold.

The simulated temperature evolutions for $T_e$ and $T_p$ are plotted for three different fluence values in panels \ref{fig:4}d, \ref{fig:4}g and \ref{fig:4}j. The \ch{Gd} layers, having a Curie temperature below room temperature, only obtain an induced magnetization via exchange coupling at the interface with \ch{Co}, hence the lower magnitude of the red lines in Fig.~\ref{fig:4}b-j. It can also be seen that the ferromagnetic layers (blue for \ch{Co} and gray for \ch{Pt}/\ch{Co}) are initialized in the negative direction in panels \ref{fig:4}b, \ref{fig:4}e and \ref{fig:4}h to mimic the effect of the static switching field. The green lines for \ch{IrMn}, representing the N\'{e}el vector, are not reversed by fluences below $F_{\mathrm{an}}$ (panels \ref{fig:4}h-j) because $T_p$ does not exceed the N\'{e}el temperature. Above $F_{\mathrm{an}}$ and with a large enough field applied, the ferromagnetic layers de- and remagnetize along the field orientation and drag the \ch{IrMn} along, bringing the \ch{IrMn} in the switched state for assisting fields (panels \ref{fig:4}b and \ref{fig:4}e) and in the original state for hindering fields (panels \ref{fig:4}d and \ref{fig:4}g). In the field-free case, a distinction is made between fluences below the AOS threshold $F_{\mathrm{th}}$ (panel \ref{fig:4}f) where no reversal takes place, and fluences above $F_{\mathrm{th}}$ (panel \ref{fig:4}c) where AOS takes place and the \ch{IrMn} is switched as well. All in all, the simulations thus qualitatively reproduce all features from the experimental data in Fig.~\ref{fig:4}a, supporting our interpretation of the behavior on the ultrafast timescales.

\section*{Discussion and outlook}
We have demonstrated the field-free ultrafast reversal of exchange bias in our optimized stacks upon excitation with a single femtosecond laser pulse. It was revealed that the magnitude of \HEB{} reduces upon the first pulse, but is maintained upon successive pulses and always satisfies the undirectionality condition $|H_{\mathrm{EB}}|>H_{c}$. We explored a state space of three parameters, namely antiferromagnetic layer thickness, laser fluence and applied magnetic field. Experimentally, we identified two threshold fluences in the reversal process: one related to the N\'{e}el temperature of the \ch{IrMn} and one to the Curie temperature of the \ch{Co}. With simulations we reproduced these threshold fluences using a microscopic model of the magnetization dynamics, and we were able to predict the magnitude of \HEB{} after temperature relaxation using an Arrhenius description.

It was found that short timescale exchange-driven processes determine whether a switch takes place or not, but long timescale heat-driven processes govern what value of \HEB{} is retained. For granular antiferromagnets such as \ch{IrMn} it is indeed known that their exchange bias may evolve slowly at elevated temperatures,  even at room temperature \cite{Migliorini2018}. This is an effect we also observed in idle samples, where remeasuring the same switched areas after a few months showed an increase in the value of \HEB{}. Thus, if the magnitude of \HEB{} after laser-induced switching is to be maximized, one may accelerate this spontaneous setting process by keeping the system at a higher elevated temperature for longer. Controlling the heat dissipation and with it the temperature relaxation rate is therefore key to further optimize the \HEB{} reversal performance. It is emphasized that despite the long relaxation time, the excitation time remains ultrashort and thus does not necessarily limit the speed of operation.

From a material perspective, even though the \ch{Co}/\ch{Gd} multilayers display the required performance for application in devices \cite{Wang2022a}, the use of rare-earth elements like \ch{Gd} is undesirable because of its scarcity \cite{Zhao2023} and rapid degradation \cite{Kools2023}. Resorting to rare-earth-free material platforms like Heusler alloys that also display helicity-independent AOS \cite{Banerjee2020} is one alternative. Further stack simplifications can be achieved owing to the recently emerging area of all-antiferromagnetic spintronics \cite{Qin2023,Qi2024} and their unexplored magneto-optic interactions \cite{Adamantopoulos2024,Kimel2024}. Even if rare-earths can not be avoided, abolishing the requirement for strong perpendicular anisotropy still relaxes some material and layer thickness constraints. Transitioning to exchange-biased in-plane \ch{Co}/\ch{Gd} systems, which also display equally efficient AOS \cite{Lin2023}, is therefore a logical next step. Nevertheless, the merits of perpendicular magnetization as targeted by the present work are strong in many areas of spintronics \cite{Dieny2020}. We are therefore confident that the technique presented in this work has merit for application in novel innovative spintronic devices.

\section*{Methods}\label{sec:methods}%
\subsection*{Thin film deposition}
Samples were prepared with DC magnetron sputtering deposition at a base pressure of \SI{e-7}{\pascal}. A shutter situated between the target and the substrate was moved at constant speed during the deposition of \ch{Ir_{0.2}Mn_{0.8}} (composition hereafter omitted) to create a linear thickness gradient ${0<t_{\ch{IrMn}}<\SI{15}{\nano\meter}}$ across the sample, which has a slope of \SI{0.75}{\nano\meter\per\milli\meter}. The stack consists of \layer{Ta}{4}/\layer{Pt}{4}/\layer{Co}{0.6}/\layer{Gd}{5.5}/\layer{Co}{1}/[\layer{Pt}{1.25}/\layer{Co}{0.6}]$_{\mathrm{x}2}$/\layer{IrMn}{$t_{\ch{IrMn}}$}/\layer{Ta}{5} grown on a thermally oxidized \ch{Si}/\layer{SiO2}{100} substrate (thicknesses in nanometers, see Fig.~\ref{fig:1}a). All layers are deposited in the presence of a static \SI{100}{\milli\tesla} field. This ensures an exchange bias of up to \SI{120}{\milli\tesla} is set by the \ch{Co} during deposition (as a function of \ch{IrMn} thickness, see Fig.~\ref{fig:1}b) and eliminates the need for post-annealing. There are two main reasons we purposefully avoid annealing. Firstly, annealing these stacks is likely to result in heavy diffusion of the mobile \ch{Gd} atoms \cite{Vorobiov2015}. Secondly, annealing would enlarge the antiferromagnetic grains, making them more stable and ultimately less susceptible to reversal. The \ch{Pt}/\ch{Co} layers serve as a buffer preventing the \ch{Gd} from contaminating the \ch{Co}/\ch{IrMn} interface and compromising the exchange bias. More details about stack considerations are provided in \ref{sec:stack}.

\subsection*{Performing the laser excitation experiments}
In our experiments, the films were subjected to one or more Gaussian laser pulses (pulse length \SI{100}{\femto\second} at half maximum, \SI{700}{\nano\meter} wavelength) focused onto the sample at normal incidence (spot diameter \mytilde\SI{26}{\micro\meter} at half maximum, see \ref{sec:spotsize}). Sweeping the position along the \ch{IrMn} wedge allowed us to parameterize the \ch{IrMn} thickness, as is illustrated in Fig.~\ref{fig:1}a. Furthermore, we parameterized the peak intensity of the Gaussian laser pulse by means of a neutral density filter and the magnitude of a constant magnetic field perpendicular to the sample plane using the field as a tool to influence the AOS process, amounting to a three-dimensional state space. Characterization of the exchange bias was carried out using Kerr microscopy, with which $M$-$H$ loops were measured all throughout the spatial extent of the switched region, as shown in Fig.~\ref{fig:1}c. The effective exchange bias field distribution \HEB{(x,y)} is then extracted from the shift in the $M$-$H$ loops as in Fig.~\ref{fig:2}c. Relating the position $(x,y)$ back to the intensity distribution of the Gaussian laser pulse $F(r)$ allowed us to investigate the behavior of the exchange bias reversal as a function of the laser fluence. This procedure is repeated for all spots, amounting to a total of \mytilde3.4 million hysteresis loops that were processed.

We verified that there is a direct relation between the apparent thresholds in the ring structure of Fig.~\ref{fig:2}b and the local laser fluence. We did this by varying the peak intensity of the pulse and observing how the ring boundaries shift to lower or higher radii. Calculating the fluence dependence for many different peak intensities and then overlaying them, allowed us to conclude that indeed the regions appear at the same fluence levels regardless of the peak intensity of the laser pulse, justifying our direct conversion from radius to fluence.

\subsection*{Modeling ultrafast and slow dynamics}
We approach modeling the exchange bias setting from two angles: one concerning magnetization dynamics on the ultrashort timescales and another concerning stochastic thermally driven processes on the longer timescales. We bridge the two perspectives by using the magnetization state and temperature after the ultrafast part as an input for the slow dynamics. Details on the implementation can be found in \ref{sec:model}.

The first regime (within \SIrange[range-phrase=\textrm{ to },range-units=single]{2}{3}{\pico\second} after laser incidence) is described by the layered microscopic three-temperature model (M3TM) framework for modeling AOS of the ferromagnetic layers \cite{Koopmans2010,Beens2019}. It implements the laser absorption as an almost instantaneous increase in the electron temperature $T_e$, followed by an equilibration through electron-phonon scattering and an increase in the phonon temperature $T_p$. The magnetization dynamics are described in this framework by \cite{Beens2019}%
\begin{equation}\label{eq:m3tmmain}%
	\frac{\mathrm{d}m_{i}}{\mathrm{d}t}=R\left(m_{i-1},m_{i},m_{i+1},T_{e},T_{p}\right),%
\end{equation}%
where the constituent parts of the function $R$ are described in \ref{sec:model}. Interaction between different sublattice magnetizations ($m_{i}$, $m_{i\pm1}$) is mediated through electron-electron angular momentum scattering processes, which ultimately result in the reversal of the \ch{Co}/\ch{Gd} layers and its propagation to the \ch{IrMn} layer via interlayer exchange coupling. The whole reversal process takes place approximately in the first \SI{10}{\pico\second}, after which the system temperature reaches the N\'{e}el temperature of the \ch{IrMn} layer and the magnetization state is fed to the second part of the simulation.

In the second temporal regime of the simulation (up to one millisecond after laser excitation), the exchange bias setting is modeled with an Arrhenius description \cite{Khamtawi2023}. The antiferromagnet is modeled as a collection of non-interacting grains with log-normal distributed volumes and uniform N\'{e}el vectors and anisotropies. Using an attempt frequency of $\tau_{0}^{-1}=\SI{e10}{\per\second}$, the rate $\tau^{-1}$ at which a grain will reverse its orientation is calculated via Arrhenius' equation as%
\begin{equation}\label{eq:transrates}%
	\tau=\tau_{0}\exp\left(\frac{\Delta E(t)}{k_{B}T(t)}\right),%
\end{equation}%
where $k_B$ is the Boltzmann constant and $\Delta E(t)$ is the energy barrier for N\'{e}el vector reversal depending on the anisotropy and exchange interaction. Both those latter quantities also have underlying dependencies on the temperature evolution $T(t)$ via the magnetization at each instance in the participating atomic layers.

\section*{Data availability}
All data generated for the current study is available from the corresponding author upon request.

\bibliography{references.bib}

%aipnum4-2.bst 2019-01-14 (MD) hand-edited version of apsrev4-1.bst
\begin{thebibliography}{60}%
\makeatletter
\providecommand \@ifxundefined [1]{%
 \@ifx{#1\undefined}
}%
\providecommand \@ifnum [1]{%
 \ifnum #1\expandafter \@firstoftwo
 \else \expandafter \@secondoftwo
 \fi
}%
\providecommand \@ifx [1]{%
 \ifx #1\expandafter \@firstoftwo
 \else \expandafter \@secondoftwo
 \fi
}%
\providecommand \natexlab [1]{#1}%
\providecommand \enquote  [1]{``#1''}%
\providecommand \bibnamefont  [1]{#1}%
\providecommand \bibfnamefont [1]{#1}%
\providecommand \citenamefont [1]{#1}%
\providecommand \href@noop [0]{\@secondoftwo}%
\providecommand \href [0]{\begingroup \@sanitize@url \@href}%
\providecommand \@href[1]{\@@startlink{#1}\@@href}%
\providecommand \@@href[1]{\endgroup#1\@@endlink}%
\providecommand \@sanitize@url [0]{\catcode `\\12\catcode `\$12\catcode
  `\&12\catcode `\#12\catcode `\^12\catcode `\_12\catcode `\%12\relax}%
\providecommand \@@startlink[1]{}%
\providecommand \@@endlink[0]{}%
\providecommand \url  [0]{\begingroup\@sanitize@url \@url }%
\providecommand \@url [1]{\endgroup\@href {#1}{\urlprefix }}%
\providecommand \urlprefix  [0]{URL }%
\providecommand \Eprint [0]{\href }%
\providecommand \doibase [0]{https://doi.org/}%
\providecommand \selectlanguage [0]{\@gobble}%
\providecommand \bibinfo  [0]{\@secondoftwo}%
\providecommand \bibfield  [0]{\@secondoftwo}%
\providecommand \translation [1]{[#1]}%
\providecommand \BibitemOpen [0]{}%
\providecommand \bibitemStop [0]{}%
\providecommand \bibitemNoStop [0]{.\EOS\space}%
\providecommand \EOS [0]{\spacefactor3000\relax}%
\providecommand \BibitemShut  [1]{\csname bibitem#1\endcsname}%
\let\auto@bib@innerbib\@empty
%</preamble>
\bibitem [{\citenamefont {Du}\ \emph {et~al.}(2023)\citenamefont {Du},
  \citenamefont {Zhu}, \citenamefont {Cao}, \citenamefont {Zhang},
  \citenamefont {Guo}, \citenamefont {Shi}, \citenamefont {Xiong},
  \citenamefont {Xiao}, \citenamefont {Cai}, \citenamefont {Yin}, \citenamefont
  {Lu}, \citenamefont {Zhang}, \citenamefont {Zhang}, \citenamefont {Luo},
  \citenamefont {Fert},\ and\ \citenamefont {Zhao}}]{Du2023}%
  \BibitemOpen
  \bibfield  {author} {\bibinfo {author} {\bibfnamefont {A.}~\bibnamefont
  {Du}}, \bibinfo {author} {\bibfnamefont {D.}~\bibnamefont {Zhu}}, \bibinfo
  {author} {\bibfnamefont {K.}~\bibnamefont {Cao}}, \bibinfo {author}
  {\bibfnamefont {Z.}~\bibnamefont {Zhang}}, \bibinfo {author} {\bibfnamefont
  {Z.}~\bibnamefont {Guo}}, \bibinfo {author} {\bibfnamefont {K.}~\bibnamefont
  {Shi}}, \bibinfo {author} {\bibfnamefont {D.}~\bibnamefont {Xiong}}, \bibinfo
  {author} {\bibfnamefont {R.}~\bibnamefont {Xiao}}, \bibinfo {author}
  {\bibfnamefont {W.}~\bibnamefont {Cai}}, \bibinfo {author} {\bibfnamefont
  {J.}~\bibnamefont {Yin}}, \bibinfo {author} {\bibfnamefont {S.}~\bibnamefont
  {Lu}}, \bibinfo {author} {\bibfnamefont {C.}~\bibnamefont {Zhang}}, \bibinfo
  {author} {\bibfnamefont {Y.}~\bibnamefont {Zhang}}, \bibinfo {author}
  {\bibfnamefont {S.}~\bibnamefont {Luo}}, \bibinfo {author} {\bibfnamefont
  {A.}~\bibnamefont {Fert}},\ and\ \bibinfo {author} {\bibfnamefont
  {W.}~\bibnamefont {Zhao}},\ }\bibfield  {title} {\enquote {\bibinfo {title}
  {{Electrical manipulation and detection of antiferromagnetism in magnetic
  tunnel junctions}},}\ }\href {https://doi.org/10.1038/s41928-023-00975-3}
  {\bibfield  {journal} {\bibinfo  {journal} {Nature Electronics}\ }\textbf
  {\bibinfo {volume} {6}},\ \bibinfo {pages} {425--433} (\bibinfo {year}
  {2023})}\BibitemShut {NoStop}%
\bibitem [{\citenamefont {Dieny}\ \emph {et~al.}(2020)\citenamefont {Dieny},
  \citenamefont {Prejbeanu}, \citenamefont {Garello}, \citenamefont
  {Gambardella}, \citenamefont {Freitas}, \citenamefont {Lehndorff},
  \citenamefont {Raberg}, \citenamefont {Ebels}, \citenamefont {Demokritov},
  \citenamefont {Akerman}, \citenamefont {Deac}, \citenamefont {Pirro},
  \citenamefont {Adelmann}, \citenamefont {Anane}, \citenamefont {Chumak},
  \citenamefont {Hirohata}, \citenamefont {Mangin}, \citenamefont {Valenzuela},
  \citenamefont {Onbaşlı}, \citenamefont {D'Aquino}, \citenamefont {Prenat},
  \citenamefont {Finocchio}, \citenamefont {Lopez-Diaz}, \citenamefont
  {Chantrell}, \citenamefont {Chubykalo-Fesenko},\ and\ \citenamefont
  {Bortolotti}}]{Dieny2020}%
  \BibitemOpen
  \bibfield  {author} {\bibinfo {author} {\bibfnamefont {B.}~\bibnamefont
  {Dieny}}, \bibinfo {author} {\bibfnamefont {I.~L.}\ \bibnamefont
  {Prejbeanu}}, \bibinfo {author} {\bibfnamefont {K.}~\bibnamefont {Garello}},
  \bibinfo {author} {\bibfnamefont {P.}~\bibnamefont {Gambardella}}, \bibinfo
  {author} {\bibfnamefont {P.}~\bibnamefont {Freitas}}, \bibinfo {author}
  {\bibfnamefont {R.}~\bibnamefont {Lehndorff}}, \bibinfo {author}
  {\bibfnamefont {W.}~\bibnamefont {Raberg}}, \bibinfo {author} {\bibfnamefont
  {U.}~\bibnamefont {Ebels}}, \bibinfo {author} {\bibfnamefont {S.~O.}\
  \bibnamefont {Demokritov}}, \bibinfo {author} {\bibfnamefont
  {J.}~\bibnamefont {Akerman}}, \bibinfo {author} {\bibfnamefont
  {A.}~\bibnamefont {Deac}}, \bibinfo {author} {\bibfnamefont {P.}~\bibnamefont
  {Pirro}}, \bibinfo {author} {\bibfnamefont {C.}~\bibnamefont {Adelmann}},
  \bibinfo {author} {\bibfnamefont {A.}~\bibnamefont {Anane}}, \bibinfo
  {author} {\bibfnamefont {A.~V.}\ \bibnamefont {Chumak}}, \bibinfo {author}
  {\bibfnamefont {A.}~\bibnamefont {Hirohata}}, \bibinfo {author}
  {\bibfnamefont {S.}~\bibnamefont {Mangin}}, \bibinfo {author} {\bibfnamefont
  {S.~O.}\ \bibnamefont {Valenzuela}}, \bibinfo {author} {\bibfnamefont
  {M.~C.}\ \bibnamefont {Onbaşlı}}, \bibinfo {author} {\bibfnamefont
  {M.}~\bibnamefont {D'Aquino}}, \bibinfo {author} {\bibfnamefont
  {G.}~\bibnamefont {Prenat}}, \bibinfo {author} {\bibfnamefont
  {G.}~\bibnamefont {Finocchio}}, \bibinfo {author} {\bibfnamefont
  {L.}~\bibnamefont {Lopez-Diaz}}, \bibinfo {author} {\bibfnamefont
  {R.}~\bibnamefont {Chantrell}}, \bibinfo {author} {\bibfnamefont
  {O.}~\bibnamefont {Chubykalo-Fesenko}},\ and\ \bibinfo {author}
  {\bibfnamefont {P.}~\bibnamefont {Bortolotti}},\ }\bibfield  {title}
  {\enquote {\bibinfo {title} {{Opportunities and challenges for spintronics in
  the microelectronics industry}},}\ }\href
  {https://doi.org/10.1038/s41928-020-0461-5} {\bibfield  {journal} {\bibinfo
  {journal} {Nature Electronics}\ }\textbf {\bibinfo {volume} {3}},\ \bibinfo
  {pages} {446--459} (\bibinfo {year} {2020})},\ \Eprint
  {https://arxiv.org/abs/1908.10584} {arXiv:1908.10584} \BibitemShut {NoStop}%
\bibitem [{\citenamefont {Becker}\ \emph {et~al.}(2022)\citenamefont {Becker},
  \citenamefont {Bao}, \citenamefont {Karnaushenko}, \citenamefont {Bandari},
  \citenamefont {Rivkin}, \citenamefont {Li}, \citenamefont {Faghih},
  \citenamefont {Karnaushenko},\ and\ \citenamefont {Schmidt}}]{Becker2022}%
  \BibitemOpen
  \bibfield  {author} {\bibinfo {author} {\bibfnamefont {C.}~\bibnamefont
  {Becker}}, \bibinfo {author} {\bibfnamefont {B.}~\bibnamefont {Bao}},
  \bibinfo {author} {\bibfnamefont {D.~D.}\ \bibnamefont {Karnaushenko}},
  \bibinfo {author} {\bibfnamefont {V.~K.}\ \bibnamefont {Bandari}}, \bibinfo
  {author} {\bibfnamefont {B.}~\bibnamefont {Rivkin}}, \bibinfo {author}
  {\bibfnamefont {Z.}~\bibnamefont {Li}}, \bibinfo {author} {\bibfnamefont
  {M.}~\bibnamefont {Faghih}}, \bibinfo {author} {\bibfnamefont
  {D.}~\bibnamefont {Karnaushenko}},\ and\ \bibinfo {author} {\bibfnamefont
  {O.~G.}\ \bibnamefont {Schmidt}},\ }\bibfield  {title} {\enquote {\bibinfo
  {title} {{A new dimension for magnetosensitive e-skins: active matrix
  integrated micro-origami sensor arrays}},}\ }\href
  {https://doi.org/10.1038/s41467-022-29802-7} {\bibfield  {journal} {\bibinfo
  {journal} {Nature Communications}\ }\textbf {\bibinfo {volume} {13}}
  (\bibinfo {year} {2022}),\ 10.1038/s41467-022-29802-7}\BibitemShut {NoStop}%
\bibitem [{\citenamefont {Becker}\ \emph {et~al.}(2019)\citenamefont {Becker},
  \citenamefont {Karnaushenko}, \citenamefont {Kang}, \citenamefont
  {Karnaushenko}, \citenamefont {Faghih}, \citenamefont {Mirhajivarzaneh},\
  and\ \citenamefont {Schmidt}}]{Becker2019}%
  \BibitemOpen
  \bibfield  {author} {\bibinfo {author} {\bibfnamefont {C.}~\bibnamefont
  {Becker}}, \bibinfo {author} {\bibfnamefont {D.}~\bibnamefont
  {Karnaushenko}}, \bibinfo {author} {\bibfnamefont {T.}~\bibnamefont {Kang}},
  \bibinfo {author} {\bibfnamefont {D.~D.}\ \bibnamefont {Karnaushenko}},
  \bibinfo {author} {\bibfnamefont {M.}~\bibnamefont {Faghih}}, \bibinfo
  {author} {\bibfnamefont {A.}~\bibnamefont {Mirhajivarzaneh}},\ and\ \bibinfo
  {author} {\bibfnamefont {O.~G.}\ \bibnamefont {Schmidt}},\ }\bibfield
  {title} {\enquote {\bibinfo {title} {{Self-assembly of highly sensitive 3D
  magnetic field vector angular encoders}},}\ }\href
  {https://doi.org/10.1126/sciadv.aay7459} {\bibfield  {journal} {\bibinfo
  {journal} {Science Advances}\ }\textbf {\bibinfo {volume} {5}} (\bibinfo
  {year} {2019}),\ 10.1126/sciadv.aay7459}\BibitemShut {NoStop}%
\bibitem [{\citenamefont {Freitas}, \citenamefont {Ferreira},\ and\
  \citenamefont {Cardoso}(2016)}]{Freitas2016}%
  \BibitemOpen
  \bibfield  {author} {\bibinfo {author} {\bibfnamefont {P.~P.}\ \bibnamefont
  {Freitas}}, \bibinfo {author} {\bibfnamefont {R.}~\bibnamefont {Ferreira}},\
  and\ \bibinfo {author} {\bibfnamefont {S.}~\bibnamefont {Cardoso}},\
  }\bibfield  {title} {\enquote {\bibinfo {title} {{Spintronic Sensors}},}\
  }\href {https://doi.org/10.1109/JPROC.2016.2578303} {\bibfield  {journal}
  {\bibinfo  {journal} {Proceedings of the IEEE}\ }\textbf {\bibinfo {volume}
  {104}},\ \bibinfo {pages} {1894--1918} (\bibinfo {year} {2016})}\BibitemShut
  {NoStop}%
\bibitem [{\citenamefont {Meiklejohn}\ and\ \citenamefont
  {Bean}(1956)}]{Meiklejohn1956}%
  \BibitemOpen
  \bibfield  {author} {\bibinfo {author} {\bibfnamefont {W.~H.}\ \bibnamefont
  {Meiklejohn}}\ and\ \bibinfo {author} {\bibfnamefont {C.~P.}\ \bibnamefont
  {Bean}},\ }\bibfield  {title} {\enquote {\bibinfo {title} {{New Magnetic
  Anisotropy}},}\ }\href {https://doi.org/10.1103/PhysRev.102.1413} {\bibfield
  {journal} {\bibinfo  {journal} {Physical Review}\ }\textbf {\bibinfo {volume}
  {102}},\ \bibinfo {pages} {1413--1414} (\bibinfo {year} {1956})}\BibitemShut
  {NoStop}%
\bibitem [{\citenamefont {Nogu{\'{e}}s}\ and\ \citenamefont
  {Schuller}(1999)}]{Nogues1999}%
  \BibitemOpen
  \bibfield  {author} {\bibinfo {author} {\bibfnamefont {J.}~\bibnamefont
  {Nogu{\'{e}}s}}\ and\ \bibinfo {author} {\bibfnamefont {I.~K.}\ \bibnamefont
  {Schuller}},\ }\bibfield  {title} {\enquote {\bibinfo {title} {{Exchange
  bias}},}\ }\href {https://doi.org/10.1016/S0304-8853(98)00266-2} {\bibfield
  {journal} {\bibinfo  {journal} {Journal of Magnetism and Magnetic Materials}\
  }\textbf {\bibinfo {volume} {192}},\ \bibinfo {pages} {203--232} (\bibinfo
  {year} {1999})}\BibitemShut {NoStop}%
\bibitem [{\citenamefont {Hasan}, \citenamefont {Kossak},\ and\ \citenamefont
  {Beach}(2023)}]{Hasan2023}%
  \BibitemOpen
  \bibfield  {author} {\bibinfo {author} {\bibfnamefont {M.~U.}\ \bibnamefont
  {Hasan}}, \bibinfo {author} {\bibfnamefont {A.~E.}\ \bibnamefont {Kossak}},\
  and\ \bibinfo {author} {\bibfnamefont {G.~S.~D.}\ \bibnamefont {Beach}},\
  }\bibfield  {title} {\enquote {\bibinfo {title} {{Large exchange bias
  enhancement and control of ferromagnetic energy landscape by solid-state
  hydrogen gating}},}\ }\href {https://doi.org/10.1038/s41467-023-43955-z}
  {\bibfield  {journal} {\bibinfo  {journal} {Nature Communications}\ }\textbf
  {\bibinfo {volume} {14}},\ \bibinfo {pages} {8510} (\bibinfo {year}
  {2023})}\BibitemShut {NoStop}%
\bibitem [{\citenamefont {Ueberschar}\ \emph {et~al.}(2015)\citenamefont
  {Ueberschar}, \citenamefont {Almeida}, \citenamefont {Matthes}, \citenamefont
  {Muller}, \citenamefont {Ecke}, \citenamefont {Ruckriem}, \citenamefont
  {Schuster}, \citenamefont {Exner},\ and\ \citenamefont
  {Schulz}}]{Ueberschar2015b}%
  \BibitemOpen
  \bibfield  {author} {\bibinfo {author} {\bibfnamefont {O.}~\bibnamefont
  {Ueberschar}}, \bibinfo {author} {\bibfnamefont {M.~J.}\ \bibnamefont
  {Almeida}}, \bibinfo {author} {\bibfnamefont {P.}~\bibnamefont {Matthes}},
  \bibinfo {author} {\bibfnamefont {M.}~\bibnamefont {Muller}}, \bibinfo
  {author} {\bibfnamefont {R.}~\bibnamefont {Ecke}}, \bibinfo {author}
  {\bibfnamefont {R.}~\bibnamefont {Ruckriem}}, \bibinfo {author}
  {\bibfnamefont {J.}~\bibnamefont {Schuster}}, \bibinfo {author}
  {\bibfnamefont {H.}~\bibnamefont {Exner}},\ and\ \bibinfo {author}
  {\bibfnamefont {S.~E.}\ \bibnamefont {Schulz}},\ }\bibfield  {title}
  {\enquote {\bibinfo {title} {{Optimized Monolithic 2-D Spin-Valve Sensor for
  High-Sensitivity Compass Applications}},}\ }\href
  {https://doi.org/10.1109/TMAG.2014.2358802} {\bibfield  {journal} {\bibinfo
  {journal} {IEEE Transactions on Magnetics}\ }\textbf {\bibinfo {volume}
  {51}},\ \bibinfo {pages} {1--4} (\bibinfo {year} {2015})}\BibitemShut
  {NoStop}%
\bibitem [{\citenamefont {Fujiwara}\ \emph {et~al.}(2018)\citenamefont
  {Fujiwara}, \citenamefont {Oogane}, \citenamefont {Kanno}, \citenamefont
  {Imada}, \citenamefont {Jono}, \citenamefont {Terauchi}, \citenamefont
  {Okuno}, \citenamefont {Aritomi}, \citenamefont {Morikawa}, \citenamefont
  {Tsuchida}, \citenamefont {Nakasato},\ and\ \citenamefont
  {Ando}}]{Fujiwara2018}%
  \BibitemOpen
  \bibfield  {author} {\bibinfo {author} {\bibfnamefont {K.}~\bibnamefont
  {Fujiwara}}, \bibinfo {author} {\bibfnamefont {M.}~\bibnamefont {Oogane}},
  \bibinfo {author} {\bibfnamefont {A.}~\bibnamefont {Kanno}}, \bibinfo
  {author} {\bibfnamefont {M.}~\bibnamefont {Imada}}, \bibinfo {author}
  {\bibfnamefont {J.}~\bibnamefont {Jono}}, \bibinfo {author} {\bibfnamefont
  {T.}~\bibnamefont {Terauchi}}, \bibinfo {author} {\bibfnamefont
  {T.}~\bibnamefont {Okuno}}, \bibinfo {author} {\bibfnamefont
  {Y.}~\bibnamefont {Aritomi}}, \bibinfo {author} {\bibfnamefont
  {M.}~\bibnamefont {Morikawa}}, \bibinfo {author} {\bibfnamefont
  {M.}~\bibnamefont {Tsuchida}}, \bibinfo {author} {\bibfnamefont
  {N.}~\bibnamefont {Nakasato}},\ and\ \bibinfo {author} {\bibfnamefont
  {Y.}~\bibnamefont {Ando}},\ }\bibfield  {title} {\enquote {\bibinfo {title}
  {{Magnetocardiography and magnetoencephalography measurements at room
  temperature using tunnel magneto-resistance sensors}},}\ }\href
  {https://doi.org/10.7567/APEX.11.023001} {\bibfield  {journal} {\bibinfo
  {journal} {Applied Physics Express}\ }\textbf {\bibinfo {volume} {11}},\
  \bibinfo {pages} {023001} (\bibinfo {year} {2018})}\BibitemShut {NoStop}%
\bibitem [{\citenamefont {Wang}\ \emph
  {et~al.}(2024{\natexlab{a}})\citenamefont {Wang}, \citenamefont {Qin},
  \citenamefont {Guan}, \citenamefont {Liu}, \citenamefont {Cai}, \citenamefont
  {Zhang}, \citenamefont {Zhou}, \citenamefont {Zhang}, \citenamefont {Wu},\
  and\ \citenamefont {Tao}}]{Wang2023a}%
  \BibitemOpen
  \bibfield  {author} {\bibinfo {author} {\bibfnamefont {S.}~\bibnamefont
  {Wang}}, \bibinfo {author} {\bibfnamefont {W.}~\bibnamefont {Qin}}, \bibinfo
  {author} {\bibfnamefont {T.}~\bibnamefont {Guan}}, \bibinfo {author}
  {\bibfnamefont {J.}~\bibnamefont {Liu}}, \bibinfo {author} {\bibfnamefont
  {Q.}~\bibnamefont {Cai}}, \bibinfo {author} {\bibfnamefont {S.}~\bibnamefont
  {Zhang}}, \bibinfo {author} {\bibfnamefont {L.}~\bibnamefont {Zhou}},
  \bibinfo {author} {\bibfnamefont {Y.}~\bibnamefont {Zhang}}, \bibinfo
  {author} {\bibfnamefont {Y.}~\bibnamefont {Wu}},\ and\ \bibinfo {author}
  {\bibfnamefont {Z.}~\bibnamefont {Tao}},\ }\bibfield  {title} {\enquote
  {\bibinfo {title} {{Flexible generation of structured terahertz fields via
  programmable exchange-biased spintronic emitters}},}\ }\href
  {https://doi.org/10.1186/s43593-024-00069-3} {\bibfield  {journal} {\bibinfo
  {journal} {eLight}\ }\textbf {\bibinfo {volume} {4}},\ \bibinfo {pages} {11}
  (\bibinfo {year} {2024}{\natexlab{a}})},\ \Eprint
  {https://arxiv.org/abs/2311.11499} {arXiv:2311.11499} \BibitemShut {NoStop}%
\bibitem [{\citenamefont {Berthold}\ \emph {et~al.}(2014)\citenamefont
  {Berthold}, \citenamefont {L{\"{o}}schner}, \citenamefont {Schille},
  \citenamefont {Ebert},\ and\ \citenamefont {Exner}}]{Berthold2014}%
  \BibitemOpen
  \bibfield  {author} {\bibinfo {author} {\bibfnamefont {I.}~\bibnamefont
  {Berthold}}, \bibinfo {author} {\bibfnamefont {U.}~\bibnamefont
  {L{\"{o}}schner}}, \bibinfo {author} {\bibfnamefont {J.}~\bibnamefont
  {Schille}}, \bibinfo {author} {\bibfnamefont {R.}~\bibnamefont {Ebert}},\
  and\ \bibinfo {author} {\bibfnamefont {H.}~\bibnamefont {Exner}},\ }\bibfield
   {title} {\enquote {\bibinfo {title} {{Exchange Bias Realignment Using a
  Laser-based Direct-write Technique}},}\ }\href
  {https://doi.org/10.1016/j.phpro.2014.08.028} {\bibfield  {journal} {\bibinfo
   {journal} {Physics Procedia}\ }\textbf {\bibinfo {volume} {56}},\ \bibinfo
  {pages} {1136--1142} (\bibinfo {year} {2014})}\BibitemShut {NoStop}%
\bibitem [{\citenamefont {Almeida}\ \emph {et~al.}(2015)\citenamefont
  {Almeida}, \citenamefont {Matthes}, \citenamefont {Uebersch{\"{a}}r},
  \citenamefont {M{\"{u}}ller}, \citenamefont {Ecke}, \citenamefont {Exner},
  \citenamefont {Albrecht},\ and\ \citenamefont {Schulz}}]{Almeida2015}%
  \BibitemOpen
  \bibfield  {author} {\bibinfo {author} {\bibfnamefont {M.}~\bibnamefont
  {Almeida}}, \bibinfo {author} {\bibfnamefont {P.}~\bibnamefont {Matthes}},
  \bibinfo {author} {\bibfnamefont {O.}~\bibnamefont {Uebersch{\"{a}}r}},
  \bibinfo {author} {\bibfnamefont {M.}~\bibnamefont {M{\"{u}}ller}}, \bibinfo
  {author} {\bibfnamefont {R.}~\bibnamefont {Ecke}}, \bibinfo {author}
  {\bibfnamefont {H.}~\bibnamefont {Exner}}, \bibinfo {author} {\bibfnamefont
  {M.}~\bibnamefont {Albrecht}},\ and\ \bibinfo {author} {\bibfnamefont
  {S.}~\bibnamefont {Schulz}},\ }\bibfield  {title} {\enquote {\bibinfo {title}
  {{Optimum Laser Exposure for Setting Exchange Bias in Spin Valve Sensors}},}\
  }\href {https://doi.org/10.1016/j.phpro.2015.12.120} {\bibfield  {journal}
  {\bibinfo  {journal} {Physics Procedia}\ }\textbf {\bibinfo {volume} {75}},\
  \bibinfo {pages} {1192--1197} (\bibinfo {year} {2015})}\BibitemShut {NoStop}%
\bibitem [{\citenamefont {Radu}\ \emph {et~al.}(2011)\citenamefont {Radu},
  \citenamefont {Vahaplar}, \citenamefont {Stamm}, \citenamefont {Kachel},
  \citenamefont {Pontius}, \citenamefont {D{\"{u}}rr}, \citenamefont {Ostler},
  \citenamefont {Barker}, \citenamefont {Evans}, \citenamefont {Chantrell},
  \citenamefont {Tsukamoto}, \citenamefont {Itoh}, \citenamefont {Kirilyuk},
  \citenamefont {Rasing},\ and\ \citenamefont {Kimel}}]{Radu2011}%
  \BibitemOpen
  \bibfield  {author} {\bibinfo {author} {\bibfnamefont {I.}~\bibnamefont
  {Radu}}, \bibinfo {author} {\bibfnamefont {K.}~\bibnamefont {Vahaplar}},
  \bibinfo {author} {\bibfnamefont {C.}~\bibnamefont {Stamm}}, \bibinfo
  {author} {\bibfnamefont {T.}~\bibnamefont {Kachel}}, \bibinfo {author}
  {\bibfnamefont {N.}~\bibnamefont {Pontius}}, \bibinfo {author} {\bibfnamefont
  {H.~A.}\ \bibnamefont {D{\"{u}}rr}}, \bibinfo {author} {\bibfnamefont
  {T.~A.}\ \bibnamefont {Ostler}}, \bibinfo {author} {\bibfnamefont
  {J.}~\bibnamefont {Barker}}, \bibinfo {author} {\bibfnamefont {R.~F.~L.}\
  \bibnamefont {Evans}}, \bibinfo {author} {\bibfnamefont {R.~W.}\ \bibnamefont
  {Chantrell}}, \bibinfo {author} {\bibfnamefont {A.}~\bibnamefont
  {Tsukamoto}}, \bibinfo {author} {\bibfnamefont {A.}~\bibnamefont {Itoh}},
  \bibinfo {author} {\bibfnamefont {A.}~\bibnamefont {Kirilyuk}}, \bibinfo
  {author} {\bibfnamefont {T.}~\bibnamefont {Rasing}},\ and\ \bibinfo {author}
  {\bibfnamefont {A.~V.}\ \bibnamefont {Kimel}},\ }\bibfield  {title} {\enquote
  {\bibinfo {title} {{Transient ferromagnetic-like state mediating ultrafast
  reversal of antiferromagnetically coupled spins}},}\ }\href
  {https://doi.org/10.1038/nature09901} {\bibfield  {journal} {\bibinfo
  {journal} {Nature}\ }\textbf {\bibinfo {volume} {472}},\ \bibinfo {pages}
  {205--208} (\bibinfo {year} {2011})}\BibitemShut {NoStop}%
\bibitem [{\citenamefont {Ostler}\ \emph {et~al.}(2012)\citenamefont {Ostler},
  \citenamefont {Barker}, \citenamefont {Evans}, \citenamefont {Chantrell},
  \citenamefont {Atxitia}, \citenamefont {Chubykalo-Fesenko}, \citenamefont
  {{El Moussaoui}}, \citenamefont {{Le Guyader}}, \citenamefont {Mengotti},
  \citenamefont {Heyderman}, \citenamefont {Nolting}, \citenamefont
  {Tsukamoto}, \citenamefont {Itoh}, \citenamefont {Afanasiev}, \citenamefont
  {Ivanov}, \citenamefont {Kalashnikova}, \citenamefont {Vahaplar},
  \citenamefont {Mentink}, \citenamefont {Kirilyuk}, \citenamefont {Rasing},\
  and\ \citenamefont {Kimel}}]{Ostler2012}%
  \BibitemOpen
  \bibfield  {author} {\bibinfo {author} {\bibfnamefont {T.}~\bibnamefont
  {Ostler}}, \bibinfo {author} {\bibfnamefont {J.}~\bibnamefont {Barker}},
  \bibinfo {author} {\bibfnamefont {R.}~\bibnamefont {Evans}}, \bibinfo
  {author} {\bibfnamefont {R.}~\bibnamefont {Chantrell}}, \bibinfo {author}
  {\bibfnamefont {U.}~\bibnamefont {Atxitia}}, \bibinfo {author} {\bibfnamefont
  {O.}~\bibnamefont {Chubykalo-Fesenko}}, \bibinfo {author} {\bibfnamefont
  {S.}~\bibnamefont {{El Moussaoui}}}, \bibinfo {author} {\bibfnamefont
  {L.}~\bibnamefont {{Le Guyader}}}, \bibinfo {author} {\bibfnamefont
  {E.}~\bibnamefont {Mengotti}}, \bibinfo {author} {\bibfnamefont
  {L.}~\bibnamefont {Heyderman}}, \bibinfo {author} {\bibfnamefont
  {F.}~\bibnamefont {Nolting}}, \bibinfo {author} {\bibfnamefont
  {A.}~\bibnamefont {Tsukamoto}}, \bibinfo {author} {\bibfnamefont
  {A.}~\bibnamefont {Itoh}}, \bibinfo {author} {\bibfnamefont {D.}~\bibnamefont
  {Afanasiev}}, \bibinfo {author} {\bibfnamefont {B.}~\bibnamefont {Ivanov}},
  \bibinfo {author} {\bibfnamefont {A.}~\bibnamefont {Kalashnikova}}, \bibinfo
  {author} {\bibfnamefont {K.}~\bibnamefont {Vahaplar}}, \bibinfo {author}
  {\bibfnamefont {J.}~\bibnamefont {Mentink}}, \bibinfo {author} {\bibfnamefont
  {A.}~\bibnamefont {Kirilyuk}}, \bibinfo {author} {\bibfnamefont
  {T.}~\bibnamefont {Rasing}},\ and\ \bibinfo {author} {\bibfnamefont
  {A.}~\bibnamefont {Kimel}},\ }\bibfield  {title} {\enquote {\bibinfo {title}
  {{Ultrafast heating as a sufficient stimulus for magnetization reversal in a
  ferrimagnet}},}\ }\href {https://doi.org/10.1038/ncomms1666} {\bibfield
  {journal} {\bibinfo  {journal} {Nature Communications}\ }\textbf {\bibinfo
  {volume} {3}},\ \bibinfo {pages} {666} (\bibinfo {year} {2012})}\BibitemShut
  {NoStop}%
\bibitem [{\citenamefont {Lalieu}\ \emph {et~al.}(2017)\citenamefont {Lalieu},
  \citenamefont {Peeters}, \citenamefont {Haenen}, \citenamefont {Lavrijsen},\
  and\ \citenamefont {Koopmans}}]{Lalieu2017}%
  \BibitemOpen
  \bibfield  {author} {\bibinfo {author} {\bibfnamefont {M.~L.~M.}\
  \bibnamefont {Lalieu}}, \bibinfo {author} {\bibfnamefont {M.~J.~G.}\
  \bibnamefont {Peeters}}, \bibinfo {author} {\bibfnamefont {S.~R.~R.}\
  \bibnamefont {Haenen}}, \bibinfo {author} {\bibfnamefont {R.}~\bibnamefont
  {Lavrijsen}},\ and\ \bibinfo {author} {\bibfnamefont {B.}~\bibnamefont
  {Koopmans}},\ }\bibfield  {title} {\enquote {\bibinfo {title} {{Deterministic
  all-optical switching of synthetic ferrimagnets using single femtosecond
  laser pulses}},}\ }\href {https://doi.org/10.1103/PhysRevB.96.220411}
  {\bibfield  {journal} {\bibinfo  {journal} {Physical Review B}\ }\textbf
  {\bibinfo {volume} {96}},\ \bibinfo {pages} {220411} (\bibinfo {year}
  {2017})}\BibitemShut {NoStop}%
\bibitem [{\citenamefont {Cao}\ and\ \citenamefont {Freitas}(2010)}]{Cao2010}%
  \BibitemOpen
  \bibfield  {author} {\bibinfo {author} {\bibfnamefont {J.}~\bibnamefont
  {Cao}}\ and\ \bibinfo {author} {\bibfnamefont {P.~P.}\ \bibnamefont
  {Freitas}},\ }\bibfield  {title} {\enquote {\bibinfo {title} {{Wheatstone
  bridge sensor composed of linear MgO magnetic tunnel junctions}},}\ }\href
  {https://doi.org/10.1063/1.3360583} {\bibfield  {journal} {\bibinfo
  {journal} {Journal of Applied Physics}\ }\textbf {\bibinfo {volume} {107}},\
  \bibinfo {pages} {09E712} (\bibinfo {year} {2010})}\BibitemShut {NoStop}%
\bibitem [{\citenamefont {Papusoi}\ \emph {et~al.}(2008)\citenamefont
  {Papusoi}, \citenamefont {Sousa}, \citenamefont {Dieny}, \citenamefont
  {Prejbeanu}, \citenamefont {Conraux}, \citenamefont {MacKay},\ and\
  \citenamefont {Noz{\`{i}}res}}]{Papusoi2008}%
  \BibitemOpen
  \bibfield  {author} {\bibinfo {author} {\bibfnamefont {C.}~\bibnamefont
  {Papusoi}}, \bibinfo {author} {\bibfnamefont {R.~C.}\ \bibnamefont {Sousa}},
  \bibinfo {author} {\bibfnamefont {B.}~\bibnamefont {Dieny}}, \bibinfo
  {author} {\bibfnamefont {I.~L.}\ \bibnamefont {Prejbeanu}}, \bibinfo {author}
  {\bibfnamefont {Y.}~\bibnamefont {Conraux}}, \bibinfo {author} {\bibfnamefont
  {K.}~\bibnamefont {MacKay}},\ and\ \bibinfo {author} {\bibfnamefont {J.~P.}\
  \bibnamefont {Noz{\`{i}}res}},\ }\bibfield  {title} {\enquote {\bibinfo
  {title} {{Reversing exchange bias in thermally assisted magnetic random
  access memory cell by electric current heating pulses}},}\ }\href
  {https://doi.org/10.1063/1.2951931} {\bibfield  {journal} {\bibinfo
  {journal} {Journal of Applied Physics}\ } (\bibinfo {year} {2008}),\
  10.1063/1.2951931}\BibitemShut {NoStop}%
\bibitem [{\citenamefont {Wang}\ \emph
  {et~al.}(2024{\natexlab{b}})\citenamefont {Wang}, \citenamefont {Chen},
  \citenamefont {Wang}, \citenamefont {Wang}, \citenamefont {Han},
  \citenamefont {Su}, \citenamefont {Hu}, \citenamefont {Wang},\ and\
  \citenamefont {Liu}}]{Wang2024}%
  \BibitemOpen
  \bibfield  {author} {\bibinfo {author} {\bibfnamefont {W.}~\bibnamefont
  {Wang}}, \bibinfo {author} {\bibfnamefont {Y.}~\bibnamefont {Chen}}, \bibinfo
  {author} {\bibfnamefont {B.}~\bibnamefont {Wang}}, \bibinfo {author}
  {\bibfnamefont {L.}~\bibnamefont {Wang}}, \bibinfo {author} {\bibfnamefont
  {Y.}~\bibnamefont {Han}}, \bibinfo {author} {\bibfnamefont {W.}~\bibnamefont
  {Su}}, \bibinfo {author} {\bibfnamefont {Z.}~\bibnamefont {Hu}}, \bibinfo
  {author} {\bibfnamefont {Z.}~\bibnamefont {Wang}},\ and\ \bibinfo {author}
  {\bibfnamefont {M.}~\bibnamefont {Liu}},\ }\bibfield  {title} {\enquote
  {\bibinfo {title} {{Locally Reconfigurable Exchange Bias for Tunable
  Spintronics by Pulsed Current Injection}},}\ }\href
  {https://doi.org/10.1002/pssr.202300303} {\bibfield  {journal} {\bibinfo
  {journal} {physica status solidi (RRL) – Rapid Research Letters}\ }\textbf
  {\bibinfo {volume} {18}} (\bibinfo {year} {2024}{\natexlab{b}}),\
  10.1002/pssr.202300303}\BibitemShut {NoStop}%
\bibitem [{\citenamefont {Wang}\ \emph
  {et~al.}(2022{\natexlab{a}})\citenamefont {Wang}, \citenamefont {Taniguchi},
  \citenamefont {Lin}, \citenamefont {Zicchino}, \citenamefont {Nickl},
  \citenamefont {Sahliger}, \citenamefont {Lai}, \citenamefont {Song},
  \citenamefont {Wu}, \citenamefont {Dai},\ and\ \citenamefont
  {Back}}]{Wang2022b}%
  \BibitemOpen
  \bibfield  {author} {\bibinfo {author} {\bibfnamefont {Y.}~\bibnamefont
  {Wang}}, \bibinfo {author} {\bibfnamefont {T.}~\bibnamefont {Taniguchi}},
  \bibinfo {author} {\bibfnamefont {P.~H.}\ \bibnamefont {Lin}}, \bibinfo
  {author} {\bibfnamefont {D.}~\bibnamefont {Zicchino}}, \bibinfo {author}
  {\bibfnamefont {A.}~\bibnamefont {Nickl}}, \bibinfo {author} {\bibfnamefont
  {J.}~\bibnamefont {Sahliger}}, \bibinfo {author} {\bibfnamefont {C.~H.}\
  \bibnamefont {Lai}}, \bibinfo {author} {\bibfnamefont {C.}~\bibnamefont
  {Song}}, \bibinfo {author} {\bibfnamefont {H.}~\bibnamefont {Wu}}, \bibinfo
  {author} {\bibfnamefont {Q.}~\bibnamefont {Dai}},\ and\ \bibinfo {author}
  {\bibfnamefont {C.~H.}\ \bibnamefont {Back}},\ }\bibfield  {title} {\enquote
  {\bibinfo {title} {{Time-resolved detection of spin-orbit torque switching of
  magnetization and exchange bias}},}\ }\href
  {https://doi.org/10.1038/s41928-022-00870-3} {\bibfield  {journal} {\bibinfo
  {journal} {Nature Electronics}\ }\textbf {\bibinfo {volume} {5}},\ \bibinfo
  {pages} {840--848} (\bibinfo {year} {2022}{\natexlab{a}})}\BibitemShut
  {NoStop}%
\bibitem [{\citenamefont {Lin}\ \emph {et~al.}(2019)\citenamefont {Lin},
  \citenamefont {Yang}, \citenamefont {Tsai}, \citenamefont {Chen},
  \citenamefont {Huang}, \citenamefont {Lin},\ and\ \citenamefont
  {Lai}}]{Lin2019}%
  \BibitemOpen
  \bibfield  {author} {\bibinfo {author} {\bibfnamefont {P.-H.}\ \bibnamefont
  {Lin}}, \bibinfo {author} {\bibfnamefont {B.-Y.}\ \bibnamefont {Yang}},
  \bibinfo {author} {\bibfnamefont {M.-H.}\ \bibnamefont {Tsai}}, \bibinfo
  {author} {\bibfnamefont {P.-C.}\ \bibnamefont {Chen}}, \bibinfo {author}
  {\bibfnamefont {K.-F.}\ \bibnamefont {Huang}}, \bibinfo {author}
  {\bibfnamefont {H.-H.}\ \bibnamefont {Lin}},\ and\ \bibinfo {author}
  {\bibfnamefont {C.-H.}\ \bibnamefont {Lai}},\ }\bibfield  {title} {\enquote
  {\bibinfo {title} {{Manipulating exchange bias by spin–orbit torque}},}\
  }\href {https://doi.org/10.1038/s41563-019-0289-4} {\bibfield  {journal}
  {\bibinfo  {journal} {Nature Materials}\ }\textbf {\bibinfo {volume} {18}},\
  \bibinfo {pages} {335--341} (\bibinfo {year} {2019})}\BibitemShut {NoStop}%
\bibitem [{\citenamefont {Yun}\ \emph {et~al.}(2024)\citenamefont {Yun},
  \citenamefont {Zhang}, \citenamefont {Xu}, \citenamefont {Guo}, \citenamefont
  {Xi},\ and\ \citenamefont {Jin}}]{Yun2024}%
  \BibitemOpen
  \bibfield  {author} {\bibinfo {author} {\bibfnamefont {J.}~\bibnamefont
  {Yun}}, \bibinfo {author} {\bibfnamefont {Q.}~\bibnamefont {Zhang}}, \bibinfo
  {author} {\bibfnamefont {H.}~\bibnamefont {Xu}}, \bibinfo {author}
  {\bibfnamefont {X.}~\bibnamefont {Guo}}, \bibinfo {author} {\bibfnamefont
  {L.}~\bibnamefont {Xi}},\ and\ \bibinfo {author} {\bibfnamefont
  {K.}~\bibnamefont {Jin}},\ }\bibfield  {title} {\enquote {\bibinfo {title}
  {{Field-free manipulation of exchange bias in perpendicularly magnetized
  Pt/Co/IrMn structures by spin-orbit torque}},}\ }\href
  {https://doi.org/10.1103/PhysRevMaterials.8.064407} {\bibfield  {journal}
  {\bibinfo  {journal} {Physical Review Materials}\ }\textbf {\bibinfo {volume}
  {8}},\ \bibinfo {pages} {064407} (\bibinfo {year} {2024})}\BibitemShut
  {NoStop}%
\bibitem [{\citenamefont {Guo}\ \emph {et~al.}(2024{\natexlab{a}})\citenamefont
  {Guo}, \citenamefont {Shi}, \citenamefont {Wang}, \citenamefont {Su},
  \citenamefont {Zhang},\ and\ \citenamefont {Tang}}]{Guo2024a}%
  \BibitemOpen
  \bibfield  {author} {\bibinfo {author} {\bibfnamefont {L.}~\bibnamefont
  {Guo}}, \bibinfo {author} {\bibfnamefont {G.}~\bibnamefont {Shi}}, \bibinfo
  {author} {\bibfnamefont {G.}~\bibnamefont {Wang}}, \bibinfo {author}
  {\bibfnamefont {H.}~\bibnamefont {Su}}, \bibinfo {author} {\bibfnamefont
  {H.}~\bibnamefont {Zhang}},\ and\ \bibinfo {author} {\bibfnamefont
  {X.}~\bibnamefont {Tang}},\ }\bibfield  {title} {\enquote {\bibinfo {title}
  {{Asymmetric Manipulation of Perpendicular Exchange Bias and Programmable
  Spin Logical Cells by Spin–Orbit Torque in a Ferromagnet/Antiferromagnet
  System}},}\ }\href {https://doi.org/10.1002/advs.202403648} {\bibfield
  {journal} {\bibinfo  {journal} {Advanced Science}\ } (\bibinfo {year}
  {2024}{\natexlab{a}}),\ 10.1002/advs.202403648}\BibitemShut {NoStop}%
\bibitem [{\citenamefont {Liu}\ \emph {et~al.}(2023)\citenamefont {Liu},
  \citenamefont {Zhang}, \citenamefont {Yuan}, \citenamefont {Lu},
  \citenamefont {Liu}, \citenamefont {Chen}, \citenamefont {Wei}, \citenamefont
  {Wu}, \citenamefont {You}, \citenamefont {Zhang},\ and\ \citenamefont
  {Du}}]{Liu2023}%
  \BibitemOpen
  \bibfield  {author} {\bibinfo {author} {\bibfnamefont {R.}~\bibnamefont
  {Liu}}, \bibinfo {author} {\bibfnamefont {Y.}~\bibnamefont {Zhang}}, \bibinfo
  {author} {\bibfnamefont {Y.}~\bibnamefont {Yuan}}, \bibinfo {author}
  {\bibfnamefont {Y.}~\bibnamefont {Lu}}, \bibinfo {author} {\bibfnamefont
  {T.}~\bibnamefont {Liu}}, \bibinfo {author} {\bibfnamefont {J.}~\bibnamefont
  {Chen}}, \bibinfo {author} {\bibfnamefont {L.}~\bibnamefont {Wei}}, \bibinfo
  {author} {\bibfnamefont {D.}~\bibnamefont {Wu}}, \bibinfo {author}
  {\bibfnamefont {B.}~\bibnamefont {You}}, \bibinfo {author} {\bibfnamefont
  {W.}~\bibnamefont {Zhang}},\ and\ \bibinfo {author} {\bibfnamefont
  {J.}~\bibnamefont {Du}},\ }\bibfield  {title} {\enquote {\bibinfo {title}
  {{Manipulating exchange bias in Ir 25 Mn 75 /CoTb bilayer through
  spin–orbit torque}},}\ }\href {https://doi.org/10.1063/5.0139997}
  {\bibfield  {journal} {\bibinfo  {journal} {Applied Physics Letters}\
  }\textbf {\bibinfo {volume} {122}},\ \bibinfo {pages} {062401} (\bibinfo
  {year} {2023})}\BibitemShut {NoStop}%
\bibitem [{\citenamefont {Chen}\ \emph {et~al.}(2023)\citenamefont {Chen},
  \citenamefont {Lin}, \citenamefont {Zhang}, \citenamefont {Cao},
  \citenamefont {Zhao}, \citenamefont {Zhou}, \citenamefont {Wang},
  \citenamefont {Yan}, \citenamefont {Du}, \citenamefont {Leng},\ and\
  \citenamefont {Yan}}]{Chen2023}%
  \BibitemOpen
  \bibfield  {author} {\bibinfo {author} {\bibfnamefont {W.}~\bibnamefont
  {Chen}}, \bibinfo {author} {\bibfnamefont {Y.}~\bibnamefont {Lin}}, \bibinfo
  {author} {\bibfnamefont {K.}~\bibnamefont {Zhang}}, \bibinfo {author}
  {\bibfnamefont {Z.}~\bibnamefont {Cao}}, \bibinfo {author} {\bibfnamefont
  {X.}~\bibnamefont {Zhao}}, \bibinfo {author} {\bibfnamefont {Z.}~\bibnamefont
  {Zhou}}, \bibinfo {author} {\bibfnamefont {X.}~\bibnamefont {Wang}}, \bibinfo
  {author} {\bibfnamefont {S.}~\bibnamefont {Yan}}, \bibinfo {author}
  {\bibfnamefont {H.}~\bibnamefont {Du}}, \bibinfo {author} {\bibfnamefont
  {Q.}~\bibnamefont {Leng}},\ and\ \bibinfo {author} {\bibfnamefont
  {S.}~\bibnamefont {Yan}},\ }\bibfield  {title} {\enquote {\bibinfo {title}
  {{Spin Orbit Torque Locally Controlling Exchange Bias to Realize High
  Detection Sensitivity of Two-dimensional Magnetic Field}},}\ }\href
  {https://doi.org/10.1016/j.fmre.2023.07.010} {\bibfield  {journal} {\bibinfo
  {journal} {Fundamental Research}\ } (\bibinfo {year} {2023}),\
  10.1016/j.fmre.2023.07.010}\BibitemShut {NoStop}%
\bibitem [{\citenamefont {Qi}\ \emph {et~al.}(2024{\natexlab{a}})\citenamefont
  {Qi}, \citenamefont {Zhao}, \citenamefont {Zhang}, \citenamefont {Yang},
  \citenamefont {Huang}, \citenamefont {Lyu}, \citenamefont {Shao},
  \citenamefont {Zhang}, \citenamefont {Li}, \citenamefont {Zhu}, \citenamefont
  {Yu}, \citenamefont {Wei}, \citenamefont {Zhou}, \citenamefont {Shen},\ and\
  \citenamefont {Wang}}]{Qi2024a}%
  \BibitemOpen
  \bibfield  {author} {\bibinfo {author} {\bibfnamefont {J.}~\bibnamefont
  {Qi}}, \bibinfo {author} {\bibfnamefont {Y.}~\bibnamefont {Zhao}}, \bibinfo
  {author} {\bibfnamefont {Y.}~\bibnamefont {Zhang}}, \bibinfo {author}
  {\bibfnamefont {G.}~\bibnamefont {Yang}}, \bibinfo {author} {\bibfnamefont
  {H.}~\bibnamefont {Huang}}, \bibinfo {author} {\bibfnamefont
  {H.}~\bibnamefont {Lyu}}, \bibinfo {author} {\bibfnamefont {B.}~\bibnamefont
  {Shao}}, \bibinfo {author} {\bibfnamefont {J.}~\bibnamefont {Zhang}},
  \bibinfo {author} {\bibfnamefont {J.}~\bibnamefont {Li}}, \bibinfo {author}
  {\bibfnamefont {T.}~\bibnamefont {Zhu}}, \bibinfo {author} {\bibfnamefont
  {G.}~\bibnamefont {Yu}}, \bibinfo {author} {\bibfnamefont {H.}~\bibnamefont
  {Wei}}, \bibinfo {author} {\bibfnamefont {S.}~\bibnamefont {Zhou}}, \bibinfo
  {author} {\bibfnamefont {B.}~\bibnamefont {Shen}},\ and\ \bibinfo {author}
  {\bibfnamefont {S.}~\bibnamefont {Wang}},\ }\bibfield  {title} {\enquote
  {\bibinfo {title} {{Full electrical manipulation of perpendicular exchange
  bias in ultrathin antiferromagnetic film with epitaxial strain}},}\ }\href
  {https://doi.org/10.1038/s41467-024-49214-z} {\bibfield  {journal} {\bibinfo
  {journal} {Nature Communications}\ }\textbf {\bibinfo {volume} {15}},\
  \bibinfo {pages} {4734} (\bibinfo {year} {2024}{\natexlab{a}})}\BibitemShut
  {NoStop}%
\bibitem [{\citenamefont {Su}\ \emph {et~al.}(2020)\citenamefont {Su},
  \citenamefont {Wang}, \citenamefont {Chen}, \citenamefont {Zhao},
  \citenamefont {Hu}, \citenamefont {Wen}, \citenamefont {Hu}, \citenamefont
  {Wu}, \citenamefont {Zhou},\ and\ \citenamefont {Liu}}]{Su2020}%
  \BibitemOpen
  \bibfield  {author} {\bibinfo {author} {\bibfnamefont {W.}~\bibnamefont
  {Su}}, \bibinfo {author} {\bibfnamefont {Z.}~\bibnamefont {Wang}}, \bibinfo
  {author} {\bibfnamefont {Y.}~\bibnamefont {Chen}}, \bibinfo {author}
  {\bibfnamefont {X.}~\bibnamefont {Zhao}}, \bibinfo {author} {\bibfnamefont
  {C.}~\bibnamefont {Hu}}, \bibinfo {author} {\bibfnamefont {T.}~\bibnamefont
  {Wen}}, \bibinfo {author} {\bibfnamefont {Z.}~\bibnamefont {Hu}}, \bibinfo
  {author} {\bibfnamefont {J.}~\bibnamefont {Wu}}, \bibinfo {author}
  {\bibfnamefont {Z.}~\bibnamefont {Zhou}},\ and\ \bibinfo {author}
  {\bibfnamefont {M.}~\bibnamefont {Liu}},\ }\bibfield  {title} {\enquote
  {\bibinfo {title} {{Reconfigurable Magnetoresistive Sensor Based on
  Magnetoelectric Coupling}},}\ }\href {https://doi.org/10.1002/aelm.201901061}
  {\bibfield  {journal} {\bibinfo  {journal} {Advanced Electronic Materials}\
  }\textbf {\bibinfo {volume} {6}} (\bibinfo {year} {2020}),\
  10.1002/aelm.201901061}\BibitemShut {NoStop}%
\bibitem [{\citenamefont {Wang}\ \emph {et~al.}(2023)\citenamefont {Wang},
  \citenamefont {Li}, \citenamefont {Lu}, \citenamefont {Xu},\ and\
  \citenamefont {Jiang}}]{Wang2023}%
  \BibitemOpen
  \bibfield  {author} {\bibinfo {author} {\bibfnamefont {M.}~\bibnamefont
  {Wang}}, \bibinfo {author} {\bibfnamefont {M.}~\bibnamefont {Li}}, \bibinfo
  {author} {\bibfnamefont {Y.}~\bibnamefont {Lu}}, \bibinfo {author}
  {\bibfnamefont {X.}~\bibnamefont {Xu}},\ and\ \bibinfo {author}
  {\bibfnamefont {Y.}~\bibnamefont {Jiang}},\ }\bibfield  {title} {\enquote
  {\bibinfo {title} {{Electric field controlled perpendicular exchange bias in
  Ta/Pt/Co/IrMn/Pt heterostructure}},}\ }\href
  {https://doi.org/10.1063/5.0160957} {\bibfield  {journal} {\bibinfo
  {journal} {Applied Physics Letters}\ }\textbf {\bibinfo {volume} {123}}
  (\bibinfo {year} {2023}),\ 10.1063/5.0160957}\BibitemShut {NoStop}%
\bibitem [{\citenamefont {Guo}\ \emph {et~al.}(2024{\natexlab{b}})\citenamefont
  {Guo}, \citenamefont {Wang}, \citenamefont {Malinowski}, \citenamefont
  {Zhang}, \citenamefont {Zhang}, \citenamefont {Wang}, \citenamefont {Lyu},
  \citenamefont {Peng}, \citenamefont {Vallobra}, \citenamefont {Xu},
  \citenamefont {Xu}, \citenamefont {Jenkins}, \citenamefont {Chantrell},
  \citenamefont {Evans}, \citenamefont {Mangin}, \citenamefont {Zhao},\ and\
  \citenamefont {Hehn}}]{Guo2024}%
  \BibitemOpen
  \bibfield  {author} {\bibinfo {author} {\bibfnamefont {Z.}~\bibnamefont
  {Guo}}, \bibinfo {author} {\bibfnamefont {J.}~\bibnamefont {Wang}}, \bibinfo
  {author} {\bibfnamefont {G.}~\bibnamefont {Malinowski}}, \bibinfo {author}
  {\bibfnamefont {B.}~\bibnamefont {Zhang}}, \bibinfo {author} {\bibfnamefont
  {W.}~\bibnamefont {Zhang}}, \bibinfo {author} {\bibfnamefont
  {H.}~\bibnamefont {Wang}}, \bibinfo {author} {\bibfnamefont {C.}~\bibnamefont
  {Lyu}}, \bibinfo {author} {\bibfnamefont {Y.}~\bibnamefont {Peng}}, \bibinfo
  {author} {\bibfnamefont {P.}~\bibnamefont {Vallobra}}, \bibinfo {author}
  {\bibfnamefont {Y.}~\bibnamefont {Xu}}, \bibinfo {author} {\bibfnamefont
  {Y.}~\bibnamefont {Xu}}, \bibinfo {author} {\bibfnamefont {S.}~\bibnamefont
  {Jenkins}}, \bibinfo {author} {\bibfnamefont {R.~W.}\ \bibnamefont
  {Chantrell}}, \bibinfo {author} {\bibfnamefont {R.~F.~L.}\ \bibnamefont
  {Evans}}, \bibinfo {author} {\bibfnamefont {S.}~\bibnamefont {Mangin}},
  \bibinfo {author} {\bibfnamefont {W.}~\bibnamefont {Zhao}},\ and\ \bibinfo
  {author} {\bibfnamefont {M.}~\bibnamefont {Hehn}},\ }\bibfield  {title}
  {\enquote {\bibinfo {title} {{Single‐Shot Laser‐Induced Switching of an
  Exchange Biased Antiferromagnet}},}\ }\href
  {https://doi.org/10.1002/adma.202311643} {\bibfield  {journal} {\bibinfo
  {journal} {Advanced Materials}\ }\textbf {\bibinfo {volume} {36}} (\bibinfo
  {year} {2024}{\natexlab{b}}),\ 10.1002/adma.202311643},\ \Eprint
  {https://arxiv.org/abs/2302.04510} {arXiv:2302.04510} \BibitemShut {NoStop}%
\bibitem [{\citenamefont {Beens}\ \emph {et~al.}(2019)\citenamefont {Beens},
  \citenamefont {Lalieu}, \citenamefont {Deenen}, \citenamefont {Duine},\ and\
  \citenamefont {Koopmans}}]{Beens2019}%
  \BibitemOpen
  \bibfield  {author} {\bibinfo {author} {\bibfnamefont {M.}~\bibnamefont
  {Beens}}, \bibinfo {author} {\bibfnamefont {M.~L.~M.}\ \bibnamefont
  {Lalieu}}, \bibinfo {author} {\bibfnamefont {A.~J.~M.}\ \bibnamefont
  {Deenen}}, \bibinfo {author} {\bibfnamefont {R.~A.}\ \bibnamefont {Duine}},\
  and\ \bibinfo {author} {\bibfnamefont {B.}~\bibnamefont {Koopmans}},\
  }\bibfield  {title} {\enquote {\bibinfo {title} {{Comparing all-optical
  switching in synthetic-ferrimagnetic multilayers and alloys}},}\ }\href
  {https://doi.org/10.1103/PhysRevB.100.220409} {\bibfield  {journal} {\bibinfo
   {journal} {Physical Review B}\ }\textbf {\bibinfo {volume} {100}},\ \bibinfo
  {pages} {220409} (\bibinfo {year} {2019})},\ \Eprint
  {https://arxiv.org/abs/1908.07292} {arXiv:1908.07292} \BibitemShut {NoStop}%
\bibitem [{\citenamefont {Wang}\ \emph
  {et~al.}(2022{\natexlab{b}})\citenamefont {Wang}, \citenamefont {Cheng},
  \citenamefont {Li}, \citenamefont {van Hees}, \citenamefont {Liu},
  \citenamefont {Cao}, \citenamefont {Lavrijsen}, \citenamefont {Lin},
  \citenamefont {Koopmans},\ and\ \citenamefont {Zhao}}]{Wang2022a}%
  \BibitemOpen
  \bibfield  {author} {\bibinfo {author} {\bibfnamefont {L.}~\bibnamefont
  {Wang}}, \bibinfo {author} {\bibfnamefont {H.}~\bibnamefont {Cheng}},
  \bibinfo {author} {\bibfnamefont {P.}~\bibnamefont {Li}}, \bibinfo {author}
  {\bibfnamefont {Y.~L.~W.}\ \bibnamefont {van Hees}}, \bibinfo {author}
  {\bibfnamefont {Y.}~\bibnamefont {Liu}}, \bibinfo {author} {\bibfnamefont
  {K.}~\bibnamefont {Cao}}, \bibinfo {author} {\bibfnamefont {R.}~\bibnamefont
  {Lavrijsen}}, \bibinfo {author} {\bibfnamefont {X.}~\bibnamefont {Lin}},
  \bibinfo {author} {\bibfnamefont {B.}~\bibnamefont {Koopmans}},\ and\
  \bibinfo {author} {\bibfnamefont {W.}~\bibnamefont {Zhao}},\ }\bibfield
  {title} {\enquote {\bibinfo {title} {{Picosecond optospintronic tunnel
  junctions}},}\ }\href {https://doi.org/10.1073/pnas.2204732119} {\bibfield
  {journal} {\bibinfo  {journal} {Proceedings of the National Academy of
  Sciences}\ }\textbf {\bibinfo {volume} {119}},\ \bibinfo {pages} {1--7}
  (\bibinfo {year} {2022}{\natexlab{b}})}\BibitemShut {NoStop}%
\bibitem [{\citenamefont {Li}\ \emph {et~al.}(2023)\citenamefont {Li},
  \citenamefont {Kools}, \citenamefont {Koopmans},\ and\ \citenamefont
  {Lavrijsen}}]{Li2023}%
  \BibitemOpen
  \bibfield  {author} {\bibinfo {author} {\bibfnamefont {P.}~\bibnamefont
  {Li}}, \bibinfo {author} {\bibfnamefont {T.~J.}\ \bibnamefont {Kools}},
  \bibinfo {author} {\bibfnamefont {B.}~\bibnamefont {Koopmans}},\ and\
  \bibinfo {author} {\bibfnamefont {R.}~\bibnamefont {Lavrijsen}},\ }\bibfield
  {title} {\enquote {\bibinfo {title} {{Ultrafast Racetrack Based on
  Compensated Co/Gd-Based Synthetic Ferrimagnet with All-Optical Switching}},}\
  }\href {https://doi.org/10.1002/aelm.202200613} {\bibfield  {journal}
  {\bibinfo  {journal} {Advanced Electronic Materials}\ } (\bibinfo {year}
  {2023}),\ 10.1002/aelm.202200613},\ \Eprint
  {https://arxiv.org/abs/2204.11595} {arXiv:2204.11595} \BibitemShut {NoStop}%
\bibitem [{\citenamefont {Yang}\ \emph {et~al.}(2019)\citenamefont {Yang},
  \citenamefont {Li}, \citenamefont {Yan}, \citenamefont {Bian}, \citenamefont
  {Zhang}, \citenamefont {Lou}, \citenamefont {Zhang}, \citenamefont {Zhang},\
  and\ \citenamefont {Jin}}]{Yang2019}%
  \BibitemOpen
  \bibfield  {author} {\bibinfo {author} {\bibfnamefont {X.~Y.}\ \bibnamefont
  {Yang}}, \bibinfo {author} {\bibfnamefont {W.}~\bibnamefont {Li}}, \bibinfo
  {author} {\bibfnamefont {J.~Q.}\ \bibnamefont {Yan}}, \bibinfo {author}
  {\bibfnamefont {Y.~J.}\ \bibnamefont {Bian}}, \bibinfo {author}
  {\bibfnamefont {Y.~Y.}\ \bibnamefont {Zhang}}, \bibinfo {author}
  {\bibfnamefont {S.~T.}\ \bibnamefont {Lou}}, \bibinfo {author} {\bibfnamefont
  {Z.~Z.}\ \bibnamefont {Zhang}}, \bibinfo {author} {\bibfnamefont {X.~L.}\
  \bibnamefont {Zhang}},\ and\ \bibinfo {author} {\bibfnamefont {Q.~Y.}\
  \bibnamefont {Jin}},\ }\bibfield  {title} {\enquote {\bibinfo {title}
  {{Magnetic damping in perpendicular [Pt/Co]3/MnIr multilayers}},}\ }\href
  {https://doi.org/10.1016/j.jmmm.2019.165286} {\bibfield  {journal} {\bibinfo
  {journal} {Journal of Magnetism and Magnetic Materials}\ }\textbf {\bibinfo
  {volume} {487}} (\bibinfo {year} {2019}),\
  10.1016/j.jmmm.2019.165286}\BibitemShut {NoStop}%
\bibitem [{\citenamefont {Jenkins}\ \emph {et~al.}(2019)\citenamefont
  {Jenkins}, \citenamefont {Chantrell}, \citenamefont {Klemmer},\ and\
  \citenamefont {Evans}}]{Jenkins2019}%
  \BibitemOpen
  \bibfield  {author} {\bibinfo {author} {\bibfnamefont {S.}~\bibnamefont
  {Jenkins}}, \bibinfo {author} {\bibfnamefont {R.~W.}\ \bibnamefont
  {Chantrell}}, \bibinfo {author} {\bibfnamefont {T.~J.}\ \bibnamefont
  {Klemmer}},\ and\ \bibinfo {author} {\bibfnamefont {R.~F.~L.}\ \bibnamefont
  {Evans}},\ }\bibfield  {title} {\enquote {\bibinfo {title} {{Magnetic
  anisotropy of the noncollinear antiferromagnet IrMn3}},}\ }\href
  {https://doi.org/10.1103/PhysRevB.100.220405} {\bibfield  {journal} {\bibinfo
   {journal} {Physical Review B}\ }\textbf {\bibinfo {volume} {100}},\ \bibinfo
  {pages} {220405} (\bibinfo {year} {2019})}\BibitemShut {NoStop}%
\bibitem [{\citenamefont {Jenkins}\ \emph {et~al.}(2020)\citenamefont
  {Jenkins}, \citenamefont {Fan}, \citenamefont {Gaina}, \citenamefont
  {Chantrell}, \citenamefont {Klemmer},\ and\ \citenamefont
  {Evans}}]{Jenkins2020}%
  \BibitemOpen
  \bibfield  {author} {\bibinfo {author} {\bibfnamefont {S.}~\bibnamefont
  {Jenkins}}, \bibinfo {author} {\bibfnamefont {W.~J.}\ \bibnamefont {Fan}},
  \bibinfo {author} {\bibfnamefont {R.}~\bibnamefont {Gaina}}, \bibinfo
  {author} {\bibfnamefont {R.~W.}\ \bibnamefont {Chantrell}}, \bibinfo {author}
  {\bibfnamefont {T.}~\bibnamefont {Klemmer}},\ and\ \bibinfo {author}
  {\bibfnamefont {R.~F.~L.}\ \bibnamefont {Evans}},\ }\bibfield  {title}
  {\enquote {\bibinfo {title} {{Atomistic origin of exchange anisotropy in
  noncollinear $\gamma$-IrMn3-CoFe bilayers}},}\ }\href
  {https://doi.org/10.1103/PhysRevB.102.140404} {\bibfield  {journal} {\bibinfo
   {journal} {Physical Review B}\ }\textbf {\bibinfo {volume} {102}},\ \bibinfo
  {pages} {140404} (\bibinfo {year} {2020})}\BibitemShut {NoStop}%
\bibitem [{\citenamefont {Jenkins}, \citenamefont {Chantrell},\ and\
  \citenamefont {Evans}(2021)}]{Jenkins2021}%
  \BibitemOpen
  \bibfield  {author} {\bibinfo {author} {\bibfnamefont {S.}~\bibnamefont
  {Jenkins}}, \bibinfo {author} {\bibfnamefont {R.~W.}\ \bibnamefont
  {Chantrell}},\ and\ \bibinfo {author} {\bibfnamefont {R.~F.}\ \bibnamefont
  {Evans}},\ }\bibfield  {title} {\enquote {\bibinfo {title} {{Atomistic
  simulations of the magnetic properties of IrxMn1-x alloys}},}\ }\href
  {https://doi.org/10.1103/PhysRevMaterials.5.034406} {\bibfield  {journal}
  {\bibinfo  {journal} {Physical Review Materials}\ } (\bibinfo {year}
  {2021}),\ 10.1103/PhysRevMaterials.5.034406}\BibitemShut {NoStop}%
\bibitem [{\citenamefont {Szunyogh}\ \emph {et~al.}(2011)\citenamefont
  {Szunyogh}, \citenamefont {Udvardi}, \citenamefont {Jackson}, \citenamefont
  {Nowak},\ and\ \citenamefont {Chantrell}}]{Szunyogh2011}%
  \BibitemOpen
  \bibfield  {author} {\bibinfo {author} {\bibfnamefont {L.}~\bibnamefont
  {Szunyogh}}, \bibinfo {author} {\bibfnamefont {L.}~\bibnamefont {Udvardi}},
  \bibinfo {author} {\bibfnamefont {J.}~\bibnamefont {Jackson}}, \bibinfo
  {author} {\bibfnamefont {U.}~\bibnamefont {Nowak}},\ and\ \bibinfo {author}
  {\bibfnamefont {R.}~\bibnamefont {Chantrell}},\ }\bibfield  {title} {\enquote
  {\bibinfo {title} {{Atomistic spin model based on a spin-cluster expansion
  technique: Application to the IrMn3/Co interface}},}\ }\href
  {https://doi.org/10.1103/PhysRevB.83.024401} {\bibfield  {journal} {\bibinfo
  {journal} {Physical Review B - Condensed Matter and Materials Physics}\ }
  (\bibinfo {year} {2011}),\ 10.1103/PhysRevB.83.024401},\ \Eprint
  {https://arxiv.org/abs/1010.2375} {arXiv:1010.2375} \BibitemShut {NoStop}%
\bibitem [{\citenamefont {Koopmans}\ \emph {et~al.}(2010)\citenamefont
  {Koopmans}, \citenamefont {Malinowski}, \citenamefont {{Dalla Longa}},
  \citenamefont {Steiauf}, \citenamefont {F{\"{a}}hnle}, \citenamefont {Roth},
  \citenamefont {Cinchetti},\ and\ \citenamefont {Aeschlimann}}]{Koopmans2010}%
  \BibitemOpen
  \bibfield  {author} {\bibinfo {author} {\bibfnamefont {B.}~\bibnamefont
  {Koopmans}}, \bibinfo {author} {\bibfnamefont {G.}~\bibnamefont
  {Malinowski}}, \bibinfo {author} {\bibfnamefont {F.}~\bibnamefont {{Dalla
  Longa}}}, \bibinfo {author} {\bibfnamefont {D.}~\bibnamefont {Steiauf}},
  \bibinfo {author} {\bibfnamefont {M.}~\bibnamefont {F{\"{a}}hnle}}, \bibinfo
  {author} {\bibfnamefont {T.}~\bibnamefont {Roth}}, \bibinfo {author}
  {\bibfnamefont {M.}~\bibnamefont {Cinchetti}},\ and\ \bibinfo {author}
  {\bibfnamefont {M.}~\bibnamefont {Aeschlimann}},\ }\bibfield  {title}
  {\enquote {\bibinfo {title} {{Explaining the paradoxical diversity of
  ultrafast laser-induced demagnetization}},}\ }\href
  {https://doi.org/10.1038/nmat2593} {\bibfield  {journal} {\bibinfo  {journal}
  {Nature Materials}\ } (\bibinfo {year} {2010}),\
  10.1038/nmat2593}\BibitemShut {NoStop}%
\bibitem [{\citenamefont {Khamtawi}\ \emph {et~al.}(2023)\citenamefont
  {Khamtawi}, \citenamefont {Daeng-am}, \citenamefont {Chureemart},
  \citenamefont {Chantrell},\ and\ \citenamefont {Chureemart}}]{Khamtawi2023}%
  \BibitemOpen
  \bibfield  {author} {\bibinfo {author} {\bibfnamefont {R.}~\bibnamefont
  {Khamtawi}}, \bibinfo {author} {\bibfnamefont {W.}~\bibnamefont {Daeng-am}},
  \bibinfo {author} {\bibfnamefont {P.}~\bibnamefont {Chureemart}}, \bibinfo
  {author} {\bibfnamefont {R.~W.}\ \bibnamefont {Chantrell}},\ and\ \bibinfo
  {author} {\bibfnamefont {J.}~\bibnamefont {Chureemart}},\ }\bibfield  {title}
  {\enquote {\bibinfo {title} {{Exchange bias model including setting process:
  Investigation of antiferromagnetic alignment fraction due to thermal
  activation}},}\ }\href {https://doi.org/10.1063/5.0136278} {\bibfield
  {journal} {\bibinfo  {journal} {Journal of Applied Physics}\ }\textbf
  {\bibinfo {volume} {133}},\ \bibinfo {pages} {023903} (\bibinfo {year}
  {2023})}\BibitemShut {NoStop}%
\bibitem [{\citenamefont {Peeters}, \citenamefont {van Ballegooie},\ and\
  \citenamefont {Koopmans}(2022)}]{Peeters2022}%
  \BibitemOpen
  \bibfield  {author} {\bibinfo {author} {\bibfnamefont {M.~J.~G.}\
  \bibnamefont {Peeters}}, \bibinfo {author} {\bibfnamefont {Y.~M.}\
  \bibnamefont {van Ballegooie}},\ and\ \bibinfo {author} {\bibfnamefont
  {B.}~\bibnamefont {Koopmans}},\ }\bibfield  {title} {\enquote {\bibinfo
  {title} {{Influence of magnetic fields on ultrafast laser-induced switching
  dynamics in Co/Gd bilayers}},}\ }\href
  {https://doi.org/10.1103/PhysRevB.105.014429} {\bibfield  {journal} {\bibinfo
   {journal} {Physical Review B}\ }\textbf {\bibinfo {volume} {105}},\ \bibinfo
  {pages} {014429} (\bibinfo {year} {2022})}\BibitemShut {NoStop}%
\bibitem [{\citenamefont {Migliorini}\ \emph {et~al.}(2018)\citenamefont
  {Migliorini}, \citenamefont {Kuerbanjiang}, \citenamefont {Huminiuc},
  \citenamefont {Kepaptsoglou}, \citenamefont {Mu{\~{n}}oz}, \citenamefont
  {Cu{\~{n}}ado}, \citenamefont {Camarero}, \citenamefont {Aroca},
  \citenamefont {Vallejo-Fern{\'{a}}ndez}, \citenamefont {Lazarov},\ and\
  \citenamefont {Prieto}}]{Migliorini2018}%
  \BibitemOpen
  \bibfield  {author} {\bibinfo {author} {\bibfnamefont {A.}~\bibnamefont
  {Migliorini}}, \bibinfo {author} {\bibfnamefont {B.}~\bibnamefont
  {Kuerbanjiang}}, \bibinfo {author} {\bibfnamefont {T.}~\bibnamefont
  {Huminiuc}}, \bibinfo {author} {\bibfnamefont {D.}~\bibnamefont
  {Kepaptsoglou}}, \bibinfo {author} {\bibfnamefont {M.}~\bibnamefont
  {Mu{\~{n}}oz}}, \bibinfo {author} {\bibfnamefont {J.~L.}\ \bibnamefont
  {Cu{\~{n}}ado}}, \bibinfo {author} {\bibfnamefont {J.}~\bibnamefont
  {Camarero}}, \bibinfo {author} {\bibfnamefont {C.}~\bibnamefont {Aroca}},
  \bibinfo {author} {\bibfnamefont {G.}~\bibnamefont
  {Vallejo-Fern{\'{a}}ndez}}, \bibinfo {author} {\bibfnamefont {V.~K.}\
  \bibnamefont {Lazarov}},\ and\ \bibinfo {author} {\bibfnamefont {J.~L.}\
  \bibnamefont {Prieto}},\ }\bibfield  {title} {\enquote {\bibinfo {title}
  {{Spontaneous exchange bias formation driven by a structural phase transition
  in the antiferromagnetic material}},}\ }\href
  {https://doi.org/10.1038/NMAT5030} {\bibfield  {journal} {\bibinfo  {journal}
  {Nature Materials}\ } (\bibinfo {year} {2018}),\
  10.1038/NMAT5030}\BibitemShut {NoStop}%
\bibitem [{\citenamefont {Zhao}\ \emph {et~al.}(2023)\citenamefont {Zhao},
  \citenamefont {Wang}, \citenamefont {Chen}, \citenamefont {Wang},
  \citenamefont {Wang},\ and\ \citenamefont {Chen}}]{Zhao2023}%
  \BibitemOpen
  \bibfield  {author} {\bibinfo {author} {\bibfnamefont {S.}~\bibnamefont
  {Zhao}}, \bibinfo {author} {\bibfnamefont {P.}~\bibnamefont {Wang}}, \bibinfo
  {author} {\bibfnamefont {W.}~\bibnamefont {Chen}}, \bibinfo {author}
  {\bibfnamefont {L.}~\bibnamefont {Wang}}, \bibinfo {author} {\bibfnamefont
  {Q.~C.}\ \bibnamefont {Wang}},\ and\ \bibinfo {author} {\bibfnamefont
  {W.~Q.}\ \bibnamefont {Chen}},\ }\bibfield  {title} {\enquote {\bibinfo
  {title} {{Supply and demand conflicts of critical heavy rare earth element:
  Lessons from gadolinium}},}\ }\href
  {https://doi.org/10.1016/j.resconrec.2023.107254} {\bibfield  {journal}
  {\bibinfo  {journal} {Resources, Conservation and Recycling}\ } (\bibinfo
  {year} {2023}),\ 10.1016/j.resconrec.2023.107254}\BibitemShut {NoStop}%
\bibitem [{\citenamefont {Kools}\ \emph {et~al.}(2023)\citenamefont {Kools},
  \citenamefont {van Hees}, \citenamefont {Poissonnier}, \citenamefont {Li},
  \citenamefont {{Barcones Campo}}, \citenamefont {Verheijen}, \citenamefont
  {Koopmans},\ and\ \citenamefont {Lavrijsen}}]{Kools2023}%
  \BibitemOpen
  \bibfield  {author} {\bibinfo {author} {\bibfnamefont {T.~J.}\ \bibnamefont
  {Kools}}, \bibinfo {author} {\bibfnamefont {Y.~L.}\ \bibnamefont {van Hees}},
  \bibinfo {author} {\bibfnamefont {K.}~\bibnamefont {Poissonnier}}, \bibinfo
  {author} {\bibfnamefont {P.}~\bibnamefont {Li}}, \bibinfo {author}
  {\bibfnamefont {B.}~\bibnamefont {{Barcones Campo}}}, \bibinfo {author}
  {\bibfnamefont {M.~A.}\ \bibnamefont {Verheijen}}, \bibinfo {author}
  {\bibfnamefont {B.}~\bibnamefont {Koopmans}},\ and\ \bibinfo {author}
  {\bibfnamefont {R.}~\bibnamefont {Lavrijsen}},\ }\bibfield  {title} {\enquote
  {\bibinfo {title} {{Aging and passivation of magnetic properties in Co/Gd
  bilayers}},}\ }\href {https://doi.org/10.1063/5.0160135} {\bibfield
  {journal} {\bibinfo  {journal} {Applied Physics Letters}\ } (\bibinfo {year}
  {2023}),\ 10.1063/5.0160135},\ \Eprint {https://arxiv.org/abs/2305.18984}
  {arXiv:2305.18984} \BibitemShut {NoStop}%
\bibitem [{\citenamefont {Banerjee}\ \emph {et~al.}(2020)\citenamefont
  {Banerjee}, \citenamefont {Teichert}, \citenamefont {Siewierska},
  \citenamefont {Gercsi}, \citenamefont {Atcheson}, \citenamefont {Stamenov},
  \citenamefont {Rode}, \citenamefont {Coey},\ and\ \citenamefont
  {Besbas}}]{Banerjee2020}%
  \BibitemOpen
  \bibfield  {author} {\bibinfo {author} {\bibfnamefont {C.}~\bibnamefont
  {Banerjee}}, \bibinfo {author} {\bibfnamefont {N.}~\bibnamefont {Teichert}},
  \bibinfo {author} {\bibfnamefont {K.~E.}\ \bibnamefont {Siewierska}},
  \bibinfo {author} {\bibfnamefont {Z.}~\bibnamefont {Gercsi}}, \bibinfo
  {author} {\bibfnamefont {G.~Y.}\ \bibnamefont {Atcheson}}, \bibinfo {author}
  {\bibfnamefont {P.}~\bibnamefont {Stamenov}}, \bibinfo {author}
  {\bibfnamefont {K.}~\bibnamefont {Rode}}, \bibinfo {author} {\bibfnamefont
  {J.~M.}\ \bibnamefont {Coey}},\ and\ \bibinfo {author} {\bibfnamefont
  {J.}~\bibnamefont {Besbas}},\ }\bibfield  {title} {\enquote {\bibinfo {title}
  {{Single pulse all-optical toggle switching of magnetization without
  gadolinium in the ferrimagnet Mn2RuxGa}},}\ }\href
  {https://doi.org/10.1038/s41467-020-18340-9} {\bibfield  {journal} {\bibinfo
  {journal} {Nature Communications}\ } (\bibinfo {year} {2020}),\
  10.1038/s41467-020-18340-9}\BibitemShut {NoStop}%
\bibitem [{\citenamefont {Qin}\ \emph {et~al.}(2023)\citenamefont {Qin},
  \citenamefont {Yan}, \citenamefont {Wang}, \citenamefont {Chen},
  \citenamefont {Meng}, \citenamefont {Dong}, \citenamefont {Zhu},
  \citenamefont {Cai}, \citenamefont {Feng}, \citenamefont {Zhou},
  \citenamefont {Liu}, \citenamefont {Zhang}, \citenamefont {Zeng},
  \citenamefont {Zhang}, \citenamefont {Jiang},\ and\ \citenamefont
  {Liu}}]{Qin2023}%
  \BibitemOpen
  \bibfield  {author} {\bibinfo {author} {\bibfnamefont {P.}~\bibnamefont
  {Qin}}, \bibinfo {author} {\bibfnamefont {H.}~\bibnamefont {Yan}}, \bibinfo
  {author} {\bibfnamefont {X.}~\bibnamefont {Wang}}, \bibinfo {author}
  {\bibfnamefont {H.}~\bibnamefont {Chen}}, \bibinfo {author} {\bibfnamefont
  {Z.}~\bibnamefont {Meng}}, \bibinfo {author} {\bibfnamefont {J.}~\bibnamefont
  {Dong}}, \bibinfo {author} {\bibfnamefont {M.}~\bibnamefont {Zhu}}, \bibinfo
  {author} {\bibfnamefont {J.}~\bibnamefont {Cai}}, \bibinfo {author}
  {\bibfnamefont {Z.}~\bibnamefont {Feng}}, \bibinfo {author} {\bibfnamefont
  {X.}~\bibnamefont {Zhou}}, \bibinfo {author} {\bibfnamefont {L.}~\bibnamefont
  {Liu}}, \bibinfo {author} {\bibfnamefont {T.}~\bibnamefont {Zhang}}, \bibinfo
  {author} {\bibfnamefont {Z.}~\bibnamefont {Zeng}}, \bibinfo {author}
  {\bibfnamefont {J.}~\bibnamefont {Zhang}}, \bibinfo {author} {\bibfnamefont
  {C.}~\bibnamefont {Jiang}},\ and\ \bibinfo {author} {\bibfnamefont
  {Z.}~\bibnamefont {Liu}},\ }\bibfield  {title} {\enquote {\bibinfo {title}
  {{Room-temperature magnetoresistance in an all-antiferromagnetic tunnel
  junction}},}\ }\href {https://doi.org/10.1038/s41586-022-05461-y} {\bibfield
  {journal} {\bibinfo  {journal} {Nature}\ }\textbf {\bibinfo {volume} {613}},\
  \bibinfo {pages} {485--489} (\bibinfo {year} {2023})}\BibitemShut {NoStop}%
\bibitem [{\citenamefont {Qi}\ \emph {et~al.}(2024{\natexlab{b}})\citenamefont
  {Qi}, \citenamefont {Zhang}, \citenamefont {Chen}, \citenamefont {Du},
  \citenamefont {Zheng}, \citenamefont {Xiao}, \citenamefont {Tian},
  \citenamefont {Hu}, \citenamefont {Shen}, \citenamefont {Sun},\ and\
  \citenamefont {Zhao}}]{Qi2024}%
  \BibitemOpen
  \bibfield  {author} {\bibinfo {author} {\bibfnamefont {W.}~\bibnamefont
  {Qi}}, \bibinfo {author} {\bibfnamefont {H.}~\bibnamefont {Zhang}}, \bibinfo
  {author} {\bibfnamefont {L.}~\bibnamefont {Chen}}, \bibinfo {author}
  {\bibfnamefont {A.}~\bibnamefont {Du}}, \bibinfo {author} {\bibfnamefont
  {D.}~\bibnamefont {Zheng}}, \bibinfo {author} {\bibfnamefont
  {Y.}~\bibnamefont {Xiao}}, \bibinfo {author} {\bibfnamefont {D.}~\bibnamefont
  {Tian}}, \bibinfo {author} {\bibfnamefont {F.}~\bibnamefont {Hu}}, \bibinfo
  {author} {\bibfnamefont {B.}~\bibnamefont {Shen}}, \bibinfo {author}
  {\bibfnamefont {J.}~\bibnamefont {Sun}},\ and\ \bibinfo {author}
  {\bibfnamefont {W.}~\bibnamefont {Zhao}},\ }\bibfield  {title} {\enquote
  {\bibinfo {title} {{Antiferromagnetic Spintronics in Magnetic Memory
  Devices}},}\ }\href {https://doi.org/10.1109/TMAT.2024.3415591} {\bibfield
  {journal} {\bibinfo  {journal} {IEEE Transactions on Materials for Electron
  Devices}\ }\textbf {\bibinfo {volume} {1}},\ \bibinfo {pages} {23--35}
  (\bibinfo {year} {2024}{\natexlab{b}})}\BibitemShut {NoStop}%
\bibitem [{\citenamefont {Adamantopoulos}\ \emph {et~al.}(2024)\citenamefont
  {Adamantopoulos}, \citenamefont {Merte}, \citenamefont {Freimuth},
  \citenamefont {Go}, \citenamefont {Zhang}, \citenamefont
  {Le{\v{z}}ai{\'{c}}}, \citenamefont {Feng}, \citenamefont {Yao},
  \citenamefont {Sinova}, \citenamefont {{\v{S}}mejkal}, \citenamefont
  {Bl{\"{u}}gel},\ and\ \citenamefont {Mokrousov}}]{Adamantopoulos2024}%
  \BibitemOpen
  \bibfield  {author} {\bibinfo {author} {\bibfnamefont {T.}~\bibnamefont
  {Adamantopoulos}}, \bibinfo {author} {\bibfnamefont {M.}~\bibnamefont
  {Merte}}, \bibinfo {author} {\bibfnamefont {F.}~\bibnamefont {Freimuth}},
  \bibinfo {author} {\bibfnamefont {D.}~\bibnamefont {Go}}, \bibinfo {author}
  {\bibfnamefont {L.}~\bibnamefont {Zhang}}, \bibinfo {author} {\bibfnamefont
  {M.}~\bibnamefont {Le{\v{z}}ai{\'{c}}}}, \bibinfo {author} {\bibfnamefont
  {W.}~\bibnamefont {Feng}}, \bibinfo {author} {\bibfnamefont {Y.}~\bibnamefont
  {Yao}}, \bibinfo {author} {\bibfnamefont {J.}~\bibnamefont {Sinova}},
  \bibinfo {author} {\bibfnamefont {L.}~\bibnamefont {{\v{S}}mejkal}}, \bibinfo
  {author} {\bibfnamefont {S.}~\bibnamefont {Bl{\"{u}}gel}},\ and\ \bibinfo
  {author} {\bibfnamefont {Y.}~\bibnamefont {Mokrousov}},\ }\bibfield  {title}
  {\enquote {\bibinfo {title} {{Spin and orbital magnetism by light in rutile
  altermagnets}},}\ }\href {https://doi.org/10.1038/s44306-024-00053-0}
  {\bibfield  {journal} {\bibinfo  {journal} {npj Spintronics}\ }\textbf
  {\bibinfo {volume} {2}},\ \bibinfo {pages} {46} (\bibinfo {year} {2024})},\
  \Eprint {https://arxiv.org/abs/2403.10235} {arXiv:2403.10235} \BibitemShut
  {NoStop}%
\bibitem [{\citenamefont {Kimel}, \citenamefont {Rasing},\ and\ \citenamefont
  {Ivanov}(2024)}]{Kimel2024}%
  \BibitemOpen
  \bibfield  {author} {\bibinfo {author} {\bibfnamefont {A.}~\bibnamefont
  {Kimel}}, \bibinfo {author} {\bibfnamefont {T.}~\bibnamefont {Rasing}},\ and\
  \bibinfo {author} {\bibfnamefont {B.}~\bibnamefont {Ivanov}},\ }\bibfield
  {title} {\enquote {\bibinfo {title} {{Optical read-out and control of
  antiferromagnetic N{\'{e}}el vector in altermagnets and beyond}},}\ }\href
  {https://doi.org/10.1016/j.jmmm.2024.172039} {\bibfield  {journal} {\bibinfo
  {journal} {Journal of Magnetism and Magnetic Materials}\ }\textbf {\bibinfo
  {volume} {598}},\ \bibinfo {pages} {172039} (\bibinfo {year}
  {2024})}\BibitemShut {NoStop}%
\bibitem [{\citenamefont {Lin}\ \emph {et~al.}(2023)\citenamefont {Lin},
  \citenamefont {Hehn}, \citenamefont {Hauet}, \citenamefont {Peng},
  \citenamefont {Igarashi}, \citenamefont {{Le Guen}}, \citenamefont {Remy},
  \citenamefont {Gorchon}, \citenamefont {Malinowski}, \citenamefont {Mangin},\
  and\ \citenamefont {Hohlfeld}}]{Lin2023}%
  \BibitemOpen
  \bibfield  {author} {\bibinfo {author} {\bibfnamefont {J.~X.}\ \bibnamefont
  {Lin}}, \bibinfo {author} {\bibfnamefont {M.}~\bibnamefont {Hehn}}, \bibinfo
  {author} {\bibfnamefont {T.}~\bibnamefont {Hauet}}, \bibinfo {author}
  {\bibfnamefont {Y.}~\bibnamefont {Peng}}, \bibinfo {author} {\bibfnamefont
  {J.}~\bibnamefont {Igarashi}}, \bibinfo {author} {\bibfnamefont
  {Y.}~\bibnamefont {{Le Guen}}}, \bibinfo {author} {\bibfnamefont
  {Q.}~\bibnamefont {Remy}}, \bibinfo {author} {\bibfnamefont {J.}~\bibnamefont
  {Gorchon}}, \bibinfo {author} {\bibfnamefont {G.}~\bibnamefont {Malinowski}},
  \bibinfo {author} {\bibfnamefont {S.}~\bibnamefont {Mangin}},\ and\ \bibinfo
  {author} {\bibfnamefont {J.}~\bibnamefont {Hohlfeld}},\ }\bibfield  {title}
  {\enquote {\bibinfo {title} {{Single laser pulse induced magnetization
  switching in in-plane magnetized GdCo alloys}},}\ }\href
  {https://doi.org/10.1103/PhysRevB.108.L220403} {\bibfield  {journal}
  {\bibinfo  {journal} {Physical Review B}\ } (\bibinfo {year} {2023}),\
  10.1103/PhysRevB.108.L220403}\BibitemShut {NoStop}%
\bibitem [{\citenamefont {Vorobiov}\ \emph {et~al.}(2015)\citenamefont
  {Vorobiov}, \citenamefont {Lytvynenko}, \citenamefont {Hauet}, \citenamefont
  {Hehn}, \citenamefont {Derecha},\ and\ \citenamefont
  {Chornous}}]{Vorobiov2015}%
  \BibitemOpen
  \bibfield  {author} {\bibinfo {author} {\bibfnamefont {S.}~\bibnamefont
  {Vorobiov}}, \bibinfo {author} {\bibfnamefont {I.}~\bibnamefont
  {Lytvynenko}}, \bibinfo {author} {\bibfnamefont {T.}~\bibnamefont {Hauet}},
  \bibinfo {author} {\bibfnamefont {M.}~\bibnamefont {Hehn}}, \bibinfo {author}
  {\bibfnamefont {D.}~\bibnamefont {Derecha}},\ and\ \bibinfo {author}
  {\bibfnamefont {A.}~\bibnamefont {Chornous}},\ }\bibfield  {title} {\enquote
  {\bibinfo {title} {{The effect of annealing on magnetic properties of Co/Gd
  multilayers}},}\ }\href {https://doi.org/10.1016/j.vacuum.2015.06.009}
  {\bibfield  {journal} {\bibinfo  {journal} {Vacuum}\ }\textbf {\bibinfo
  {volume} {120}},\ \bibinfo {pages} {9--12} (\bibinfo {year}
  {2015})}\BibitemShut {NoStop}%
\bibitem [{\citenamefont {Ourdani}\ \emph {et~al.}(2021)\citenamefont
  {Ourdani}, \citenamefont {Roussign{\'{e}}}, \citenamefont {Ch{\'{e}}rif},
  \citenamefont {Gabor},\ and\ \citenamefont {Belmeguenai}}]{Ourdani2021}%
  \BibitemOpen
  \bibfield  {author} {\bibinfo {author} {\bibfnamefont {D.}~\bibnamefont
  {Ourdani}}, \bibinfo {author} {\bibfnamefont {Y.}~\bibnamefont
  {Roussign{\'{e}}}}, \bibinfo {author} {\bibfnamefont {S.~M.}\ \bibnamefont
  {Ch{\'{e}}rif}}, \bibinfo {author} {\bibfnamefont {M.~S.}\ \bibnamefont
  {Gabor}},\ and\ \bibinfo {author} {\bibfnamefont {M.}~\bibnamefont
  {Belmeguenai}},\ }\bibfield  {title} {\enquote {\bibinfo {title} {{Dependence
  of the interfacial Dzyaloshinskii-Moriya interaction, perpendicular magnetic
  anisotropy, and damping in Co-based systems on the thickness of Pt and Ir
  layers}},}\ }\href {https://doi.org/10.1103/PhysRevB.104.104421} {\bibfield
  {journal} {\bibinfo  {journal} {Physical Review B}\ }\textbf {\bibinfo
  {volume} {104}},\ \bibinfo {pages} {104421} (\bibinfo {year}
  {2021})}\BibitemShut {NoStop}%
\bibitem [{\citenamefont {Ali}, \citenamefont {Marrows},\ and\ \citenamefont
  {Hickey}(2008)}]{Ali2008}%
  \BibitemOpen
  \bibfield  {author} {\bibinfo {author} {\bibfnamefont {M.}~\bibnamefont
  {Ali}}, \bibinfo {author} {\bibfnamefont {C.~H.}\ \bibnamefont {Marrows}},\
  and\ \bibinfo {author} {\bibfnamefont {B.~J.}\ \bibnamefont {Hickey}},\
  }\bibfield  {title} {\enquote {\bibinfo {title} {{Controlled enhancement or
  suppression of exchange biasing using impurity $\delta$ layers}},}\ }\href
  {https://doi.org/10.1103/PhysRevB.77.134401} {\bibfield  {journal} {\bibinfo
  {journal} {Physical Review B}\ }\textbf {\bibinfo {volume} {77}},\ \bibinfo
  {pages} {134401} (\bibinfo {year} {2008})}\BibitemShut {NoStop}%
\bibitem [{\citenamefont {Sort}\ \emph {et~al.}(2005)\citenamefont {Sort},
  \citenamefont {Baltz}, \citenamefont {Garcia}, \citenamefont {Rodmacq},\ and\
  \citenamefont {Dieny}}]{Sort2005}%
  \BibitemOpen
  \bibfield  {author} {\bibinfo {author} {\bibfnamefont {J.}~\bibnamefont
  {Sort}}, \bibinfo {author} {\bibfnamefont {V.}~\bibnamefont {Baltz}},
  \bibinfo {author} {\bibfnamefont {F.}~\bibnamefont {Garcia}}, \bibinfo
  {author} {\bibfnamefont {B.}~\bibnamefont {Rodmacq}},\ and\ \bibinfo {author}
  {\bibfnamefont {B.}~\bibnamefont {Dieny}},\ }\bibfield  {title} {\enquote
  {\bibinfo {title} {{Tailoring perpendicular exchange bias in [Pt/Co]-IrMn
  multilayers}},}\ }\href {https://doi.org/10.1103/PhysRevB.71.054411}
  {\bibfield  {journal} {\bibinfo  {journal} {Physical Review B}\ }\textbf
  {\bibinfo {volume} {71}},\ \bibinfo {pages} {054411} (\bibinfo {year}
  {2005})}\BibitemShut {NoStop}%
\bibitem [{\citenamefont {Zhang}\ and\ \citenamefont
  {Willis}(2001)}]{Zhang2001}%
  \BibitemOpen
  \bibfield  {author} {\bibinfo {author} {\bibfnamefont {R.}~\bibnamefont
  {Zhang}}\ and\ \bibinfo {author} {\bibfnamefont {R.~F.}\ \bibnamefont
  {Willis}},\ }\bibfield  {title} {\enquote {\bibinfo {title}
  {{Thickness-Dependent Curie Temperatures of Ultrathin Magnetic Films: Effect
  of the Range of Spin-Spin Interactions}},}\ }\href
  {https://doi.org/10.1103/PhysRevLett.86.2665} {\bibfield  {journal} {\bibinfo
   {journal} {Physical Review Letters}\ }\textbf {\bibinfo {volume} {86}},\
  \bibinfo {pages} {2665--2668} (\bibinfo {year} {2001})}\BibitemShut {NoStop}%
\bibitem [{\citenamefont {Coey}(2001)}]{Coey2001}%
  \BibitemOpen
  \bibfield  {author} {\bibinfo {author} {\bibfnamefont {J.~M.~D.}\
  \bibnamefont {Coey}},\ }\href {https://doi.org/10.1017/CBO9780511845000}
  {\emph {\bibinfo {title} {{Magnetism and Magnetic Materials}}}}\ (\bibinfo
  {publisher} {Cambridge University Press},\ \bibinfo {year}
  {2001})\BibitemShut {NoStop}%
\bibitem [{\citenamefont {{Dalla Longa}}\ \emph {et~al.}(2007)\citenamefont
  {{Dalla Longa}}, \citenamefont {Kohlhepp}, \citenamefont {{De Jonge}},\ and\
  \citenamefont {Koopmans}}]{DallaLonga2007}%
  \BibitemOpen
  \bibfield  {author} {\bibinfo {author} {\bibfnamefont {F.}~\bibnamefont
  {{Dalla Longa}}}, \bibinfo {author} {\bibfnamefont {J.~T.}\ \bibnamefont
  {Kohlhepp}}, \bibinfo {author} {\bibfnamefont {W.~J.}\ \bibnamefont {{De
  Jonge}}},\ and\ \bibinfo {author} {\bibfnamefont {B.}~\bibnamefont
  {Koopmans}},\ }\bibfield  {title} {\enquote {\bibinfo {title} {{Influence of
  photon angular momentum on ultrafast demagnetization in nickel}},}\ }\href
  {https://doi.org/10.1103/PhysRevB.75.224431} {\bibfield  {journal} {\bibinfo
  {journal} {Physical Review B - Condensed Matter and Materials Physics}\ }
  (\bibinfo {year} {2007}),\ 10.1103/PhysRevB.75.224431}\BibitemShut {NoStop}%
\bibitem [{\citenamefont {Gerlach}\ \emph {et~al.}(2017)\citenamefont
  {Gerlach}, \citenamefont {Oroszlany}, \citenamefont {Hinzke}, \citenamefont
  {Sievering}, \citenamefont {Wienholdt}, \citenamefont {Szunyogh},\ and\
  \citenamefont {Nowak}}]{Gerlach2017}%
  \BibitemOpen
  \bibfield  {author} {\bibinfo {author} {\bibfnamefont {S.}~\bibnamefont
  {Gerlach}}, \bibinfo {author} {\bibfnamefont {L.}~\bibnamefont {Oroszlany}},
  \bibinfo {author} {\bibfnamefont {D.}~\bibnamefont {Hinzke}}, \bibinfo
  {author} {\bibfnamefont {S.}~\bibnamefont {Sievering}}, \bibinfo {author}
  {\bibfnamefont {S.}~\bibnamefont {Wienholdt}}, \bibinfo {author}
  {\bibfnamefont {L.}~\bibnamefont {Szunyogh}},\ and\ \bibinfo {author}
  {\bibfnamefont {U.}~\bibnamefont {Nowak}},\ }\bibfield  {title} {\enquote
  {\bibinfo {title} {{Modeling ultrafast all-optical switching in synthetic
  ferrimagnets}},}\ }\href {https://doi.org/10.1103/PhysRevB.95.224435}
  {\bibfield  {journal} {\bibinfo  {journal} {Physical Review B}\ } (\bibinfo
  {year} {2017}),\ 10.1103/PhysRevB.95.224435},\ \Eprint
  {https://arxiv.org/abs/1703.05220} {arXiv:1703.05220} \BibitemShut {NoStop}%
\bibitem [{\citenamefont {Bovensiepen}(2007)}]{Bovensiepen2007}%
  \BibitemOpen
  \bibfield  {author} {\bibinfo {author} {\bibfnamefont {U.}~\bibnamefont
  {Bovensiepen}},\ }\bibfield  {title} {\enquote {\bibinfo {title} {{Coherent
  and incoherent excitations of the Gd(0001) surface on ultrafast
  timescales}},}\ }\href {https://doi.org/10.1088/0953-8984/19/8/083201}
  {\bibfield  {journal} {\bibinfo  {journal} {Journal of Physics Condensed
  Matter}\ } (\bibinfo {year} {2007}),\
  10.1088/0953-8984/19/8/083201}\BibitemShut {NoStop}%
\bibitem [{\citenamefont {Pettersson}, \citenamefont {Roman},\ and\
  \citenamefont {Ingan{\"{a}}s}(1999)}]{Pettersson1999}%
  \BibitemOpen
  \bibfield  {author} {\bibinfo {author} {\bibfnamefont {L.~A.~A.}\
  \bibnamefont {Pettersson}}, \bibinfo {author} {\bibfnamefont {L.~S.}\
  \bibnamefont {Roman}},\ and\ \bibinfo {author} {\bibfnamefont
  {O.}~\bibnamefont {Ingan{\"{a}}s}},\ }\bibfield  {title} {\enquote {\bibinfo
  {title} {{Modeling photocurrent action spectra of photovoltaic devices based
  on organic thin films}},}\ }\href {https://doi.org/10.1063/1.370757}
  {\bibfield  {journal} {\bibinfo  {journal} {Journal of Applied Physics}\
  }\textbf {\bibinfo {volume} {86}},\ \bibinfo {pages} {487--496} (\bibinfo
  {year} {1999})}\BibitemShut {NoStop}%
\bibitem [{\citenamefont {Lang}, \citenamefont {Zheng},\ and\ \citenamefont
  {Jiang}(2007)}]{Lang2007}%
  \BibitemOpen
  \bibfield  {author} {\bibinfo {author} {\bibfnamefont {X.~Y.}\ \bibnamefont
  {Lang}}, \bibinfo {author} {\bibfnamefont {W.~T.}\ \bibnamefont {Zheng}},\
  and\ \bibinfo {author} {\bibfnamefont {Q.}~\bibnamefont {Jiang}},\ }\bibfield
   {title} {\enquote {\bibinfo {title} {{Dependence of the blocking temperature
  in exchange biased ferromagnetic/antiferromagnetic bilayers on the thickness
  of the antiferromagnetic layer}},}\ }\href
  {https://doi.org/10.1088/0957-4484/18/15/155701} {\bibfield  {journal}
  {\bibinfo  {journal} {Nanotechnology}\ }\textbf {\bibinfo {volume} {18}},\
  \bibinfo {pages} {155701} (\bibinfo {year} {2007})}\BibitemShut {NoStop}%
\end{thebibliography}%

\section*{Funding}
This work is financially supported by the Eindhoven Hendrik Casimir Institute.

\section*{Author contributions}
F.J.F.v.R. wrote the manuscript, performed the experiments and analyzed the data, S.M.V. performed and analyzed the blocking temperature measurements, B.K. and D.C.L. supervised the work and revised the manuscript.

\section*{Competing interests}
The authors declare no competing interests.

\appendix

\onecolumngrid
\clearpage
\begin{center}
	\textbf{\large Supplemental Information}
\end{center}
%%%%%%%%%% Merge with supplemental materials %%%%%%%%%%
%%%%%%%%%% Prefix a "S" to all equations, figures, tables and reset the counter %%%%%%%%%%
\setcounter{section}{0}
\setcounter{equation}{0}
\setcounter{figure}{0}
\setcounter{table}{0}
\setcounter{page}{1}
\makeatletter
\renewcommand{\thesection}{Supplementary Note \arabic{section}}
\renewcommand{\theequation}{S\arabic{equation}}
\renewcommand{\thefigure}{S\arabic{figure}}
\renewcommand{\bibnumfmt}[1]{[#1]}
\renewcommand{\citenumfont}[1]{#1}
%%%%%%%%%% Prefix a "S" to all equations, figures, tables and reset the counter %%%%%%%%%%

\section{Considerations for stack engineering}\label{sec:stack}
The stacks we use in our experiments have been carefully engineered for maximized performance of \HEB{} reversal. At the bottom of the stack, a \SI{4}{\nano\meter} \ch{Ta} layer acts as both an adhesion layer between the substrate and the remaining thin films, as well as a buffer layer that sets the $(111)$ texture in the initial \ch{Pt}/\ch{Co} layers.

\textbf{Bottom-pinned versus top-pinned} - For the remainder of the stack a choice had to be made between a bottom-pinned or top-pinned configuration. Bottom-pinned has the advantage that the optically active layers can be put closer to the surface, which reduces the amount of energy needed for switching compared to when those layers are buried at the bottom of the stack.

The fact that our top-pinned stacks had strong perpendicular anisotropy and were capable of producing a remarkably high exchange bias of above \SI{100}{\milli\tesla} without annealing (only depositing in a magnetic field) led us to settle on top-pinned stacks without further pursuing bottom-pinned stacks.

\textbf{Enabling AOS} - The top-pinned arrangement means that directly on top of the \ch{Ta} buffer layer we grow our optically active layers. These start with a \ch{Pt}/\ch{Co} bilayer that induces the PMA. A \ch{Pt} layer thickness of \SI{4}{\nano\meter} is enough to maximize the PMA in \ch{Pt}/\ch{Co} bilayers \cite{Ourdani2021}. We kept the \ch{Co} layer relatively thin at \SI{0.6}{\nano\meter}, which is enough to pass the percolation limit and be ferromagnetic at room temperature, while still keeping the total magnetic moment limited. After all, for optical switching to take place the entire magnetic moment throughout the stack has to reverse, so making an effort to minimize the total magnetic moment is beneficial for the switching process. On top of the \ch{Co} layer a \ch{Gd}/\ch{Co} bilayer is grown, whose thicknesses we optimized for maximum reversal (see below). The motivation for putting an extra \ch{Co} layer on top was to have a well-defined \ch{Co} interface for growing the \ch{IrMn} on top of (a \ch{Gd}/\ch{IrMn} interface is expected to have negligible exchange coupling \cite{Ali2008}). Moreover, its thickness is another parameter that we can tune to optimize our stack.

\textbf{Inducing exchange bias} - Instead of directly pinning the \ch{Co}/\ch{Gd}/\ch{Co} trilayer to \ch{IrMn} we opted for a [\ch{Pt}/\ch{Co}]$_\mathrm{x2}$ multilayer to mediate the coupling. This allowed us a much wider window for tuning the \ch{Gd} and \ch{Co} layer thicknesses. The \ch{Pt} layers were \SI{1.25}{\nano\meter} thick and the \ch{Co} layers \SI{0.6}{\nano\meter} thick. These values and the number of repeats were obtained from literature to produce maximized exchange bias \cite{Sort2005,Yang2019}.

As mentioned in the main text, the \ch{IrMn} thickness is considered a parameter that we are able to continuously vary by means of a wedged film. In the range between \SIrange[range-units=single,range-phrase=\textrm{ and }]{0}{15}{\nano\meter} the onset of exchange bias is captured, as well as the saturation of \HEB{} as the thickness increases.

\begin{figure}%
	\includegraphics[width=0.5\columnwidth]{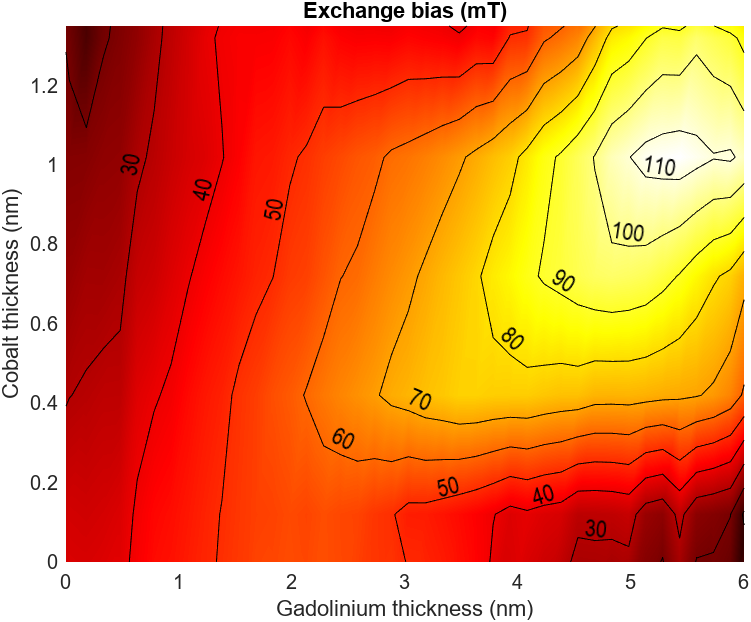}%
	\caption{\label{fig:cogdmap}Map of the exchange bias field for varying thicknesses of \ch{Gd} and \ch{Co}.}%
\end{figure}

\textbf{Optimizing \ch{Co}/\ch{Gd} thicknesses} - For optimizing the thicknesses of the \ch{Gd} and \ch{Co} layers comprising the optically switchable layers, a sample similar to the one from the main text (Fig.~\ref{fig:1}) was grown, but where the \ch{Gd}(5.5)/\ch{Co}(1) layers were wedged. The \ch{Gd} thickness along the wedge varied from \SIrange[range-units=single,range-phrase=\textrm{ to }]{0}{6}{\nano\meter}. The \ch{Co} wedge was oriented at \SI{90}{\degree} with respect to the \ch{Gd} wedge, such that each point on the sample corresponds to a unique combination of \ch{Gd} and \ch{Co} thicknesses. This allowed us to map \HEB{} as a function of both \ch{Gd} and \ch{Co} layer thickness. The measurement was carried out with a polar MOKE setup which measured the ($M$-$H$) hysteresis curves at a grid of positions. Subsequently, each hysteresis curve was used to extract the loop shift and hence estimate \HEB{}. The map is shown in Fig.~\ref{fig:cogdmap}. It can be seen that there is a clear peak visible at values of \SI{5.5}{\nano\meter} \ch{Gd} thickness and \SI{1.0}{\nano\meter} \ch{Co} thickness. These were the thicknesses that were used in our final stack%
.

It may be seen as surprising that changing the \ch{Gd} thickness beyond a few nanometers still changes the magnetostatic properties, since the bulk Curie temperature of \ch{Gd} lies just below room temperature at $T_C=\SI{292}{\kelvin}$ and hence all of its magnetic moment is induced via the \ch{Co}. Partially, this behavior may be explained as intermixing of \ch{Co} sputtered on top of \ch{Gd}, since in that case the antiferromagnetic coupling between the $3d$-{\ch{Co}} and $5d$-{\ch{Gd}} orbitals could give rise to a higher overall $T_C$ of the intermixed region. For very thick \ch{Gd} layers this interpretation based purely on the effect of direct exchange is insufficient though. It is speculated the effects of higher order excitations, which also play an essential role in describing finite size effects in this type of ultrathin system\cite{Zhang2001}, need to be taken into account to fully describe the effect of the \ch{Gd} thickness on $T_C$. However, a full discussion of this topic is beyond the scope of this manuscript.
%
%, similar to earlier work on coupled ferromagnets with different $T_C$ [1,2,3,4]. Recent evidence of an RKKY-like longer-range exchange may also play an important role here [5,6]. %

\section{Spot size characterization}\label{sec:spotsize}
\begin{figure}%
	\includegraphics[width=0.7\columnwidth]{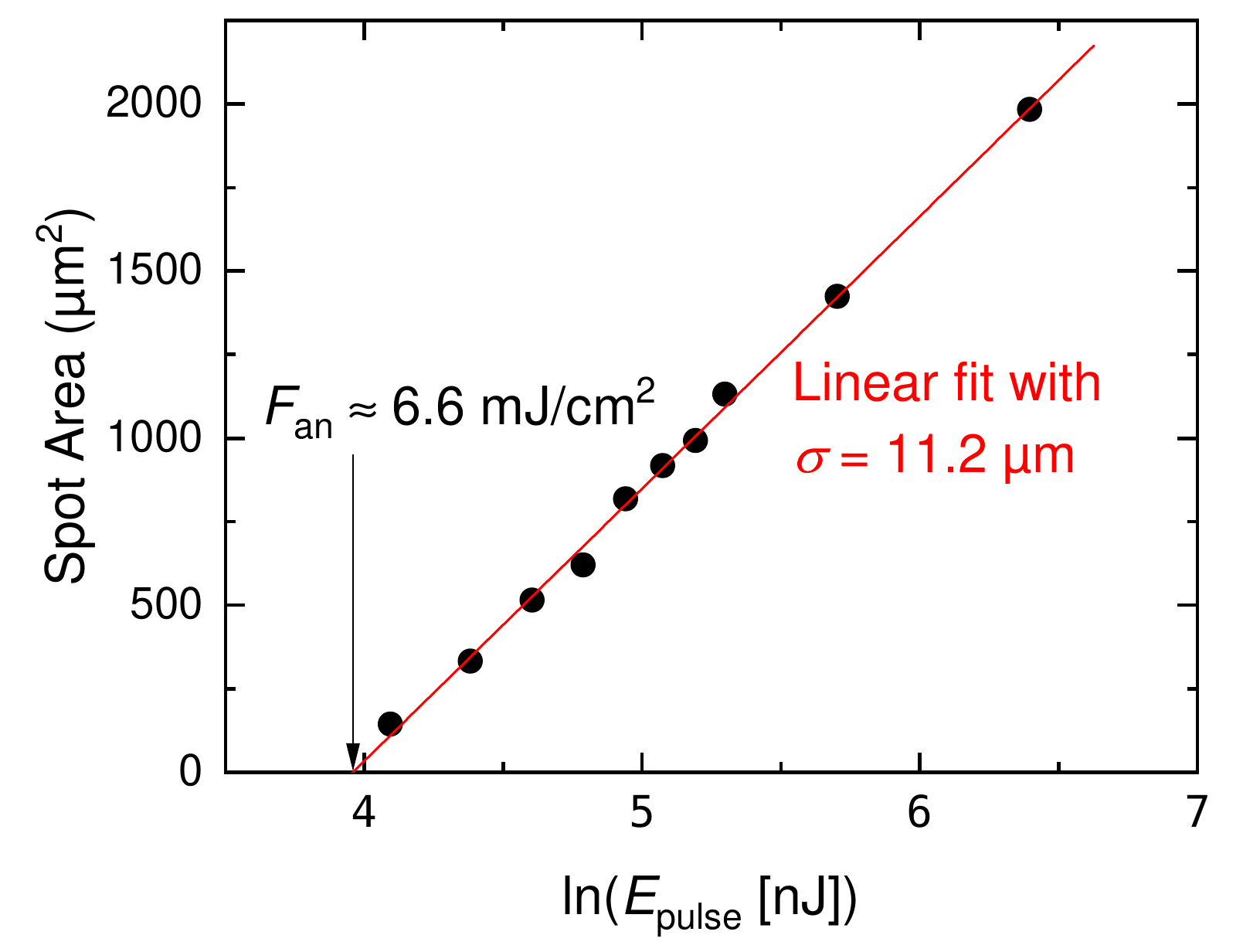}%
	\caption{\label{fig:spotsize}Linear fit of the spot area at zero field as a function of the natural logarithm of the laser pulse energy $E_{\mathrm{pulse}}$. The slope is equal to $2\pi\sigma^2$ with $\sigma=\SI{11.2}{\micro\meter}$ and the zero-crossing gives a value of $F_{\mathrm{an}}=\SI{6.6}{\milli\joule\per\square\centi\meter}$ for the annealing threshold.}%
\end{figure}%
The spot size of the laser pulse can be determined from Kerr microscopy images after exciting the sample with various laser pulse energies $E_{\mathrm{pulse}}$ (related to the measured power $P$ by the repetition rate $\nu=P/E_{\mathrm{pulse}}$). If a Gaussian with standard deviation $\sigma$ is assumed for the spatial distribution of the fluence
\begin{equation}
	F(r)=\frac{E_{\mathrm{pulse}}}{2\pi\sigma^2}\exp\left(-\frac{r^2}{2\sigma^2}\right),
\end{equation}
then it is readily seen that the area $\pi r^2$ above a certain threshold fluence depends linearly on the natural logarithm of the power $P$, and the slope is equal to $2\pi\sigma^2$. An example of such a linear dependency is shown in Fig.~\ref{fig:spotsize} and gives a value of $\sigma=\SI{11.2}{\micro\meter}$. This corresponds to a full width of $\SI{26}{\micro\meter}$ at half maximum. For the data in Fig.~\ref{fig:spotsize}, the Kerr images of the spots are taken at zero field after negative saturation. Note that it is important to keep the background field constant, because the spot grows or shrinks when the field changes. This in turn is because of the inhomogeneous distribution of switching fields as explained in Fig.~\ref{fig:2} of the main text. If the spots are imaged at zero field after negative saturation, the boundary of the switched area would correspond with the annealing threshold $F_{\mathrm{an}}$, which for the data in Fig.~\ref{fig:spotsize} gives a value of around \SI{6.6}{\milli\joule\per\square\centi\meter}. This matches with the experimental data (compare for example with Fig.~\ref{fig:4}a of the main text).

%\section{Mapping exchange bias}
%Show series of images from Kerr microscope with contrast rel. to H=0. Take images, %add image at H=0 and subtract the image at H=Hmax. Then show EB extraction procedure.
%
\section{Implementation of the model}\label{sec:model}
Our model of exchange bias reversal describes magnetization dynamics on two distinct timescales. The first part describes longitudinal switching of the ferro-/ferrimagnetic layers directly after the excitation by the laser. We run the simulation until the system temperature drops below the N\'{e}el temperature of the antiferromagnet. Then, in the second part, we simulate the stochastic setting process of the exchange bias through thermal excitations between distinct energy levels. For this, we take the polycrystalline nature of the antiferromagnet into account by estimating the grain size distribution and calculate the energy barrier for each particular grain size. The equations and parameters are worked out in more detail below.

\subsection{Part I: Longitudinal switching}\label{sec:modelpart1}
The implementation of magnetization dynamics after laser pulse excitation closely follows the layered microscopic three-temperature model (M3TM) from Beens et al.\cite{Beens2019}. The system is modeled as a collection of interacting monoatomic layers (subsystems), each with magnetic properties representing the materials that are used in the experiments. Two types of interactions are modeled that govern the magnetization dynamics. Firstly, spin angular momentum relaxation to the lattice occurs via Elliot-Yafet (EY) electron-phonon scattering processes with a certain spin-flip probability. Secondly, nearest-neighbor exchange scattering (EX) is taken into account, which describes scattering events between two electrons exchanging spin angular momentum.

Heat enters the system via the electron bath by optical excitation from the laser pulse (fluence $P_{0}$, pulse width $\Gamma$), and it relaxes predominantly through phonon heat dissipation to the substrate with a characteristic time scale $\tau_{D}$. To include these highly non-equilibrium electron and phonon states into the model, the temperatures of the electrons ($T_{e}(t)$) and phonons ($T_{p}(t)$) are modeled separately. The two temperature baths exchange heat at a rate parameterized by $g_{ep}$. The temperature evolution then follows from a two-temperature model
\begin{eqnarray}\label{eq:2tm}
	\gamma T_{e}(t)\frac{\mathrm{d}T_{e}(t)}{\mathrm{d}t}=-g_{ep}(T_{e}(t)-T_{p}(t))+\frac{P_0}{\Gamma\sqrt{\pi}}\exp\left(-\frac{t^2}{\Gamma^2}\right),\\
	C_{p}\frac{\mathrm{d}T_{p}(t)}{\mathrm{d}t}=g_{ep}(T_{e}(t)-T_{p}(t))-C_{p}\frac{T_{\mathrm{amb}}-T_{p}(t)}{\tau_{D}},
\end{eqnarray}
where $\gamma T_{e}(t)$ and $C_{p}$ are the electron and phonon heat capacities (respectively) and $T_{\mathrm{amb}}$ is the ambient temperature. It is assumed that every layer experiences the same volumetric fluence, a claim that is backed up by calculations of the absorption per layer in \ref{sec:skindepth}.

\begin{figure}%
	\includegraphics[width=0.5\columnwidth]{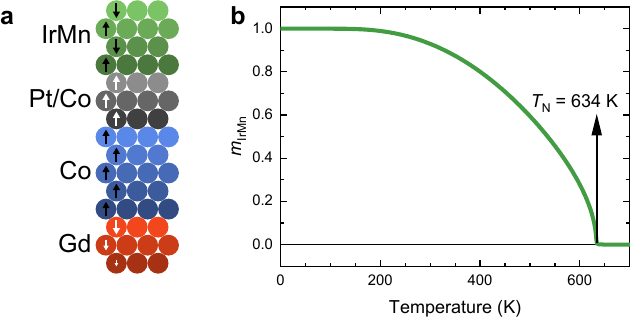}%
	\caption{\label{fig:weiss}\textbf{a} Overview of the layer structure and magnetization directions as used in the M3TM simulations. \textbf{b} The sublattice magnetization of \ch{IrMn} versus temperature from the layered Weiss model.}%
\end{figure}

Because \ch{Gd} is classified as a ferromagnet with behavior corresponding to a spin quantum number of $S_{\ch{Gd}}=7/2$ \cite{Coey2001} (whereas for \ch{Co} and \ch{Mn} the spin quantum number is best modeled as $S_{\ch{Co}}=S_{\ch{Mn}}=1/2$), each of the $2S_{i}+1$ spin levels $s\in\lbrace-S_{i},-S_{i}+1,\ldots,S_{i}\rbrace$ must be simulated separately to account for all possible transitions during exchange scattering processes. The occupation fraction of spin level $s$ for subsystem $i$ is $f_{s,i}$, and the normalized magnetization is given by the weighted sum
\begin{equation}
	m_{i}=-\frac{1}{S_{i}}\sum_{s=-S_{i}}^{S_{i}}sf_{s,i}\quad\textrm{with}\quad\sum_{s=-S_{i}}^{S_{i}}f_{s,i}=1.
\end{equation}
The spin levels have energies separated by an exchange splitting $\Delta_{i}$, i.e. the energy difference between level $s$ and $s+1$. Each subsystem $i$ experiences a magnetic exchange interaction with its neighboring subsystems ($j_{i,i-1}$ and $j_{i,i+1}$) and also from within its own subsystem ($j_{i,i}$). Each $j$ is chosen in accordance with the Weiss model and the Curie temperatures $T_{C,i}$ associated with bulk systems of the different materials. If a closed-packed structure with $(111)$ texture is assumed, then each atom has 6 nearest-neighbors within the same layer and 3 nearest-neighbors in each adjacent layer. Altogether, the exchange splitting then becomes
\begin{equation}
	\Delta_{i}=\frac{j_{i,i-1}}{4}m_{i-1}+\frac{j_{i,i}}{2}m_{i}+\frac{j_{i,i+1}}{4}m_{i+1}+2\mu_{\mathrm{B}}B,
\end{equation}
where the last term accounts for the Zeeman splitting (Bohr magneton $\mu_{\mathrm{B}}$ and magnetic flux density $B$). The full rate equation as derived by Koopmans et al.\cite{Koopmans2010} and Beens et al.\cite{Beens2019} is given by
\begin{equation}\label{eq:m3tm}
	\frac{\mathrm{d}f_{s,i}}{\mathrm{d}t}=\left.\frac{\mathrm{d}f_{s,i}}{\mathrm{d}t}\right|_{\mathrm{EY}}+\left.\frac{\mathrm{d}f_{s,i}}{\mathrm{d}t}\right|_{\mathrm{EX},i-1}+\left.\frac{\mathrm{d}f_{s,i}}{\mathrm{d}t}\right|_{\mathrm{EX},i+1},
\end{equation}
where the full expressions of the terms on the right hand side are given in the supplementary materials of the respective references. This leads to an equation of the form $\mathrm{d}m_{i}/\mathrm{d}t=R(m_{i-1},m_{i},m_{i+1},T_{e},T_{p})$, Eq.~\ref{eq:m3tmmain} in the main paper. The initial conditions $m_{i,0}$ for solving Eq.~\ref{eq:m3tm} follow from Boltzmann statistics and are obtained from solving
\begin{equation}\label{eq:initial}
	m_{i,0}=\frac{1}{S_{i}}\frac{\sum_{s=-S_{i}}^{S_{i}}s\exp\left(s\frac{\Delta_{i}}{k_{\mathrm{B}}T_{\mathrm{amb}}}\right)}{\sum_{s=-S_{i}}^{S_{i}}\exp\left(s\frac{\Delta_{i}}{k_{\mathrm{B}}T_{\mathrm{amb}}}\right)}.
\end{equation}
The stack we simulate closely resembles the one used in the experiments and encompasses $15$ atomic layers in total (see Fig.~\ref{fig:weiss}a). The antiferromagnet \ch{IrMn} is modeled as a layered ferromagnet with negative interlayer exchange coupling $j_{\ch{IrMn},\ch{IrMn}}<0$, such that $m_{i}$ alternates from positive to negative between layers. Four atomic layers of \ch{IrMn} are simulated in order to have a decent separation between the exchange bias interface and the edge of the system. Zeeman splitting is set to zero for all \ch{IrMn} layers. The \ch{Pt}/\ch{Co} multilayer in direct contact with the \ch{IrMn} is modeled as a single ferromagnet (labeled \ch{PtCo}) with three atomic layers, based on \SI{0.6}{\nano\meter} thickness with a typical atomic radius of \SI{0.2}{\nano\meter}. The \ch{PtCo} system is coupled by a weaker direct exchange interaction to the \SI{1}{\nano\meter} \ch{Co} layer below, which comprises five atomic layers. At the bottom, three \ch{Gd} layers are simulated whose magnetization will be induced by the coupling at the \ch{Co} interface. Because this induced magnetization decays very rapidly away from the \ch{Co} interface, adding more than three layers of \ch{Gd} does not significantly change the results.

Coercivity or anisotropy is not encapsulated in this model. To still account for the influence of $B$ on $m_{i,0}$, it is assumed that a ferromagnetic layer prefers to orient itself along the field direction if the Zeeman energy overcomes the interlayer exchange coupling for fully saturated layers with $m_{i}=1$, i.e., assuming a positive orientation for $\Delta_i>0$ and negative for $\Delta_i<0$. Subsequent equilibration of the system energy by Eq.~\ref{eq:initial} ensures that the lowest energy state is chosen from the two possible initial states, i.e., with \ch{Co} up and \ch{Gd} down or vice versa.

The M3TM simulation may be run on its own to describe the magnetization dynamics. For describing the exchange bias setting process, the state of the M3TM simulation when $T_{e}(t)$ drops below the N\'{e}el temperature of the \ch{IrMn} layers is used as the initial state for the stochastic heat-driven processes. From the layered Weiss model and Eq.~\ref{eq:initial}, the N\'{e}el temperature amounts to \SI{634}{\kelvin} (see Fig.~\ref{fig:weiss}b%
). At this point, magnetic order is established in the \ch{IrMn} and the slower thermally assisted exchange bias setting process commences.

\subsection{Part II: Exchange bias setting}\label{sec:modelpart2}
\begin{figure}%
	\includegraphics[width=\columnwidth]{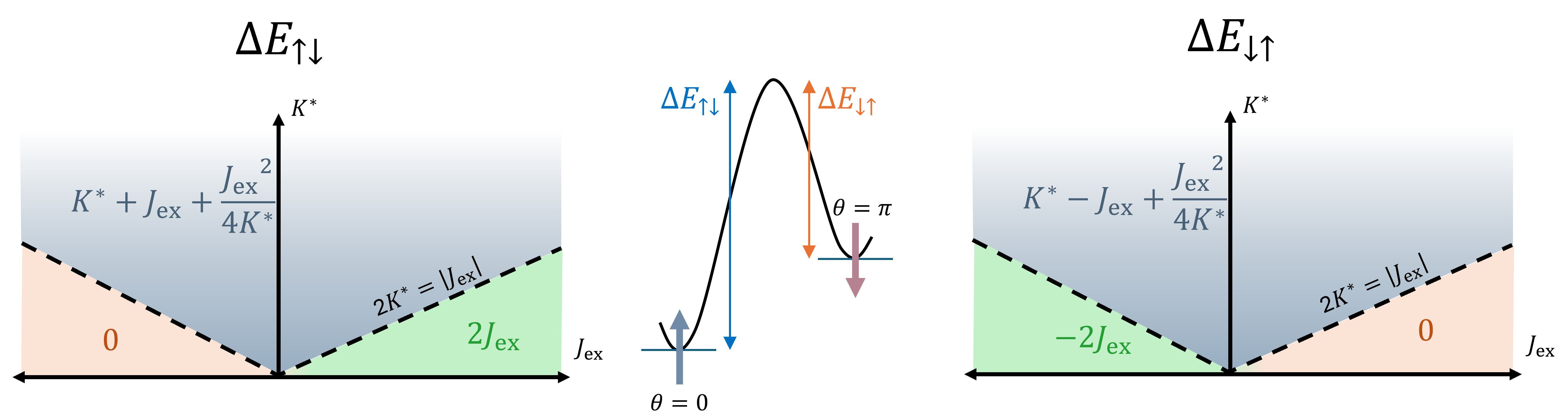}%
	\caption{\label{fig:deltae}Energy barrier calculation, following Eq.~\ref{eq:afmenergy} with $K^{*}\equiv K_{\mathrm{AFM}}t_{\ch{IrMn}}$.}%
\end{figure}%
In this part of the simulation, the antiferromagnet is modeled by the parameter $m_{\mathrm{AFM}}(t,A)\in[-1,1]$ as a collection of columnar grains with distributed areas $A$ and thickness $t_{\ch{IrMn}}$. The time $t$ and area $A$ dependence will henceforth be dropped for readability, but are always implicitly assumed. $m_{\mathrm{AFM}}$ is a population proportion and can be written as $m_{\mathrm{AFM}}=N_{\uparrow}-N_{\downarrow}$, where $N_{\uparrow}$ and $N_{\downarrow}$ are the proportions of grains with area $A$ that respectively have a positive or negative contribution to the perpendicular exchange bias. These proportions change over time under the influence of thermal excitations, with transition rates given by Eq.~(1) in the Methods. The rate equations for $N_{\uparrow}$ and $N_{\downarrow}$ can be summed into a single rate equation for $m_{\mathrm{AFM}}$ given by
\begin{equation}\label{eq:afmrateeq}
	\frac{\mathrm{d}m_{\mathrm{AFM}}}{\mathrm{d}t}=\frac{2}{\tau_{0}}\left[\frac{1-m_{\mathrm{AFM}}}{2}\exp\left(-\frac{\Delta E_{\downarrow\uparrow}}{k_{\mathrm{B}}T}\right)-\frac{1+m_{\mathrm{AFM}}}{2}\exp\left(-\frac{\Delta E_{\uparrow\downarrow}}{k_{\mathrm{B}}T}\right)\right].
\end{equation}
The energy barrier $\Delta E$ is calculated for a grain with area $A$ by taking into account the perpendicular antiferromagnetic anisotropy $K_{\mathrm{AFM}}$ and the exchange coupling $J_{\mathrm{ex}}$ with the ferromagnet. If $\theta$ is the polar angle of the exchange bias direction, then an expression for the total energy per unit area becomes
\begin{equation}\label{eq:afmenergy}
	\frac{E_{\mathrm{AFM}}}{A}=-K_{\mathrm{AFM}}t_{\ch{IrMn}}\cos^{2}\theta-J_{\mathrm{ex}}\cos\theta.
\end{equation}
Figure~\ref{fig:deltae} shows how the energy barrier changes for various combinations of $K_{\mathrm{AFM}}$ and $J_{\mathrm{ex}}$. The parameters $K_{\mathrm{AFM}}$ and $J_{\mathrm{ex}}$ themselves depend on the state (temperature and magnetization) that the system is in. Heat is assumed to dissipate on the long timescales (i.e., long after $T_{e}$ and $T_{p}$ have equilibrated) according to a one-dimensional heat kernel ($\propto t^{-1/2}$). At any time $t$, we evaluate the temperature by\cite{DallaLonga2007}
\begin{equation}\label{eq:heatkernel}
	T(t)=T_{\mathrm{amb}}+\frac{T_{f}-T_{\mathrm{amb}}}{\sqrt{\frac{t-t_{0}}{\tau_{D}}+1}},
\end{equation}
with $t_{0}$ such that $T(t=0)=T_{N}$ and $T_{f}$ the peak temperature reached directly after equilibration of $T_{e}$ and $T_{p}$. $T_{f}$ can be worked out by analytically solving Eq.~\ref{eq:2tm} in the limit $\Gamma\ll\tau_{D}$, and $t_{0}$ by solving $T(0)=T_{N}$. This gives
\begin{eqnarray}
	T_{f}=\frac{C_{p}}{\gamma}\left[\sqrt{\left(\frac{\gamma T_{\mathrm{amb}}}{C_{p}}\right)^{2}+\frac{2\gamma P_{0}}{C_{p}^{2}}}-1\right],\\
	t_{0}=\tau_{D}\left[1-\left(\frac{T_{f}-T_{\mathrm{amb}}}{T_{N}-T_{\mathrm{amb}}}\right)^2\right].
\end{eqnarray}
The temperature $T(t)$ is fed back into the layered Weiss model (Eq.~\ref{eq:initial}) at every simulation time step to work out the equilibrium magnetization $m_{i,0}$ at that particular temperature. The anisotropy and exchange coupling are also adapted accordingly, for which we assume
\begin{equation}
	K_{\mathrm{AFM}}=K_{\mathrm{AFM},0}\cdot m_{\ch{IrMn},0}^{2}\qquad\textrm{and}\qquad J_{\mathrm{ex}}=J_{\mathrm{\mathrm{ex}},0}\cdot m_{\mathrm{F}}\cdot m_{\ch{IrMn},0}\cdot m_{\ch{PtCo},0},
\end{equation}
where $K_{\mathrm{AFM},0}$ is the antiferromagnetic anisotropy at $T=0$ and $J_{\mathrm{ex},0}$ is the exchange coupling for a fully saturated ferromagnet and antiferromagnet at $T=0$. $m_{F}=\pm1$ captures the state of the ferromagnet after Part I of the simulation.

\begin{figure}%
	\includegraphics[width=0.9\columnwidth]{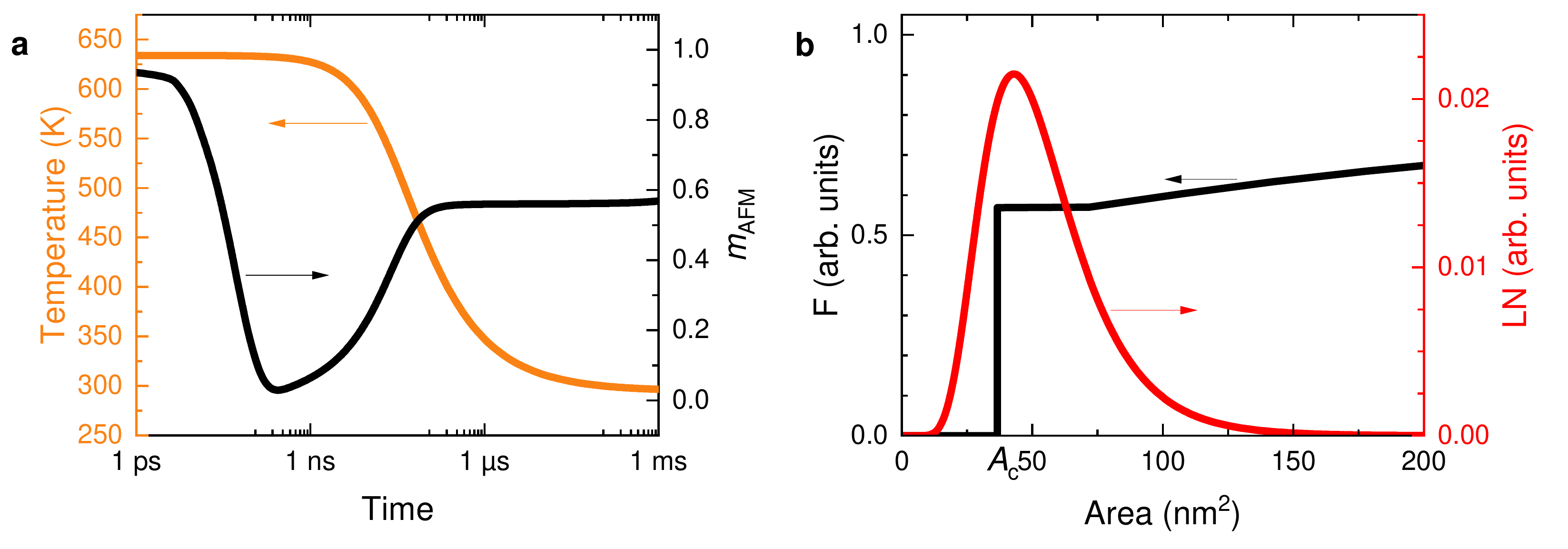}%
	\caption{\label{fig:simpart2}\textbf{a} Plot of the temperature evolution from Eq.~\ref{eq:heatkernel} in orange and the simulated parameter $m_{\mathrm{AFM}}$ in black according to Eq.~\ref{eq:afmrateeq}, for $t_{\ch{IrMn}}=\SI{5}{\nano\meter}$ and a grain area equal to the critical area $A_c$. \textbf{b} Examples of the grain area distribution function $\mathrm{LN}(A)$ in red and the switching fraction distribution function $F(A)$ in black for $t_{\ch{IrMn}}=\SI{5}{\nano\meter}$ as used in Eq.~\ref{eq:integral} for working out the \HEB{} reversal fraction. For instance, the value of $F(A)$ at $A_c$ corresponds to the value of $m_{\mathrm{AFM}}$ at $t=\SI{1}{\milli\second}$ in \textbf{a}. The laser fluence used in both \textbf{a} and \textbf{b} is $P_0=\SI{40e8}{\joule\per\cubic\meter}$.}%
\end{figure}

The simulation (Eq.~\ref{eq:afmrateeq}) is run over a time span of \SI{1}{\milli\second}, after which the temperature has relaxed sufficiently close to room temperature and $m_{\mathrm{AFM}}$ is no longer significantly changing. An example of such a simulation for $P_0=\SI{40e8}{\joule\per\cubic\meter}$, $t_{\ch{IrMn}}=\SI{5}{\nano\meter}$ and $A=\SI{36.6}{\square\nano\meter}$ is shown in Fig.~\ref{fig:simpart2}a. The figure shows the evolution for the initial value condition $m_{\mathrm{AFM}}(t=0)=1$ under the assumption that the ferromagnetic orientation $m_{F}$ directly imprints a preferential direction into the antiferromagnet when the temperature drops below $T_N$. However, we found that the initial condition does not significantly affect the outcome at $t=\SI{1}{\milli\second}$ due to the fact that the temperature still exceeds $T_b$. In the evolution of $m_{\mathrm{AFM}}$, this is visible as a quick drop towards zero followed by an increase as a response to the lowering temperature. If $m_{\mathrm{AFM}}(t=0)=0$ was chosen, the difference with $m_{\mathrm{AFM}}(t=0)=1$ will be negligible after \mytilde\SI{400}{\pico\second}.

The state at $t=\SI{1}{\milli\second}$ is the switching fraction for this particular grain area and is recorded for all possible grain areas $A$, producing a distribution of switching fractions which we will call $F(A)$. An example for $t_{\ch{IrMn}}=\SI{5}{\nano\meter}$ is shown in Fig.~\ref{fig:simpart2}b. Following Ref. \citenum{Khamtawi2023}, it is assumed that the grain areas are distributed according to a log-normal distribution $\mathrm{LN}(A)$ with median area $\mu=\pi(\SI{4}{\nano\meter})^{2}$ and standard deviation $\sigma_{\ln A}=0.4$ \cite{Khamtawi2023},
\begin{equation}
	\mathrm{LN}(A)=\frac{1}{A\sigma_{\ln A}\sqrt{2\pi}}\exp\left(-\frac{\ln(A/\mu)}{2\sigma_{\ln A}^{2}}\right),
\end{equation}
also plotted in Fig.~\ref{fig:simpart2}b. It must be taken into account that the smallest grains are not able to contribute to the exchange bias because they are superparamagnetic. The critical grain area that corresponds to this is given by
\begin{equation}
	A_{c}=25\frac{k_{\mathrm{B}}T_{\mathrm{amb}}}{K_{\mathrm{AFM}}t_{\ch{IrMn}}}.
\end{equation}
Finally, taking the grain size distribution into account, the fraction of exchange bias remaining (relative to the maximum value $H_{\mathrm{EB,max}}$ for the case where $F(A)\equiv1$ for all $A$) after the laser pulse excitation is given by
\begin{equation}\label{eq:integral}
	\frac{H_{\mathrm{EB}}}{H_{\mathrm{EB,max}}}=\int_{A_{c}}^{\infty}\mathrm{LN}(A)F(A)\mathrm{d}A.
\end{equation}
The quantity on the left hand side is the output of the simulation and may then be further examined, e.g., as a function of fluence or magnetic field. Figure~\ref{fig:simpart2}b %
shows examples of the grain area distribution function $\mathrm{LN}(A)$ and the fraction distribution function $F(A)$ as simulated for a fluence of $P_0=\SI{40e8}{\joule\per\cubic\meter}$.

To conclude this modeling section, Tables~\ref{table:params}, \ref{table:params2} and \ref{table:params3} %
\begin{table*}%
	\caption{\label{table:params}Parameters as used for the two-temperature model (Eq.~\ref{eq:2tm}) in Part I of the simulation. The same numerical values are used as in Ref.~\citenum{Beens2019}.}%
	\begin{ruledtabular}%
		\begin{tabular}{llllll}%
			$\Gamma$ [\si{\pico\second}] & $C_p$ [\si{\milli\joule\per\cubic\centi\meter\per\kelvin}] & $\gamma$ [\si{\milli\joule\per\cubic\centi\meter\per\square\kelvin}] & $g_{ep}$ [\si{\milli\joule\per\cubic\centi\meter\per\kelvin\per\pico\second}] & $T_{\mathrm{amb}}$ [\si{\kelvin}] & $\tau_{D}$ [\si{\pico\second}] \\ \hline%
			\num{0.05} & \num{4e3} & \num{2.0} & \num{4.05e3} & \num{295} & \num{20} \\%
		\end{tabular}%
	\end{ruledtabular}%
	\caption{\label{table:params2}Parameters as used for Part I (Sec.~\ref{sec:modelpart1}) of the simulation, for each material system in the stack. $\mu_{\mathrm{at}}$ is the mean magnetic moment per atom and $R$ is the transition rate constant for EY scattering. These quantities are used in Eq.~\ref{eq:m3tm} (for the full expressions, see Ref.~\citenum{Beens2019}). The exchange couplings $j_{i,i}$ are directly calculated from $T_{C}$ according to the Weiss model.}%
	\begin{ruledtabular}%
		\begin{tabular}{lllllllll}%
			$i$ & $R$ [\si{\per\pico\second}] & $\mu_{\mathrm{at}}$ [$\mu_{\mathrm{B}}$] & $T_{C}$ [\si{\kelvin}] & $S$ & $j_{i,\ch{Gd}}$ [\si{\milli\joule}] & $j_{i,\ch{Co}}$ [\si{\milli\joule}] & $j_{i,\ch{PtCo}}$ [\si{\milli\joule}] & $j_{i,\ch{IrMn}}$ [\si{\milli\joule}] \\ \hline%
			\ch{Gd} & \num{0.092}\footnotemark[1] & \num{7.55}\footnotemark[7] & \num{292}\footnotemark[2] & $7/2$\footnotemark[2] & \num{2.69e-18} & \num{-3.39e-18}~\footnotemark[6] & - & - \\%
			\ch{Co} & \num{25.3}\footnotemark[1] & \num{1.72}\footnotemark[1] & \num{1388}\footnotemark[2] & $1/2$\footnotemark[2] & \num{-1.49e-17}~\footnotemark[6] & \num{3.83e-17} & \num{2.76e-19}~\footnotemark[5] & - \\%
			\ch{PtCo} & \num{25.3}\footnotemark[3] & \num{1.72}\footnotemark[3] & \num{1000}\footnotemark[4] & $1/2$\footnotemark[3] & - & \num{2.76e-19}~\footnotemark[5] & \num{2.76e-17} & \num{1.24e-20}~\footnotemark[8] \\%
			\ch{IrMn} & \num{3.2}\footnotemark[9] & \num{2.60}\footnotemark[10] & \num{700}\footnotemark[11] & $1/2$\footnotemark[10] & - & - & \num{8.23e-21}~\footnotemark[8] & \num{1.96e-17} \\%
		\end{tabular}%
	\end{ruledtabular}%
	\footnotetext[1]{From Ref.~\citenum{Koopmans2010}.}%
	\footnotetext[2]{From Ref.~\citenum{Beens2019}.}%
	\footnotetext[3]{Chosen the same as for \ch{Co}.}%
	\footnotetext[4]{The \ch{PtCo} Curie temperature is chosen slightly below the bulk value to incorporate influences from size effects.}%
	\footnotetext[5]{Based on an interlayer exchange coupling strength that is \SI{1}{\percent} of the \ch{PtCo} intralayer exchange coupling.}%
	\footnotetext[6]{$j_{\ch{Co},\ch{Gd}}=-0.388j_{\ch{Co},\ch{Co}}$, $j_{\ch{Gd},\ch{Co}}=j_{\ch{Co},\ch{Gd}}\cdot\mu_{\mathrm{at},\ch{Co}}/\mu_{\mathrm{at},\ch{Gd}}$.\cite{Gerlach2017}}%
	\footnotetext[7]{From Ref.~\citenum{Bovensiepen2007}}%
	\footnotetext[8]{From experimental values: $H_{\mathrm{EB}}=\SI{50}{\milli\tesla}$ for $3$ atomic \ch{Co} layers, assuming an fcc ordering with $z=12$ nearest neighbors of which $3$ are \ch{Co} for each \ch{Mn} atom. Then: $j_{\ch{PtCo},\ch{IrMn}}=\frac{z}{3}\cdot(3\mu_{\mathrm{at},\ch{Co}}\mu_{\mathrm{B}}\mu_{0}H_{\mathrm{EB}})\cdot\mu_{\ch{IrMn}}/2$, $j_{\ch{IrMn},\ch{PtCo}}=j_{\ch{PtCo},\ch{IrMn}}\cdot\mu_{\mathrm{at},\ch{PtCo}}/\mu_{\mathrm{at},\ch{IrMn}}$.}%
	\footnotetext[9]{Based on a material density of \SI{10835}{\kilogram\per\cubic\meter}, an atomic mass of \SI{54.938}{u} for \ch{Mn}, similar electronic properties (density of states and spin-flip probability) as for \ch{Co}. Using parameters from the table and Eq.~19 from the supplementary of Ref. \citenum{Koopmans2010} gives the presented value.}%
	\footnotetext[10]{From Ref.~\citenum{Jenkins2021}.}%
	\footnotetext[11]{N\'{e}el temperature of bulk \ch{Ir_{0.2}Mn_{0.8}}.}%
	\caption{\label{table:params3}Parameters as used for the exchange bias setting in Part II (Sec.~\ref{sec:modelpart2}) of the simulation.}%
	\begin{ruledtabular}%
		\begin{tabular}{lllll}%
			$\tau_{0}$ [\si{\pico\second}] & $\tau_{D}$ [\si{\pico\second}] & $T_{N}$ [\si{\kelvin}] & $K_{\mathrm{AFM},0}$ [\si{\milli\joule\per\cubic\centi\meter}] & $J_{\mathrm{ex},0}$ [\si{\milli\joule\per\square\centi\meter}] \\ \hline%
			\num{100} & \num{5000} & \num{634}\footnotemark[1] & \num{5.56e8}~\footnotemark[2] & \num{1.9e-5}~\footnotemark[3] \\%
		\end{tabular}%
	\end{ruledtabular}%
	\footnotetext[1]{From Fig.~\ref{fig:weiss}b.}%
	\footnotetext[2]{From Ref.~\citenum{Khamtawi2023}.}%
	\footnotetext[3]{From Ref.~\citenum{Coey2001}.}%
\end{table*}%
respectively list the thermodynamic, micromagnetic and exchange bias setting parameters (constant or derived) as used in the two parts of the simulations.

\subsection{Linking between Part I and Part II}\label{sec:modellink}
To summarize, Part I of the simulation models the longitudinal ultrafast deterministic AOS process with the M3TM, and Part II models the stochastic thermally-assisted setting process of the exchange bias following the Arrhenius law. The two parts are linked in the sense that we use the output of Part I as an input for Part II. Mainly, this comes down to the value of $T_{f}$, which depends on the laser power and the heat capacities of the layers, and the value of $m_{\mathrm{F}}$, which is assumed to be $\pm1$ with the sign depending on whether or not AOS has taken place. On the other hand, the initial condition of $m_{\mathrm{AFM}}$ has no significant effect on the long term as it is automatically equilibrated under the influence of temperature. Since the simulation is forced to start only when the temperature has dropped below the N\'{e}el temperature of the \ch{IrMn}, we assume there is already some preference imprinted into the antiferromagnet given by the ferromagnetic state $m_{F}$. Therefore, we choose to set $m_{\mathrm{AFM}}=m_{\mathrm{F}}$ if $T_f>T_N$, otherwise we assume no effect on the \ch{IrMn} and we set $m_{\mathrm{AFM}}=1$ to emulate the as-deposited fully aligned orientation of the grains.

\section{Optical absorption effects}\label{sec:skindepth}
As evident from Fig.~\ref{fig:3}a of the main text, laser light attenuation effects are relevant for determining the absorbed energy in each of the layers. This can be estimated by the transfer matrix method \cite{Pettersson1999}. We use a wavelength of \SI{700}{\nano\meter} and refractive indices for the different materials as given in table \ref{table:rindex}. %
\begin{table}%
	\caption{\label{table:rindex}Refractive indices as used for the Transfer Matrix Method calculation.}%
	\begin{tabular}{ll}%
		\toprule%
		Material [\si{\nano\meter}] & Refractive index \\%
		\colrule%
		\ch{Si} & $3.70 + 0.005 i$ \\%
		\ch{SiO2} & $1.46 + 0.00 i$ \\%
		\ch{Ta} & $1.09 + 3.06 i$ \\%
		\ch{Pt} & $2.85 + 4.96 i$ \\%
		\ch{Co} & $2.50 + 4.84 i$ \\%
		\ch{Gd} & $2.40 + 2.88 i$ \\%
		\ch{IrMn} & $2.57 + 3.71 i$ \\%
		\botrule%
	\end{tabular}%
\end{table}%
The stack and substrate are identical to what is presented in the Methods section of the main text. For an \ch{IrMn} thickness of \SI{5}{\nano\meter}, the outcome of the calculated absorption rate is plotted in %
\begin{figure}%
	\includegraphics[width=0.7\columnwidth]{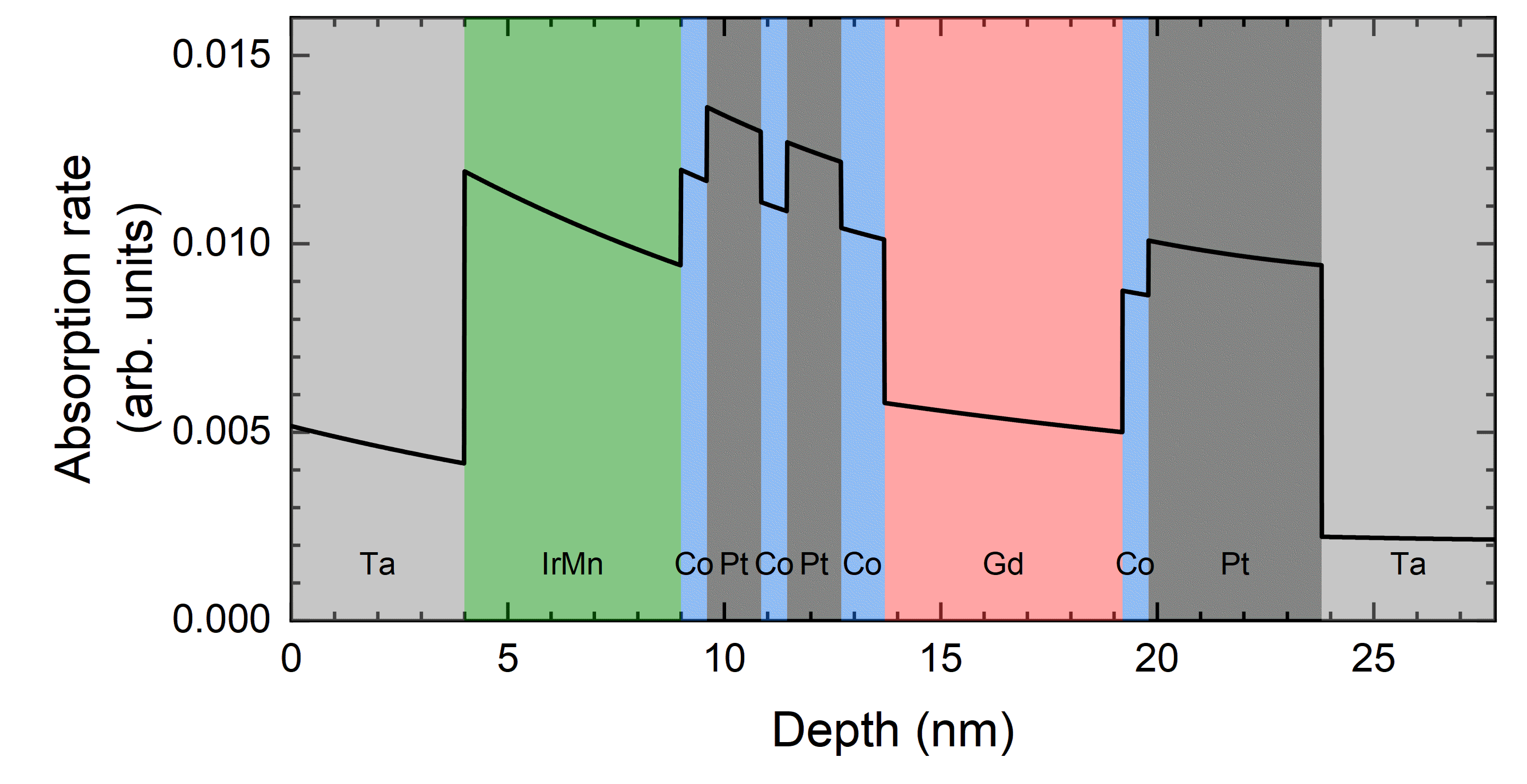}%
	\caption{\label{fig:tmm}Calculation of the absorption rate (with dimensions of reciprocal length) in each of the layers via the Transfer Matrix Method \cite{Pettersson1999}.}%
\end{figure}%
Fig.~\ref{fig:tmm}. Relative to each other, the absorptions per atomic layer lie within \mytilde\SI{10}{\percent}, as can be deduced from the absorption rates in Fig.~\ref{fig:tmm}. For this reason, it is valid to approximate the temperature as uniform throughout the stack at all times in the simulation.

\begin{figure}%
	\includegraphics[width=0.5\columnwidth]{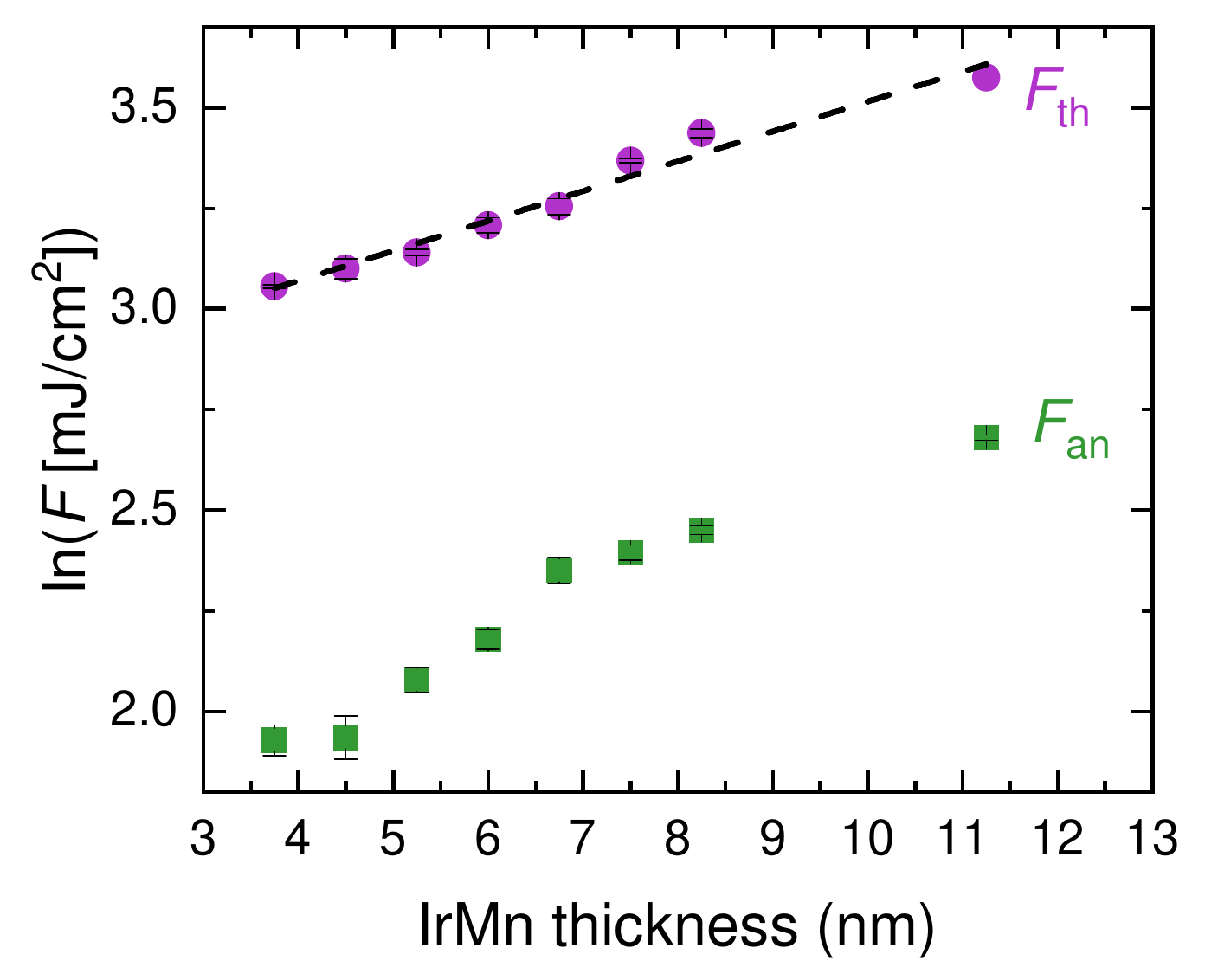}%
	\caption{\label{fig:attenuation}A plot of the natural logarithms of the threshold fluences $F_{\mathrm{an}}$ (green squares) and $F_{\mathrm{th}}$ (purple circles) as introduced in the main text, as a function of the \ch{IrMn} thickness $t_{\ch{IrMn}}$. The black dashed line is a linear fit of the trend in $\ln(F_{\mathrm{th}})$. It has a slope of \SI{0.074+-0.005}{\per\nano\meter}.}%
\end{figure}

The light attenuation is not only relevant for the simulations, but also for the experiments. In the main text (Fig.~\ref{fig:3}a) it is shown that the threshold fluence $F_{\mathrm{th}}$ increases for thicker \ch{IrMn} layers due to absorption. This is further quantified by a logarithmic plot of $F_{\mathrm{th}}$ versus \ch{IrMn} thickness in Fig.~\ref{fig:attenuation}. %
From a linear fit with slope \SI{0.074+-0.005}{\per\nano\meter}, the correction factor we extract to filter out the attenuation effect is given by%
\begin{equation}\label{eq:fan}%
	F_{\mathrm{an}}^{*}=F_{\mathrm{an}}\exp(-\SI{0.074}{\per\nano\meter}\cdot t_{\ch{IrMn}}).%
\end{equation}%
The error bars $\Delta_{\mathrm{an}}^{*}$ in the inset of Fig.~\ref{fig:3}a in the main text are calculated with%
\begin{equation}%
	\Delta_{\mathrm{an}}^{*}=\sqrt{\left[\Delta_{\mathrm{an}}\exp\left(-\SI{0.074}{\per\nano\meter}\cdot t_{\ch{IrMn}}\right)\right]^{2}+\left[\SI{0.005}{\per\nano\meter}\cdot F_{\mathrm{an}}^{*}t_{\ch{IrMn}}\right]^{2}},%
\end{equation}%
where $\pm\Delta_{\mathrm{an}}$ is the \SI{68}{\percent} uncertainty in the extraction of $F_{\mathrm{an}}$.

% 4*pi/(700nm)*3.71 = 0.067 nm^-1
Note that the value for the absorption coefficient as expected based on the complex refractive index for \ch{IrMn} in Table~\ref{table:rindex} (amounting to \SI{0.067}{\per\nano\meter} for a wavelength of \SI{700}{\nano\meter}) agrees well with the fitted value used in Eq.~\ref{eq:fan}.

\section{Blocking temperature}\label{sec:block}
\begin{figure}%
	\includegraphics[width=0.5\columnwidth]{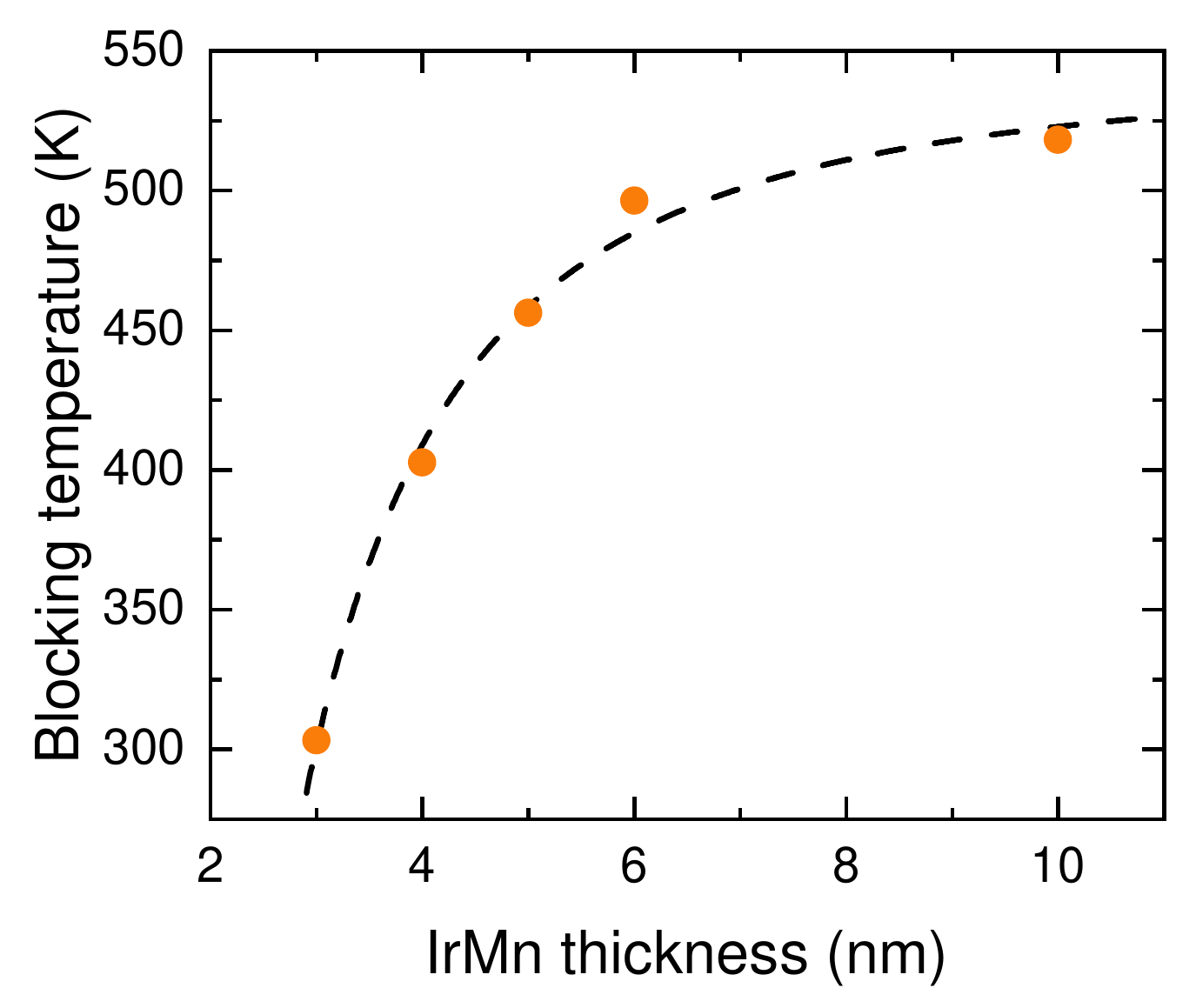}% Here is how to import EPS art
	\caption{\label{fig:tb}Blocking temperature of \ch{IrMn} as a function of $t_{\ch{IrMn}}$ The orange circles are measured data points, the black dashed line is a fit with Eq.~\ref{eq:tb}.}%
\end{figure}

Figure~3b of the main text claims a relation between $F_{\mathrm{an}}$ and the blocking temperature $T_{b}$. The measurement of $T_{b}$ was carried out with vibrating sample magnetometry (VSM) at various temperatures to locate the point at which the exchange bias vanishes. Because the stack from the main paper is engineered to have a very low magnetic moment, a different stack was measured with VSM to ensure a high enough signal. Five full film stacks were grown (with the same conditions as in the main text) consisting of \layer{Ta}{4}/\layer{Pt}{4}/\layer{Co}{0.6}/[\layer{Pt}{1.25}/\layer{Co}{0.6}]$_{\mathrm{x}9}$/\layer{IrMn}{$t_{\ch{IrMn}}$}/\layer{Ta}{7}, with $t_{\ch{IrMn}}=\SIlist[list-units=single,list-final-separator=\textrm{ or }, list-pair-separator=\textrm{, }]
{3;4;5;6;10}{\nano\meter}$. The data is shown in Fig.~\ref{fig:tb} %
along with a fit according to a phenomenological power law with variable exponent given by \cite{Lang2007}
\begin{equation}\label{eq:tb}
	T_{b}(t_{\ch{IrMn}})=T_{b}(\infty)\left[1-\left(\frac{\xi}{t_{\ch{IrMn}}}\right)^{\lambda}\right].
\end{equation}
Here, $T_{b}(\infty)$ is the theoretical limit of the blocking temperature for an infinitely thick antiferromagnetic layer and $\xi$ and $\lambda$ respectively control the width and steepness of the $T_{b}(t_{\ch{IrMn}})$ trend. Fitted numerical values are given by $T_{b}(\infty)=\SI{270+-19}{\degreeCelsius}$, $\xi=\SI{2.8+-0.1}{\nano\meter}$ and $\lambda=\num{2.1+-0.4}$.

\end{document}